\documentclass{tlp}
\usepackage{aopmath}
\newcommand{\ignore}[1]{}

\newtheorem{definition}{Definition}[section]
\newtheorem{example}{Example}[section]
\newtheorem{propositionS}{Proposition}[section]
\newtheorem{corollaryS}{Corollary}[section]

\newtheorem{remark}{Remark}[section]
\newtheorem{exampleApp}{Example}
\newtheorem{definitionApp}{Definition}

\usepackage{graphicx}
\usepackage{latexsym}
\usepackage[latin1]{inputenc}
\usepackage{color}
\usepackage{multicol}
\usepackage{amssymb,xspace}
\usepackage{epsfig}
\usepackage{subfigure}
\usepackage{enumerate}
\usepackage{hhline}
\usepackage[tworuled,noend,figure,linesnumbered,titlenotnumbered,inoutnumbered]{algorithm2e}

\usepackage{enumfrom}

\newcommand{\ImportSolution}{{\sf ImportSolution}\xspace}

\usepackage{tikz}
\usetikzlibrary{trees}
\usetikzlibrary{decorations}
\usetikzlibrary{arrows}
\usetikzlibrary{shapes}
\usetikzlibrary{fit}
\usetikzlibrary{backgrounds}
\usetikzlibrary{patterns}

\pgfdeclarelayer{background}
\pgfdeclarelayer{foreground}
\pgfsetlayers{background,main,foreground}

\newcommand{\nit}[1]{{\it #1}}
\newcommand{\boxtheorem}{\hfill $\Box$}
\newcommand{\U}{{\cal U}}
\newcommand{\Pe}{{\cal P}}
\renewcommand{\P}{{\cal P}}
\newcommand{\Ps}{{\mathfrak{P}}}

\newcommand{\G}{{\cal G}}

\renewcommand{\S}{{\cal S}}
\newcommand{\N}{{\cal N}}

\newcommand{\AC}{{\cal AC}}
\newcommand{\DB}{\nit{D}}
\newcommand{\D}{\nit{D}}
\newcommand{\IC}{\nit{IC}}
\newcommand{\n}{~{\it not}~}
\newcommand{\nn}{\nit{null}\xspace}
\newcommand{\IsN}{\nit{IsNull}}
\renewcommand{\ni}{\noindent}
\newcommand{\mm}{\mathcal{M}}
\newcommand{\schn}{\models_{_N}}

\newcommand{\sm}{\smallsetminus}
\newcommand{\peer}[1]{\texttt{#1}}
\newcommand{\p}[1]{\texttt{#1}}
\renewcommand{\sp}[1]{\small \texttt{#1}}
\newcommand{\trust}{{\it Trust}\xspace}

\newcommand{\same}{{\it same}\xspace}
\newcommand{\less}{{\it less}\xspace}
\newcommand{\NS}{\nit{N\!S}\xspace}

\newcommand{\mf}[1]{\mathfrak{#1}}

\newcommand{\aux}{{\it aux}}
\newcommand{\lea}{\leftarrow}
\newcommand{\lp}{${\cal L}(\peer{P})$}
\newcommand{\lpq}{${\cal L}(\peer{P},\p{Q})$}
\newcommand{\dbic}{\p{P},D}

\newcommand{\f}{\mathbf{f}}
\newcommand{\tr}{\mathbf{t}}
\newcommand{\ta}{\mathbf{t}}
\newcommand{\fa}{\mathbf{f}}
\newcommand{\trs}{\tr^\star}
\newcommand{\fs}{\f^\star}
\newcommand{\tss}{\tr^{\star\star}}

\newcommand{\wrt}{with respect to\xspace}
\newcommand{\rch}{r\mbox{-}\nit{Chase}^{\nn}}

\newcommand{\rs}{\ensuremath{\!\!\downharpoonright\!}}

\usepackage{enumitem}
\setlist{topsep=0pt,noitemsep}

\newcommand{\ox}{(\bar{x})}

\newcommand{\oxpy}{(\bar{x}',y)}

\newcommand{\oaob}{(\bar{a},\bar{b})}
\newcommand{\oa}{(\bar{a})}

\newcommand{\oatstar}{(\bar{a},\mathbf{t}^{\star})}
\newcommand{\oatstarr}{(\bar{a},\mathbf{t}^{\star\star})}
\newcommand{\oaobtstarr}{\!\_(\bar{a}',\bar{b},\mathbf{t}^{\star\star})}
\newcommand{\oxi}{(\bar{x_{i}})}
\newcommand{\oyj}{(\bar{y_{j}})}

\newcommand{\opa}{(\bar{a}')}

\newcommand{\oaobtstar}{\!\_(\bar{a}',\bar{b},\mathbf{t}^{\star})}

\newcommand{\nulo}{{\mbox{\tt \scriptsize null}}\!}
\newcommand{\mc}[1]{\mathcal{ #1}}
\newcommand{\sql}{{\mbox{\tt \scriptsize sql}}\!}

\newcommand{\red}[1]{{#1}}
\newcommand{\blue}[1]{\textcolor[rgb]{0.00,0.00,1.00}{#1}}
\newcommand{\green}[1]{\textcolor[rgb]{0.00,0.50,0.00}{#1}}
\newcommand{\burg}[1]{\textcolor[rgb]{0.5, 0.0, 0.13}{#1}}
\newcommand{\capri}[1]{{#1}}

\newcommand{\comlb}[1]{{\vspace{2mm}\noindent \bf \textcolor[rgb]{0.00,0.00,1.00}{COMM(LEO):}}~
{ #1 }\hfill{\bf \textcolor[rgb]{0.00,0.00,1.00}{END.}}\\}

\newcommand{\comlore}[1]{{\vspace{2mm}\noindent \bf \textcolor[rgb]{0.00,0.5,0.00}{COMM(Lore)}:}~
{ #1}\hfill {\bf \textcolor[rgb]{0.00,0.5,0.00}{END.}}\\}

\newcommand{\dproof}[1]{{\noindent\bf {\sc Proof}:\
}#1 \boxtheorem \vspace{3mm}}

\newcommand{\ICc}{\mc{C}}
\newcommand{\Mdd}{M^{\star}_{\ICc}(D,D')}

\newcommand{\defproof}[2]{{\noindent {\sc Proof of} {\it #1}: \vspace{2mm} \newline
} #2 \boxtheorem\\}

\begin{document}

\long\def\comment#1{}

\title{Consistency and Trust in Peer Data Exchange Systems}

\author[L. Bertossi, L. Bravo]
{LEOPOLDO BERTOSSI\thanks{Contact author.}\\ Carleton University, School of Computer Science,
Ottawa, Canada.\\ bertossi@scs.carleton.ca\\
\and
LORETO BRAVO\\ Universidad del Desarrollo, Facultad de Ingenier\'ia, Santiago,
Chile.\\ bravo@udd.cl}

\pagerange{\pageref{firstpage}--\pageref{lastpage}}
\volume{\textbf{10} (3):}
\jdate{March 2002}
\setcounter{page}{1}
\pubyear{2002}

\maketitle

\label{firstpage}

\begin{abstract}
We propose and investigate a semantics
for  {\em peer data exchange systems}  where different peers are  related  by
 data exchange constraints and trust relationships. These two
elements plus the data at the peers' sites and their local integrity
constraints  are made compatible via a
semantics that characterizes sets of  {\em solution instances} for the peers. They are the intended -possibly virtual- instances for a peer
 that are obtained through a data repair semantics that we introduce and investigate. The semantically
correct answers from a peer to a query, the so-called  {\em peer
consistent answers}, are defined as those answers that are invariant
under all its different solution instances. We show that solution
instances can be specified as the models of logic programs with a
stable model semantics. The repair semantics  is based on  null values as used in SQL databases,
and is also of independent interest for repairs of single databases \wrt integrity constraints.\\

\noindent {\em To appear in Theory and Practice of Logic
Programming (TPLP).}
\end{abstract}

\newcommand{\BibTeX}{{\rm B\kern-.05em{\sc i\kern-.025em b}\kern-.08em
    T\kern-.1667em\lower.7ex\hbox{E}\kern-.125emX}}

\markboth{L. Bertossi and L. Bravo}{Consistency and Trust in Peer Data Exchange Systems}

\section{Introduction}

A {\em peer data exchange system} (PDES) (also known as a peer data management
system) can be conceived as a finite set
${\cal P} = \{\peer{P}_1, \ldots, \peer{P}_n\}$ of peers, each
of them owning a local relational schema and a database instance.
 When a peer \p{P} receives a query, in
order to answer it, \p{P}'s data is completed or modified according
to relevant data that other peers may have. For this to be possible,
$\peer{P}$ has to  be directly related to some neighbor peers $\peer{Q}$ through sets, $\Sigma(\peer{P},\peer{Q})$, of {\em logical mappings} or {\em data exchange constraints} (DECs), from \p{P} to \p{Q}.
 They are first-order sentences
expressed in terms of the  elements of the database schemas of $\p{P}$ and $\p{Q}$. In their turn,
\p{P}'s neighbors may have their own neighbors, who through their data may, transitively and eventually,
have an impact on \p{P}'s data as well.

DECs
between two peers are expected to be satisfied by the combination of
the two local instances. In this sense, DECs act as integrity constraints (ICs)
on the combination of two schemas and their instances. However, they are not forced to
be satisfied, and there is no mechanism in place for maintaining them. Actually, \red{\em it is the inconsistency
of a DEC that will enable the movement of data between peers, at query-answering time}: \red{When a peer \p{P} receives a query, it examines its DECs, on that basis poses queries to its neighbors, who return
{\em consistent} data to \p{P}. \ \p{P} also
queries its own database. If its DECs are not satisfied by the collected data, \p{P}, after a -possibly virtual- consistency repair process, composes a {\em consistent} answer to the original query.}
A peer's answering of a query received from an external user or another peer triggers an iterative process of interleaved consistency checking
of  DECs \wrt the data at hand, and the repair of the latter if necessary.

This movement of data among peers is  used for query answering, but not for updating the local physical instances
with the purpose of having the DECs satisfied. Of course, this could be done by a peer who receives data from other peers, if it decides to do so.
However,  it is not the goal of data exchange in such a system to have all the instances synchronized and globally satisfying the DECs. Peers are autonomous,
 their instances are possibly being independently updated, and other peers are not expected to be aware of that. DECs constrain and trigger data
exchange through their inconsistency, which is detected and resolved locally, by a single peer, namely
the one who owns those DECs. {\em We do not assume any kind of central or global coordination; nor that any two neighbors interact for
DEC evaluation}.

A peer
\p{P} that is answering a query may, at query-answering time, import data from
its neighbors, to complement its data and/or ignore part of its own
data. The way a peer uses the imported data of course depends on its DECs, but also on the {\em trust relationships} to
its neighbors: A peer \peer{P} may trust its data the same as or less
than a neighbor's data. As a  consequence, {\em in our PDESs, data exchange is driven by inconsistency
and constrained by trust}.


\begin{example} \label{ex:intro0} Peers \p{P1} and \p{P2}
have relational schemas ${\cal S}(\p{P}1) = \{R^1(\cdot,\cdot,\cdot), S^1(\cdot)\}$, and $
{\cal S}(\p{P}2) = \{R^2(\cdot,\cdot),S^2(\cdot,\cdot)\}$. Here, $\p{P}1$ is
connected to peer $\p{P}2$ by \p{P1}'s set of DECs:
$$\Sigma(\p{P}1,\p{P}2) =
\{\forall x \forall y(R^2(x,y) \wedge S^2(y,z) \rightarrow
R^1(x,y,z)), \  \forall x (S^1(x) \rightarrow S^2(5,x))\}.$$ These DECs belong
to \p{P1}; and  \p{P2} may not even be aware of them. Let's assume that
\p{P1} trusts \p{P2} more than itself. The existence of DECs from \p{P1} to \p{P2}, and the trust
relationship are indicated in Figure \ref{fig:fig0} by the labeled arrow from one instance to the other.

We can see that the DECs are not satisfied
by the combined instance, which is perfectly acceptable; we
do not have to do anything.

\begin{figure}
\hspace*{5mm}\includegraphics[width=9.5cm]{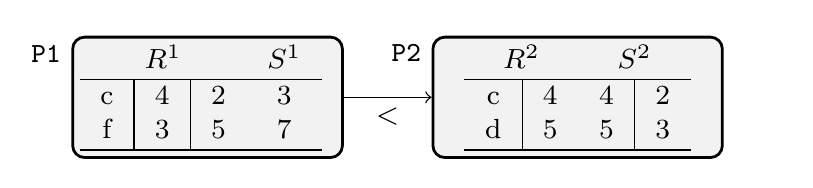}
\vspace{-3mm}\caption{A simple PDES} \label{fig:fig0}
  \end{figure}

  Now, assume that the  query \ $\mathcal{Q}(x): \exists y z
R^1(x,y,z)$ \ is posed to \p{P1}. In order to answer it, \p{P1} notices through the DECs that \p{P2}'s schema is related to its predicate $R^1$.
Accordingly, \p{P1}
has to adjust its own
data, so that the DECs with  \p{P2} are satisfied. In order to do this, \p{P1} requests  \p{P2}'s data,
by issuing to \p{P2} atomic queries in \p{P2}'s language, namely ${\cal Q}_1(x,y)\!: R^2(x,y)$ and ${\cal Q}_2(x,y)\!: S^2(x,y)$.\footnote{
Actually, \p{P1} does not need the whole extensions of \p{P2}'s tables. However, asking for \p{P2}'s whole tables allow us to better
illustrate our general approach.}
Notice that, at this stage and from the point of view of \p{P2}, the queries ${\cal Q}_1$ and ${\cal Q}_2$ can be considered
as queries posed by an external user, so as query ${\cal Q}$ was posed to \p{P1}.

Since \p{P2} has neither DECs to other peers nor local ICs, it will return to
\p{P1} its physical data, without any modification, as query answers. \ignore{In this case, \p{P2}'s current instance is its only solution instance.}
Now \p{P1}, confronted with a combined instance, can detect that the first  DEC is not satisfied, because the tuple $R^1(d,5,3)$ is not in $R^1$.
Nor the second DEC, because the tuple $S^2(5, 7)$ is not in $S^2$.

Since \p{P1} trusts \p{P2}'s data more than its own, the inconsistencies are locally solved by \p{P1} by inserting $R^1(d,5,3)$ into $R^1$, and deleting tuple
$S^1(7)$ from $S^1$. These changes could be only virtual, and for the purpose of answering the query at hand. In this case, \p{P1} has a single
{\em neighborhood solution} (instance), that is, an instance for the combined schema that satisfies the DECs, and, in some sense, {\em minimally
departs} from the (inconsistent) instance that had been formed at \p{P1}' neighborhood. This neighborhood solution for \p{P1} is shown in
Figure \ref{fig:fig1}.
 The restriction of this instance to \p{P1}'s schema is  {\em a solution} (instance) for
\p{P1}, and is used to answer the initial query posed to
\p{P1}. Thus, the answers to  $\mathcal{Q}$ are $\langle c\rangle,\langle f\rangle,\langle d\rangle$. \boxtheorem

\begin{figure}
\hspace*{5mm}\includegraphics[width=5.8cm]{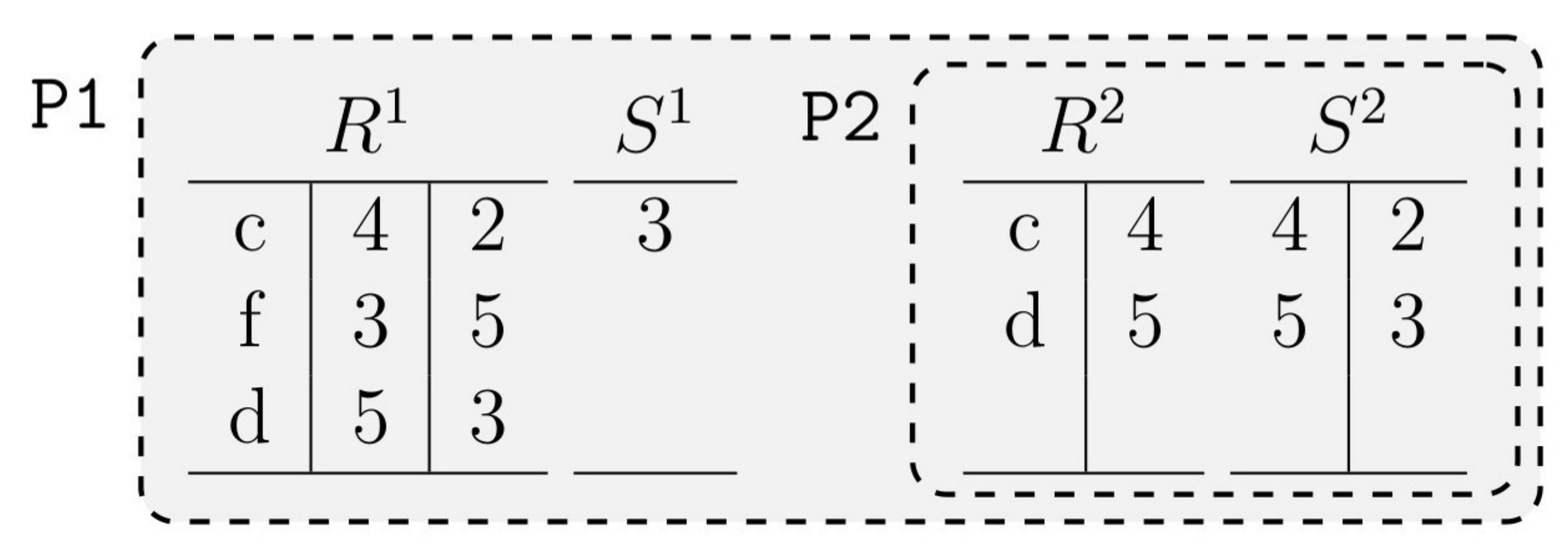}
\vspace{-3mm}\caption{A neighborhood solution instance for \p{P1}}\label{fig:fig1}
\end{figure}

\end{example}

In the example, for illustration purposes,  we used a particular and well-known {\em repair semantics}, i.e. a particular way of restoring consistency on a database that does not satisfy a given set of ICs,\footnote{We
emphasize that DECs can be seen as ICs on combined schemas and instances.} while still staying ``as close as possible" to the given, inconsistent instance.
In this case, repairs minimize, under set inclusion, the set of tuples that are deleted or inserted  \cite{ABC99}. Repair semantics come with their minimality criteria. (Cf. \cite{bertossi11} for
a general discussion of repair semantics, and for references to different  repair semantics.)
In this work we will introduce, investigate and use a variation of the  repair semantics used in the example.

In some cases, a peer \p{P} may have
several neighborhood solutions, leading to also possibly several (local) solution
instances. The answers returned by \p{P} (to an external user or another peer) will be the {\em peer consistent answers},
i.e. those that are shared (or returned) by all the different solution
instances for \p{P}. That is, a {\em cautious} (also known as {\em skeptical} or {\em certain})
semantics is applied to query answering \cite{asp,DLV}. The same do \p{P}'s neighbors, who may have
solutions of their own, when they hand over their data to \p{P}: They give away their own peer-consistent data.

The notion of solution instance for a peer
is used as an auxiliary notion, to characterize the semantically
correct answers from \peer{P}'s point of view. When trying to correctly
answer a query, we try to avoid,
as much as possible, the generation of material solutions
instances. Ideally, a peer would compute its
peer consistent answers to a user query just  by querying the already available
local instances, and dealing with the involved DECs on-the-fly, at query answering time.

In this work we formalize these ideas, by first providing a general semantic framework for PDESs and their solution-semantics. It captures in abstract terms, and in this order, the notions of: (a) PDES, including a broad and useful syntactic class of DECs; (b) neighborhood solution for a peer, appealing to an {\em abstract
repair-semantics} that involves trust relationships; (c) solution for a peer, a recursive concept due to a possible (always finite) chain of peers building each, one after another, neighborhood solutions; and (d)
 peer-consistent answer.

The result is
a formal semantics for  a system of peers who exchange data
for query answering\ignore{, characterizing in precise terms the
intended and correct answers to a query posed to and answered by a
peer in the system}. It is a model-theoretic {\em possible-world semantics} that  characterizes the {\em class of intended instances for a peer}, the above mentioned {\em solutions}.
The expected answers from a
peer are  {\em certain} \wrt that class, i.e. true in all of them.

Next, as a second main subject, we instantiate the general semantic framework, and introduce and investigate a specific repair (or solution) semantics
for PDESs. We do so by making some specific commitments and assumptions: \ (a) Database instances may have null values as those used in SQL relational databases. \
(b) Those null values behave and are handled as in DBMSs that (at least partially) comply with the SQL Standard, in particular \wrt IC satisfaction and query answering. \ (c)
The null-based solution semantics that we introduce  heavily depends on the presence of nulls \`a la SQL, and on the use of nulls to restore consistency. \ (d) This null-based repair semantics comes with its own minimality criterion. \ (e) Query answering, and DEC and IC satisfaction follow a {\em a logic} in databases with null \`a la SQL that we formalize for conjunctive queries and a
 broad class of DECs and ICs.

For the gist, the null-based repair semantics we adopt does not allow, for example, to introduce
a null value to satisfy a  join, such as that with the existential $z$ in the DEC $\forall x \forall y(R(x,y) \rightarrow \exists z (S(x,z) \wedge T(z,y)))$. In this case, a tuple deletion from  $R$
will be preferred.
However, a null can be used for variable $z$ in a tuple inserted to satisfy the DEC  $\forall x\forall y(R(x,y) \rightarrow \exists z S(x,z))$ \ (a deletion from $R$ is also
possible).

Our decision to consider null values, as used in SQL relational database systems, for  single database repairs and solutions in PDESs is based upon and motivated by the facts that: (a) In this way our research becomes close to database practice; (b) it can be made compatible with and applicable
in that practice; (c) SQL DBMSs implicitly implement a sort of null-based repair semantics (when maintaining ICs); (d) that ``repair semantics" deserves a theoretical investigation;
and (e) at least a sizeable part of the SQL standard and practice can be be put on solid logical grounds.
Actually, we are able to bring into this research essentially all the classes of
integrity constraints and logical mappings that are used in both database practice and research, and to treat them according to the just mentioned null-based SQL semantics.

We emphasize, however, that
our abstract framework is sufficiently modular and flexible to adopt alternative repair semantics for dealing with inconsistency and incompleteness of data.

We see our work also as a contribution to the subjects of repairs and consistent query answering for single relational databases. Repairs based on the use of
nulls; actually a single one with a semantics
in the style of SQL -which we also logically formalize in this work- had  been presented  in preliminary form in  \cite{iidb06,Bravo07}. Here, we provide an extended and definitive formulation, and we apply it
 to peer data exchange.

We also investigate complexity-theoretic and algorithmic issues related to the solution semantics. More specifically,
we show that
deciding if a neighborhood instance is a neighborhood solution for a peer is
$\nit{coNP}$-complete in data, and deciding if a tuple is a  peer consistent  answer to a local query is $\Pi^P_2$-complete in data.\ignore{\footnote{All the complexity statements in this paper are about {\em data complexity},
i.e., in terms of the sizes of the database instances involved.}}  We investigate a particular, important and common case of the PDES semantics,
\ignore{  kind of combination of DECs and trust relationships } where peers just import data from other peers to complete their own data set, without giving semantic priority to the latter. This
 leads to a reduction in data complexity for the solution checking problem.

Finally, we show that the  null-based, model-theoretic semantics for PDESs can be captured by means of disjunctive logic programs with
stable model semantics \cite{GL91}, also known as {\em answer set programs} (ASPs) \cite{asp}. More specifically, we establish an explicit correspondence between the solutions for a
peer and the stable models of the program, obtaining, in this way, a
declarative semantics for PDESs that can also be made executable.

In relation to most immediately related work, the idea of peer consistent query answering resembles that of {\em consistent
query answering} (CQA) in databases \cite{ABC99}: Consistent answers to a query posed to a database that
may be inconsistent \wrt certain ICs are
invariant under all {\em repairs}, i.e. the minimally repaired
and consistent versions of the original instance. Furthermore, there are
mechanisms for computing consistent answers that avoid or
minimize the physical generation of repairs.
Logic programs
for  specifying database repairs  as their stable models have
been proposed and investigated in
\cite{arenas03,greco03,barcelo03,CB10}. Cf. \cite{bertossi11,sigmodRec06,chom07} for recent surveys of CQA.

\ignore{The mappings between peers are specified as logical formulas, which allow us to
consider a logical semantics for the system. In this way, the intended instances correspond to distinguished models of a logical theory. The intended answers
from a peer to a query are then naturally defined as those that hold in the intended instances. Of course, different logics can be considered for this
task. In our case, it turns out to be a non-monotonic semantics associated to specifications written as logic programs.  }

Our work can be classified among those on {\em semantic and schema-based approaches to peer-to-peer
 data management systems}
\cite{HIST03,HIM+04,CGL+04,FKL+03,BB04a,CDL+05,Fux06}. The emphasis is on specifying the intended and legal database instances behind a system
of peers who are connected with each other by means of schema mappings.\ignore{\footnote{In this paper we neither address subjects like
topologies or protocols for P2P overlay networks, nor
distributed data structures for P2P file-sharing systems
\cite{lua05,spinellis04}.}}

Different notions and forms of trust have been considered in P2P information
sharing systems \red{\cite{demolombe,sabater05,artz07,gm06} (cf. the special issue \cite{compNets06})}, but not  much in
the context of semantic PDESs, where the trust
relationships between peers are used to essentially modulate or qualify the
use of the data exchange constraints between peers.  \ignore{Nor are they used to determine the most
reliable peer to which a query could be routed.} Our emphasis is on the
integration of trust and data exchange constraints, with a notion of trust that is closer to the notion of relative reputation or reliability of a peer
as a source of information, in relation to the quality of its data \cite{gm06}.  In this work, trust
relationships are given, and  not computed or updated  \cite{marsh06,josang}.

 \ignore{In this work we first identify the main principles and assumptions in relation to the semantics
 of PDESs.  As a particular case of this general semantic framework, we introduce and further investigate
 the null-based semantics presented in \cite{lpar07}. We also investigate in detail complexity issues around the null-based semantics, special cases of this semantics, and the impact of the  assumptions made
 about the general and special semantics.} 

This paper is structured as follows. Section \ref{sec:framework} provides
technical preliminaries and basic elements of PDESs. Section
\ref{sec:semantics} introduces a general semantic framework for PDESs.
Section \ref{sec:nulls}, in preparation for the introduction of what will be the official, concrete
peer-solution semantics, discusses and formalizes database instances with null values as used in SQL. It also formulates query answering and
integrity satisfaction in those databases. Section \ref{sec:nullsRep}  instantiates the abstract semantic framework proposed in Section
\ref{sec:semantics} by
introducing a particular consistency restoration semantics based on the use
of null values. This section also investigates the complexity of some computational problems.
Section \ref{sec:P2Ptransit} captures the semantics of the
previous section in terms of logic programs with stable model semantics.
  In Section \ref{sec:related} we discuss related work. In Section \ref{sec:concl} some final conclusions are
drawn.   \burg{The electronic  \ref{app:disc} discusses several additional and alternative approaches and issues
\wrt the previously introduced general and specific semantics. The electronic \ref{app:proofs} gives proofs of our main results.}
This work considerably extends
and develops the semantics for PDESs first suggested in \cite{BB04a}, and further developed
in \cite{lpar07}.

\section{The Basic PDES Scenario} \label{sec:framework}

\red{Every peer in a PDES will have a local relational schema with a local relational instance. For this reason, we recall first some basic notions from relational databases.}
A relational schema $\mathcal{S}$ consists of a set of relational predicates. A relational predicate $R \in \mc{S}$ with arity $n$ and attributes $A_1, ...,A_n$, is commonly denoted with $R(A_1,\ldots,A_n)$, or
sometimes simply $R(\cdot,\ldots,\cdot)$. Each attribute $A$ of (a predicate in) $\mc{S}$ has a possibly infinite data domain, $\nit{Dom}(A)$.  In general, we will denote with $\mc{U}$
the union of the attribute domains, obtaining a single, possibly infinite {\em data domain}.

In addition to the {\em database predicates} in a relational schema $\mathcal{S}$, we have a
 set $\mc{B}$ of built-in predicates (that have a fixed semantics),  e.g. comparison predicates, such as $=, \neq, <$, etc. We assume that $\mc{B}$ contains the propositional predicate $\mathbf{false}$ that is
 always false. \ (Later on, we will introduce some additional, specific, built-in predicates.)

\red{The  predicates of the relational schema ${\cal S}$ plus those in $\mc{B}$ (that we leave implicit), and the constants in $\mc{U}$ determine a language ${\cal L}({\cal S})$ of first-order predicate logic. The schema may also contain {\em integrity constraints}, that are sentences  of ${\cal L}({\cal S})$.}

An instance $D$ for a schema $\mathcal{S}$ is a finite set of  a ground atoms of the form $R(\bar{c})$, where $R(A_1,\ldots,A_n) \in \mathcal{S}$ has  some arity $n$, and $\bar{c}=\langle c_1, \ldots, c_n\rangle$, with
 $c_i \in \nit{Dom}(A_i)$. For each $n$-ary predicate $R \in \mc{S}$, an instance $D$ for $\mc{S}$ determines an  extension for $R$ that is a finite $n$-ary relation over the data domain.  If $t\in D$ we denote by ${t}[\bar{A}]$
the sequence of values in $\bar{t}$ for attributes $A \in \bar{A}$. Given an instance $D$, the {\em active domain} of $D$, denoted, $\nit{Adom}(D)$, is the finite subset of $\bigcup_A \nit{Dom}(A)$ that contains all the constants that appear
in relations in $D$.

\red{If ${\cal S}'$ is a subschema of ${\cal S}$, i.e. contains some of the relational predicates in ${\cal S}$, and $D$ is an instance for $\mc{S}$, then  $D\rs {\cal S}'$ denotes the  restriction of $D$ to ${\cal S}'$, i.e.
 $D\rs {\cal S}'= \{R(\bar{t}) \ | \ R \in {\cal S}'\mbox{ and } R(\bar{t}) \in D\}$.}

A database instance $D$ for the schema ${\cal S}$ serves then as an interpretation for the language ${\cal L}({\cal S})$.
If $D$ is an instance for ${\cal S}$,  and $\Psi$ is a set of sentences of ${\cal L}({\cal S})$, then $D \models \Psi$ denotes that $D$ satisfies (makes true)
    all the sentences in $\Psi$.  If  $\Psi$ is the set
    of integrity constraints that comes with the schema ${\cal S}$ and $D \models \Psi$, we say that $D$ is {\em consistent}. Otherwise, $D$ is
    {\em inconsistent}.\footnote{In this work, whenever we consider sets $\Sigma$ of integrity constraints, we assume that $\Sigma$ is {\em logically consistent}.}

    \red{A query $\mc{Q}(\bar{x})$
    is a formula of a language ${\cal L}({\cal S})$, where $\bar{x}$ is the list of free variables. A sequence of constants $\bar{c}$ is an answer to  $\mc{Q}(\bar{x})$ in instance $D$ for $\mc{S}$
    if $D \models \mc{Q}[\bar{c}]$, i.e. the formula becomes true in $D$ when the variables in $\bar{x}$ are replaced by the corresponding constants in $\bar{c}$. When $\bar{x}$ is empty,
    the query $\mc{Q}$ is a {\em Boolean query}, i.e. a sentence. In this case, the answer in  $D$ can be {\em yes} or {\em no} depending on whether it is true or not in $D$, denoted
    $D \models \mc{Q}$, resp. $D \not \models \mc{Q}$. }

\red{Now we introduce some of the elements that will form a {\it peer data exchange system} (PDES). After introducing them we will give the formal definition of a PDES (cf. Definitions \ref{def:locICs} and
\ref{def:elements} below).}

    A   (PDES) contains a
     finite set ${\cal P} = \{\p{P1}, \ldots, \p{Pn}\}$ of peers. Each peer \p{P} owns a
     relational database schema ${\cal S}(\peer{P})$, and a database instance
    $D(\peer{P})$  for the schema ${\cal S}(\peer{P})$. \red{We denote with $\mf{S}$ the set of all schemas of a PDES, that is, $\mf{S}=\{{\cal S}(\peer{P1}), \ldots, {\cal S}(\peer{Pn})\}$}.
    We assume, to simplify the presentation and,
    without loss of generality, that the peers' schemas are mutually
    disjoint, but share a common, possibly infinite database domain $\U$.
     Each $D(\p{P})$ can be seen as a finite set of ground atoms over $\U$, with predicates
    in ${\cal S}(\p{P})$. \ignore{\red{The (finite) {\em active domain}, $\nit{act}(D(\peer{P}))$, of peer \peer{P}
contains all the constants in $D(\peer{P})$.}} It holds,  $\nit{Adom}(D(\peer{P})) \subseteq {\mathcal U}$.

The peers' schemas,  or unions thereof, determine first-order languages, e.g. \lp, \lpq, which are ${\cal L}({\cal S}(\p{P}))$ and
    ${\cal L}({\cal S}(\p{P}) \cup {\cal S}(\p{Q}))$, resp.
    A {\em
data exchange constraint} (DEC) between peers \p{P},
\p{Q} is an ${\cal
L}(\p{P},\p{Q})$-sentence.
\ We will consider the following two syntactic classes of DECs: \vspace{2mm}

\begin{enumerate}[label=(\emph{\alph*})]
\item  A {\em
universal data exchange constraint} (UDEC) between peers \p{P},
\p{Q} is an ${\cal L}(\p{P},\p{Q})$-sentence of the form:

\begin{equation}\label{eq:UDEC}
 {\forall}\bar{x}(\bigwedge_{i = 1}^{n} R_i(\bar{x}_i) ~\longrightarrow~
\bigwedge_{k=1}^l(\bigvee_{j=1}^{m} Q_{kj}(\bar{y}_{kj}))),
\end{equation}

where the $R_i$ are predicates in ${\cal S}(\p{P}) \cup {\cal
S}(\p{Q})$, the $Q_{kj}$ are  atomic formulas with predicates  in ${\cal S}(\p{P}) \cup {\cal
S}(\p{Q})$ or atoms with  predicates  in $\mc{B}$, $\bar{x}=\bigcup_{i=1}^n \bar{x}_i$,
and $\bar{y}_{kj} \subseteq \bar{x}$.

\vspace{1mm}
 \item  A {\em referential data
exchange constraint} (RDEC) between peers \p{P}, \p{Q} is an ${\cal
L}(\p{P},\p{Q})$-sentence of the form:

\begin{equation}\label{eq:RDEC} \forall \bar{x}(\bigwedge_{i = 1}^{n} R_i(\bar{x}_i) \ \
~\longrightarrow~ \ \ \exists \bar{y}(\bigwedge_{k=1}^l
Q_{k}(\bar{x}_{k},\bar{y}_k) \wedge \varphi_k(\bar{x}_{k}',\bar{y}_k')) \ \red{\vee \varphi(\bar{x})}),
\end{equation}

where the $R_i, Q_{k}$ are predicates in ${\cal S}(\p{P}) \cup {\cal
S}(\p{Q})$, $\varphi_k$ \red{and $\varphi$ are a conjunction, resp. a disjunction,} of atoms with predicates in $\mc{B}$, and $\bar{x}_i, \bar{x}_{k}
\subseteq \bar{x}$, $\bar{x}_k' \subseteq \bar{x}_k$, $\bar{y}_k' \subseteq \bar{y}_k$, and $\bar{y}=\bigcup_{k=1}^l \bar{y}_{k} \neq
\emptyset$.

\red{The formulas $\varphi_k(\bar{x}_{k}',\bar{y}_k'))$ are used to impose conditions on the existential values, and $\varphi$ for conditions on the values for $\bar{x}$ in the antecedent.}
\end{enumerate}


\vspace{1mm}
\red{The classes of exchange constraints that we are considering are broad enough to capture all the relevant integrity constraints and logical mappings found in data exchange and virtual
data integration that are usually considered in the theoretical, technical  and applied literature on data management.
In particular,  UDECs  can be used to express {\em equality-generating dependencies} (egds), and RDECs can express general {\em tuple-generating dependencies} (tgds) \cite{AHV1995}.}

Each peer \peer{P} of $\Pe$ has a possibly empty collection  of sets of DECs, $\Sigma(\peer{P},\peer{Q})$, between \p{P} and  peers $\p{Q} \in \Pe$, with at most one $\Sigma(\peer{P},\peer{Q})$ for each
peer \peer{Q}.  Each $\Sigma(\peer{P},\peer{Q})$ is finite and logically consistent.
    Due to the local nature of PDESs systems, it is possible for $\Sigma(\peer{P},\peer{Q})$ (which is owned by \p{P}) to be different from $\Sigma(\peer{Q},\peer{P})$ (which is owned by \p{Q}).
 The DECs in $\Sigma(\p{P},\p{P})$ are the integrity constraints for (instances of) peer $\p{P}$.  We denote with $\Sigma$ the class formed by of all non-empty sets $\Sigma(\peer{P},\peer{Q})$ of a PDES.

A PDES also has a
    relation $\trust \subseteq {\cal P} \times \{\less, \same\}
    \times {\cal P}$, with exactly one triple of the form $(\p{P}, \cdot, \p{Q})$ for
    each $\Sigma(\p{P},\p{Q}) \in \Sigma$.
The intended semantics of
    $(\peer{P}, \less, \peer{Q}) \in \trust$ is that peer $\peer{P}$
    trusts itself less than $\peer{Q}$; while $(\peer{P}, \same,
    \peer{Q}) \in \trust$ indicates that $\peer{P}$ trusts
    itself the same as $\peer{Q}$.\footnote{We do not consider the case when a peer
\p{P} trusts itself more than another peer, since the information of the
latter should be irrelevant to $\p{P}$.} The trust relation is not necessarily symmetric, e.g. it could
    hold $(\peer{P}, \less, \peer{Q}), (\peer{Q}, \same, \peer{P})  \in \trust$. For a peer \p{P}, when $\Sigma(\p{P},\p{P})\neq \emptyset$, we assume $(\p{P},\same,\p{P}) \in
    \trust$.

\red{\begin{definition}\label{def:locICs}
\begin{enumerate}[label=(\emph{\alph*}),leftmargin=2em]
\item \red{The {\em schema of a PDES} $\mathfrak{P}$ is a sequence $\langle
\mc{P}, \mf{S}, \Sigma,\trust\rangle$, where $\mc{P}$ is a set of peers, $\mf{S}$ is a corresponding set
of  peer database schemas},  $\Sigma$ the set of DECs,  and
\trust the set of trust relationships.
\item
An instance $\mathfrak{D}$ of a PDES schema $\mathfrak{P}=\langle
\mc{P}, \mf{S}, \Sigma,\trust\rangle$ is  a set containing one database instance $D(\p{P})$ for the schema $S(\p{P})\in \mf{S}$, for each peer $\p{P} \in \Pe$. \boxtheorem
\end{enumerate}
\end{definition}}
Intuitively, and due to the locality of peers, a peer \p{P} will be aware
    only of the sets $\Sigma(\p{P},\p{Q})$ and elements of \trust ~whose first argument is \p{P}. More precisely,
a peer stores its own database schema and instance, its DECs
to other peers, and its trust relationships. Furthermore, for an instance $\mathfrak{D}$ of schema $\mathfrak{P}$,  the instances $D(\p{P})  \in \mathfrak{D}$  are not
required to (jointly) satisfy the DECs in $\Sigma$. Consistency and consistency restoration is a concern only when queries are posed to peers. \ignore{This will be explicitly required
when desired. In particular, a peer's instance combined with those of its neighbors, may not satisfy its DECs. Accordingly, the
peer's instance has to be (possibly virtually) updated to satisfy this requirement, leading to {\em solution instances} for the peer.

A peer's solutions will be determined  by the peer's stored information and the data obtained from other neighbors through database queries. The answers from the neighboring peers will
be determined, in their turn, by their own neighbors, and so forth. The following definition formalizes the basic elements to capture these ideas (which will be done in
Section \ref{sec:semantics}).}

\begin{definition}\label{def:elements}
Given a PDES
 schema $\mathfrak{P}=\langle
\mc{P}, \mf{S}, \Sigma,\trust\rangle$:
\begin{enumerate}[label=(\emph{\alph*}),leftmargin=2em]
\item  \red{The  {\em \red{schema of a peer} \p{P}} $\in \mc{P}$} is $\mathfrak{P}(\p{P})=\langle \mc{S}(\p{P}), \Sigma(\p{P}),\trust(\p{P})\rangle$, where $\mc{S}(\p{P})\in \mf{S}$, $\Sigma(\p{P})=\{\varphi \ | \ \varphi \in \Sigma(\peer{P},\peer{Q}),$  $\Sigma(\peer{P},\peer{Q}) \in \Sigma, \mbox{ and } \p{Q} \in \mc{P}\}$; and
    $\trust(\p{P})=\{(\p{P},t,\p{Q}) \ | \ (\p{P},t,\p{Q})\in \trust, \mbox{ and } \p{Q} \in \P\}$. \ (We can safely identify a peer \p{P} with its schema $\mathfrak{P}(\p{P})$.)

\item  The {\em accessibility graph} $\G(\mathfrak{P} )$
 contains a vertex for each $\p{P} \in {\cal P}$,
and a
 directed
 edge from $\p{P}$ to $\p{Q}$ \red{if $\peer{P}\neq \peer{Q}$}
and $\Sigma(\peer{P},\peer{Q})\neq \emptyset$. An edge from $\p{P}$
to $\p{Q}$ is labeled with ``$<$" when
$(\peer{P},\less,\peer{Q}) \in \trust$, and with ``$=$"  when
$(\peer{P},\same,\peer{Q}) \in \trust$.
\item Peer \peer{P'} is {\em
accessible} from \peer{P} if there is a path in $\G(\mathfrak{P})$
from \peer{P} to \peer{P'} or if \peer{P'}=\peer{P}.  ${\cal AC}(\peer{P})$
denotes the set of peers accessible from \p{P}. \ \red{$\G(\p{P})$ denotes
the restriction of $\G(\mathfrak{P})$ to the peers in ${\cal AC}(\p{P})$ (i.e. it contains as vertices only the peers
in ${\cal AC}(\p{P})$, and the edges between them are those inherited from $\G(\mathfrak{P})$).}
\item
Peer
\peer{P'} is a {\em neighbor} of \peer{P} if there is an edge from
$\p{P}$ to $\p{P'}$ in $\G(\mathfrak{P})$, or if $\p{P'}=\p{P}$. We denote with
${\cal N}(\peer{P})$ the set
of neighbors of
\peer{P}; and with ${\cal S}({\cal N}(\p{P}))$ the union of their schemas, i.e. $\bigcup_{\sp{Q} \in {\cal N}(\sp{P})}\!{\cal S}(\p{Q})$.
Furthermore,
${\cal N}^\circ(\peer{P}):={\cal N}(\peer{P}) \smallsetminus \{\p{P}\}$.\boxtheorem
\end{enumerate}
\end{definition}

\begin{figure}
\hspace*{2mm}\includegraphics[width=12cm]{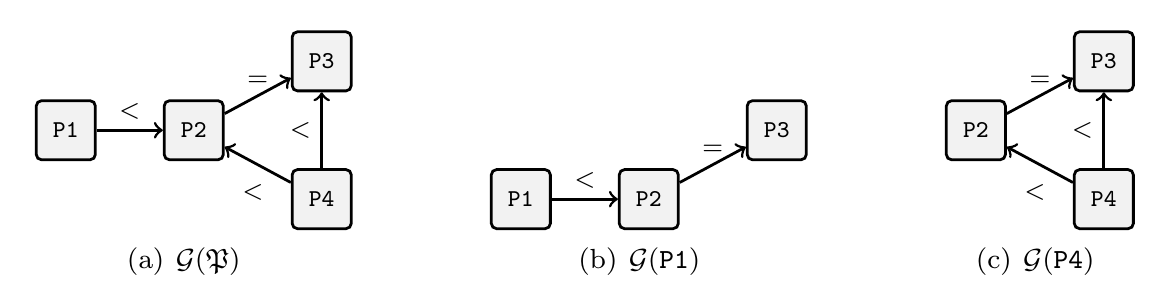} 
\vspace{-3mm}\caption{\label{fig:AG}Graphs for Example
\ref{ex:trans0}}
\end{figure}

\ignore{
\begin{figure}
\input{fig_AG}
\caption{\label{fig:AG} Graphs for Example
\ref{ex:trans0}}
\end{figure}
}

\begin{example} \label{ex:trans0} Consider $\mathfrak{P}=\langle
\mc{P}, \mc{S}, \Sigma,\trust\rangle$  with $\Pe = \{\p{P1}, \p{P2},
\p{P3}, \p{P4}\}$, $\mf{S}=\{{\cal S}(\peer{P1}),$ ${\cal S}(\peer{P2}),$ ${\cal S}(\peer{P3}),{\cal S}(\peer{P4})\}$, ${\cal S}(\peer{P1}) = \{R^1(\cdot,\cdot)\}$, ${\cal S}(\peer{P2}) =
\{R^2(\cdot,\cdot),$ $S^2(\cdot,\cdot)\}$, ${\cal S}(\peer{P3}) =
\{R^3(\cdot,\cdot)\}$,  ${\cal S}(\peer{P4}) =
\{R^4(\cdot,\cdot,\cdot)\}$. \ Furthermore,  $\trust$=$\{(\peer{P1},\less,\peer{P2}),
(\peer{P2},\same,$ $\peer{P3}),$ $(\peer{P4},\less,$ $\peer{P2}),
(\peer{P4},\less,\peer{P3})\}$, and $\Sigma$ contains:
\begin{eqnarray}
\Sigma(\peer{P1},\peer{P2}) &=& \{\forall x y(R^2(x,y) \rightarrow
R^1(x,y))\}, \label{eq:uno1}\\
\Sigma(\p{P2},\p{P2}) &=&  \{\forall x \forall y ( R^2(x,y) \wedge S^2(x,y)
\rightarrow {\bf false}),\label{eq:dos} \\
&&~~~~~~~~~~~~~~~~~~~~~~~~~~~~~~~~~~~\forall x \forall y \forall z(R^2(x,y) \wedge R^2(x,z) \rightarrow y = z)\}, \nonumber\\
\Sigma(\peer{P2},\peer{P3}) &=&\{\forall x  y ( R^2(x,y) \wedge
R^3(x,y) \rightarrow {\bf false})\}, \label{eq:four}\\
\Sigma(\peer{P4},\peer{P2}) &=&\{\forall x y  z(R^2(x,y) \wedge
S^2(y,z) \rightarrow R^4(x,y,z))\}, \label{eq:tres1}\\
\Sigma(\peer{P4},\peer{P3}) &=& \{\forall x y ( R^3(x,y) \rightarrow
\exists z R^4(x,y,z))\}. \label{eq:cuatro1}
\end{eqnarray}
The built-in atom {\bf false} in (\ref{eq:dos}) and (\ref{eq:four})
is false in every instance. The DECs in $\Sigma(\p{P2},\p{P2})$ are the local integrity constraints for \p{P2}, \red{in this case a denial constraint (i.e. a prohibited join of
positive atoms) and a functional dependency, requiring that (the values for) the first attribute functionally determines (determine the values for) the second one.}
Here, $\Sigma(\p{P4}) = \{\forall x y  z(R^2(x,y) \wedge
S^2(y,z) \rightarrow R^4(x,y,z)), \linebreak \forall x y ( R^3(x,y) \rightarrow
\exists z R^4(x,y,z))\}$.

The DEC in (\ref{eq:cuatro1}) is an RDEC, and all the others are UDECs.
Some of the accessibility graphs are shown  in Figure \ref{fig:AG}. Notice that ${\cal
AC}(\peer{P1}) $ $=\{\peer{P1},\peer{P2}, \peer{P3}\}$, ${\cal
AC}(\peer{P2})$ $=\{\peer{P2},\peer{P3}\}$, ${\cal
AC}(\peer{P3})=\{\peer{P3}\}$, ${\cal AC}(\peer{P4})=\{\peer{P2},
\peer{P3},\peer{P4}\}$, ${\cal N}(\peer{P1})$ $=$
$\{\p{P1},\peer{P2}\}$, ${\cal N}(\peer{P2})=\{\p{P2},\peer{P3}\}$,
${\cal N}(\peer{P3})=\{\p{P3}\}$, ${\cal
N}(\peer{P4})=\{\peer{P2}, \peer{P3},\p{P4}\}$, ${\cal N}^{\circ}(\peer{P1})$ $=$
$\{\peer{P2}\}$, ${\cal N}^{\circ}(\peer{P2})=\{\peer{P3}\}$,
${\cal N}^{\circ}(\peer{P3})=\emptyset$, and ${\cal
N}^{\circ}(\peer{P4})=\{\peer{P2}, \peer{P3}\}$.

  As an example, peer $\p{P4}$ only knows its schema and database, the DECs from it to  peers \p{P2} and \p{P3}, and how much it trusts them. More precisely, $\p{P4}$  knows its schema $\mathfrak{P}(\p{P4})=\langle \mc{S}(\p{P4}), \Sigma(\p{P4}),\trust(\p{P4})\rangle$, with $\Sigma(\p{P4})=\{\forall x y  z(R^2(x,y) \wedge
S^2(y,z) \rightarrow R^4(x,y,z)),\forall x y ( R^3(x,y) \rightarrow
\exists z R^4(x,y,z))\}$, and $\trust(\p{P4})$=$\{(\peer{P4},\less,$ $\peer{P2}),
(\peer{P4},\less,\peer{P3})\}$. This peer also has its database instance $D(\p{P4})$.

The schema of \p{P1} is  $\mathfrak{P}(\p{P1})=\langle \{R^1(\cdot,\cdot)\}, \{\forall x y(R^2(x,y) \rightarrow
R^1(x,y))\},$ $\{(\peer{P1},\less,$ $\peer{P2})\}\rangle$.
As will be determined by the semantics later on, the combination of \p{P1}'s DEC in (\ref{eq:uno1}) and
the trust relationship to \p{P2}, will make \p{P1} import
all the missing  data from the extension of its neighbor's $R^2$ into the extension of its own
$R^1$.  \boxtheorem
\end{example}

\begin{example}\label{ex:intro1} (example \ref{ex:intro0} cont.) \
The PDES  can be formalized through the schema $\mathfrak{P}=\langle
\mc{P}, \mf{S}, \Sigma,$ $\trust\rangle$, where $\mc{P}=\{\p{P1},\p{P2}\}$, $\mf{S}=\{\mc{S}(\p{P1}),\mc{S}(\p{P2})\}$, $\Sigma=\{\Sigma(\p{P1},\p{P2})\}$, $\Sigma(\p{P1},\p{P2})=\{\forall x y(R^2(x,y)$ $\wedge$ $S^2(y,z)$ $\rightarrow$
$R^1(x,y,z)),$ $\forall x (S^1(x)$ $\rightarrow$ $S^2(5,x))\}$, and $\trust=\{(\p{P1},\less,$ $\p{P2})\}$.

The schema of \p{P1} is
$\mathfrak{P}(\p{P1})=\langle \mc{S}(\p{P1}),$ $\Sigma(\p{P1},\p{P2}),$ $\{(\p{P1},\less,$ $\p{P2})\} \rangle$, and that of \p{P2} is $\mathfrak{P}(\p{P2})=\langle \mc{S}(\p{P2}), \emptyset, \emptyset \rangle$.
The instance for the PDES, $\mathfrak{D}=\{D(\p{P1}),D(\p{P2})\}$, contains the two instances shown in Figure \ref{fig:fig0}. \boxtheorem
\end{example}
\begin{remark}\label{rem:cycles}
 From now on we assume that the graph $\G(\mathfrak{P})$ is
acyclic. \ As a  consequence,  for each particular peer \p{P}, $\G(\p{P})$ is also
acyclic. \capri{This is a global assumption, not local to any particular peer, that will be used to define the semantics of a PDES.}
\burg{(A discussion around this and related assumptions can be found in electronic 
\ref{app:disc}}.)
\boxtheorem
\end{remark}

\red{Notice that according to its definition, the accessibility graph has no self-loops. The finiteness of the set of peers and the acyclicity of the accessibility graph imply
that there must be ``sink" (or terminal) peers, without any outgoing edges. Also notice that we are not making the assumption that
for a peer \p{P} its initial instance $D(\p{P})$ has to satisfy the local constraints in $\Sigma(\p{P},\p{P})$. This applies in particular to sink peers
in the accessibility graph. They can have  local ICs, which may be violated.}

\ignore{
\comlore{I still think we should modify this last definition. I would write it in a more general way:

\blue{(b) A set $\Sigma$ of DECs may contain cyclic subsets formed by inclusion dependencies, in particular some
involving existential quantifiers. In some cases we will make the assumption that
the DECs in a set $\Sigma$ are {\em ref-acyclic} \cite{iidb06}, in the sense that they contain no cycles
involving RDECs.}}
}

\section{A General Semantic Framework for a PDES} \label{sec:semantics}

Before presenting the formal semantics of PDESs, we describe the intended semantics in intuitive and operational terms.

A query ${\cal Q} \in {\cal L}(\p{P})$  is posed to a peer \p{P} by a certain user \p{U}, who may be an external user or
another peer in $\Pe$. Now \p{P}, \red{depending on the query predicates,} inspects the DECs in $\Sigma(\p{P})$ to identify data owned by other neighboring peers \p{Q} that may be related to its
own data. If predicate $R^{\sp{Q}} \in {\cal S}(\p{Q})$ appears in $\Sigma(\p{P})$, \red{then \p{P} requests from
$\p{Q}$ the (contents of the) relation $R^{\sp{Q}}$. \ \p{Q} returns to \p{P} a possible modified version of its relation, \ignore{Of course, the query sent from \p{P} to
\p{Q} could be more focused or restricted, depending upon the DECs in $\Sigma(\p{P})$.}
because \p{Q} may have to take into account data from its own neighbors (in the same way \p{P} did when it received the user query), which could produce local inconsistencies at \p{Q}'s level, and they
have to be repaired; and so on.}

\ignore{\comlb{See new parts in green. Basically no queries anymore.}
\comlore{Perfect!}
}

For the purpose of presenting the semantics
of the system, that should be query-independent, we will assume for the moment that \p{P} receives whole relation instances from its neighbors. \red{We emphasize
that, in their turn, \p{P}'s neighboring peers \p{Q} consider data from their own neighbors, and so on; data that may also be eventually used by \p{P},
by transitivity (cf. details below).}

\ignore{
\comlore{The reviewer notes that this ``intuitive" explanation is misleading because the formalizations below do not treat the data obtained from neighboring peers as answers to queries and also because in one we pose the query over all the solutions (within the peer) and by the definition we have that the answer to the query to neighbor is from the intersections of its solution. It is dificult to explain here, everything at once. What do you think is better? (1) Add more explanation that shows that, for a query ${\cal Q}_{\sp{Q}}$: $R^{\sp{Q}}(\bar{x})$, the answers obtained from the core of a peer are the same as the ones obtained as the certain answers from all solutions of that peer. (2) Remove from here all reference to queries between peers and leave that for the discussion section where this is explained in detail? I get the impression that it might be too much information to make the connection between the core of a neighbor and the certain answers to a query that requests the whole table... }  }

The data \p{P} receives from another peer \p{Q} is  defined in a recursive manner, because \p{Q} may have connections to other peers who may contribute with data of their own. Eventually, after
receiving the requested relation instances from its neighbors,  \p{P} has now a database instance $D$
for  the expanded database schema ${\cal S}({\cal N}(\p{P}))$, which extends the initial instance $D(\p{P})$ by adding predicates and data from its neighbors. \ $D$ can be used to interpret the
DECs in $\Sigma(\p{P})$. Most likely, $D$ will not satisfy $\Sigma(\p{P})$.

When this extended instance $D$ is inconsistent with respect to $\Sigma(\p{P})$,
different alternatives to restore consistency \wrt  $\Sigma(\p{P})$ can be considered, but \red{the (possibly virtual) updates performed on the extended instance $D$ will have to both respect the trust relationships and
make sure that the consistent alternative instances stay
{\em as close as possible} to instance $D$ (for which a form of distance has to be introduced, as we do later in this section).}
In this way, a collection of consistent, and possibly virtual, extended instances for  \p{P}'s neighborhood emerges, the {\em neighborhood instances}. By restricting those instances to the schema of peer
\p{P}, the {\em solution instances} for \p{P} are obtained.

Now, we give the precise definition of neighborhood solution. In order to do so, we will assume that it is possible to compare arbitrary instances $D_1, D_2$ \wrt a fixed instance $D$ by
 means of a preorder relation $\preceq_D$ (i.e. a reflexive and transitive binary relation). If $D, D_1, D_2$ are database instances for the same
schema, the intuition behind the relationship $D_1 \preceq_D D_2$ is that {\em $D_1$ is at least as close to $D$ as $D_2$ (is to $D$)}.  We can define $D_1 \prec_D D_2$ iff $D_1 \preceq_D D_2$, but not
$D_2 \preceq_D D_1$, with the intuition that $D_1$ is closer to $D$ than $D_2$.

In our semantics, the participating instances will always be
for a neighborhood schema $\mc{S}(\mc{N}(\p{P}))$ of a peer \p{P}.
Even more, the preorder relation will also depend on $\Sigma(\p{P})$ (which has to be satisfied). \ignore{In Section \ref{sec:nullsRep} we will propose a concrete class of  relations $\preceq_D^{\Sigma(\sp{P})}$, which will lead to a special repair semantics.}

{\em As a consequence,
we will assume, for  $D$ a fixed instance for $\mc{S}(\mc{N}(\p{P}))$, the existence of a preorder relation  $\preceq_D^{\Sigma(\p{P})}$ on instances for the schema $\mc{S}(\mc{N}(\p{P}))$. }


\begin{definition} \label{def:localsolution}  \red{Given a PDES schema $\mathfrak{P} = \langle
\mc{P}, \mf{S}, \Sigma,\trust\rangle$,  a peer
\peer{P} $\in \Pe$, and an instance $\bar{D}$ for  the schema ${\cal S}({\cal N}(\p{P}))$:}
\begin{enumerate}[label=(\emph{\alph*}),leftmargin=2em]
 \item An instance $D'$ for the schema ${\cal S}({\cal N}(\p{P}))$ is a
{\em neighborhood solution} for $\p{P}$ and $\bar{D}$  if:
\begin{enumerate}[label=(\emph{\roman*}),leftmargin=2em]
\item $D'
\models
\Sigma(\peer{P})$. \item There is no instance $D''$
satisfying {\em (i)},  and $D'' \prec_{\bar{D}}^{\Sigma(\sp{P})}D'$.\ignore{, i.e.
$D'' \preceq_{\bar{D}}^{\Sigma(\sp{P})}
D'$, but not $D' \preceq_{\bar{D}}^{\Sigma(\sp{P})}
D''$.
}

\item $D'\rs \{R\} = \bar{D}\rs \{R\}$ for every
predicate $R \in {\cal S}(\peer{Q})$ with $(\p{P}, \nit{less},
\p{Q})$ $\in \trust$.
\end{enumerate}
\item  $\nit{N\!S}^\mathfrak{P}\!(\p{P},\bar{D})$ denotes the
set of neighborhood solutions for $\p{P}$ and $\bar{D}$.

When clear from the context, we will simply use {$\nit{N\!S}(\p{P},\bar{D})$}. \boxtheorem
\end{enumerate}
\end{definition}
A neighborhood solution for \p{P} is a database for its whole
neighborhood that satisfies \p{P}'s  DECs (including its local constraints), and respects its trust relationships.
A neighborhood solution also stays as close as possible to the original neighborhood instance, while staying the same for trustable peers. In operational and intuitive terms, \red{a minimal data set \wrt
Definition \ref{def:localsolution} is imported
or given up to satisfy the DECs}.


Notice that Definition \ref{def:localsolution} introduces an abstract {\em repair semantics}, similar to those used to handle inconsistency in single databases  \cite{bertossi11}.
\red{For example, a well-studied repair semantics for single databases was introduced in \cite{ABC99}. It is based on insertions or deletions of whole tuples into/from the original
inconsistent instance $D$, and the ``distances" of two instances $D_1, D_2$ from $D$ are compared using the symmetric differences: \ $D_1 \preceq_D^{\Delta,\Sigma(\sp{P})} D_2 :\Leftrightarrow \Delta(D,D_1) \subseteq
\Delta(D,D_2)$.\footnote{For sets $S_1$ and $S_2$, \ $\Delta(S_1,S_2) := (S_1 \smallsetminus S_2) \cup (S_2 \smallsetminus S_1)$.}}

For  a  peer \p{P} in isolation, i.e. with ${\cal N}^{\circ}(\p{P})=\emptyset$, a neighborhood solution of its original instance $D(\p{P})$ will simply be an instance $D'$ for its schema $\S(\p{P})$, such that:
(a) $D'
\models
\Sigma(\peer{P},\p{P})$ (i.e. \p{P}'s own integrity constraints); and (b) There is no instance $D''$
satisfying (a) and  $D'' \prec_{D(\p{P})}^{\Sigma(\sp{P},\p{P})}
D'$. \ This defines a class of
{\em repairs} of instance $D(\p{P})$ \wrt \red{$\Sigma(\sp{P},\p{P})$.\footnote{\red{We are using abstract preorder relations to define repair semantics. It should also be possible
to use abstract ``distance measures" to define repairs and repair semantics. Cf. \cite{Arieli}.}}} Thus, this  neighborhood-solution semantics for PDESs naturally
extends the notion of repair to neighborhood solutions.

\ignore{
\comlb{Agregue un footnote sobre las distancias. Si se te ocurre algo mas inteligente ... En todo caso, la respuesta es mas o menos como
la de por que no usamos FOL mas completion. Puse lo de la dif. simetrica arriba, y la saque del comienzo.}
\comlore{La distancia presentada en el paper de Denecker no permite modelar set inclusion. It says: ``preferences based on set inclusion cannot be simulated by distance functions". Por lo tanto tampoco es más general. Este comentario lo pondre en la respuesta, pero no creo que valga la pena aca en al articulo}
}

\begin{example} \label{ex:intro2} (examples \ref{ex:intro0} and \ref{ex:intro1} cont.) \
The elements of the PDES schema are all as before,  in particular,
\begin{equation}\label{eq:uds}
\hspace*{-6mm} \Sigma(\p{P1},\p{P2})=\{\forall x y(R^2(x,y) \wedge S^2(y,z) \rightarrow
R^1(x,y,z)), \ \forall x (S^1(x) \rightarrow S^2(5,x))\},
\end{equation} but this time consider $\{(\p{P1},\same,$ $\p{P2})\}\in \trust$.

Now assume, in this example, for illustration purposes,  that
the preorder relations  are defined by $D_1 \preceq_D^{\Delta,\Sigma(\sp{P})} D_2$ iff $\Delta(D,D_1) \subseteq \Delta(D,D_2)$.

If a query is posed to
\p{P1}, it will have to compute (possibly virtually) its neighborhood solutions. Peer \p{P1}'s knowledge of the schema is limited to its own schema $\mathfrak{P}(\p{P1})=\langle \mc{S}(\p{P1}), \Sigma(\p{P1}),\trust(\p{P1})\rangle$ where ${\cal S}(\p{P}1) = \{R^1(\cdot,\cdot,\cdot), S^1(\cdot)\}$, $\Sigma(\p{P1})=\Sigma(\p{P1},\p{P2})$, and
$\trust(\p{P}1)=\{(\p{P1},\same,$ $\p{P2})\}$.

In order to enforce $\Sigma(\p{P1})$, \p{P1} poses two atomic queries to peer \p{P2}: \ ${\cal Q}_1(x,y)$: $R^{2}(x,y)$ and ${\cal Q}_2(x,y)$: $S^{2}(x,y)$. Since $\p{P2}$ is not related to any other peer, and has no local ICs,
it will provide as answers the content of those relations in $D(\p{P2})$.
 In this way, \p{P1} has an extended instance $D$, shown in Figure \ref{fig:Dp1p2}, that corresponds to the union of the two instances \red{in
Figure \ref{fig:fig0}}, that is, $D=D(\p{P1}) \cup D(\p{P2})$. Since $D$ does not satisfy $\Sigma(\p{P1})$, $D$ has to be repaired.

\begin{figure}
\begin{minipage}{2cm}
        \hspace*{-3.2cm} \fbox{
        \begin{tabular}{c|c|c} 
        \multicolumn{3}{c}{${R^1}$}\\  \hline
         c &  4 & 2 \\
        f & 3 & 5 \\ \hline
        \end{tabular}\begin{tabular}{c} 
        \multicolumn{1}{c}{$S^1$}\\ \hline
         3 \\
       7 \\ \hline
        \end{tabular}~~
        \begin{tabular}{c|c}
        \multicolumn{2}{c}{$R^2$}\\  \hline
         c & 4 \\
        d & 5 \\   \hline
        \end{tabular}\begin{tabular}{c|c}
        \multicolumn{2}{c}{$S^2$}\\  \hline
        4  &  2  \\
         5 & 3 \\ \hline
        \end{tabular} }
        \end{minipage}
  \caption{A database instance $D$ for schema ${\cal S}({\cal N}(\p{P1}))$}
\label{fig:Dp1p2}
\end{figure}

\ignore{
\begin{figure}
\hspace*{5mm}\includegraphics[width=5cm]{FigEj3}
\caption{A database instance $D$ for schema ${\cal S}({\cal N}(\p{P}))$}
\label{fig:Dp1p2}
\end{figure}
}

Since $(\p{P1},\same,$ $\p{P2})$, there are several neighborhood solutions for \p{P1}, which are
obtained by virtually modifying both peers' data. For example, the
inconsistencies \wrt the first UDEC in (\ref{eq:uds}) can  be (minimally) solved by either
removing $R^2(d,5)$ from $R^2$, or removing $S^2(5,3)$ from $S^2$, or inserting $R^1(d,5,3)$ into
$R^1$.  If $S^2(5,3)$
were removed from $S^2$, a  new inconsistency is created \wrt  the second UDEC in (\ref{eq:uds}). This one can be
solved by removing $S^1(3)$ from $S^1$. The inconsistencies \wrt the second DEC can be
solved by either removing $S^1(7)$ from $S^1$ or
inserting $S^2(5,7)$ into $S^2$.

Figure \ref{fig:varias} shows the six neighborhood solutions in {$\nit{N\!S}^\mathfrak{P}\!(\p{P1},D)$}. All these neighborhood instances
are {\em repairs} in the sense of \cite{ABC99} of instance $D$ \wrt \p{P1}'s DECs. If we are locally interested only in \p{P1}, we consider
their  restrictions to \p{P1}' schema $\mc{S}(\p{P1})$. \boxtheorem
\end{example}

\begin{figure}
\hspace*{1cm}\includegraphics[width=11cm]{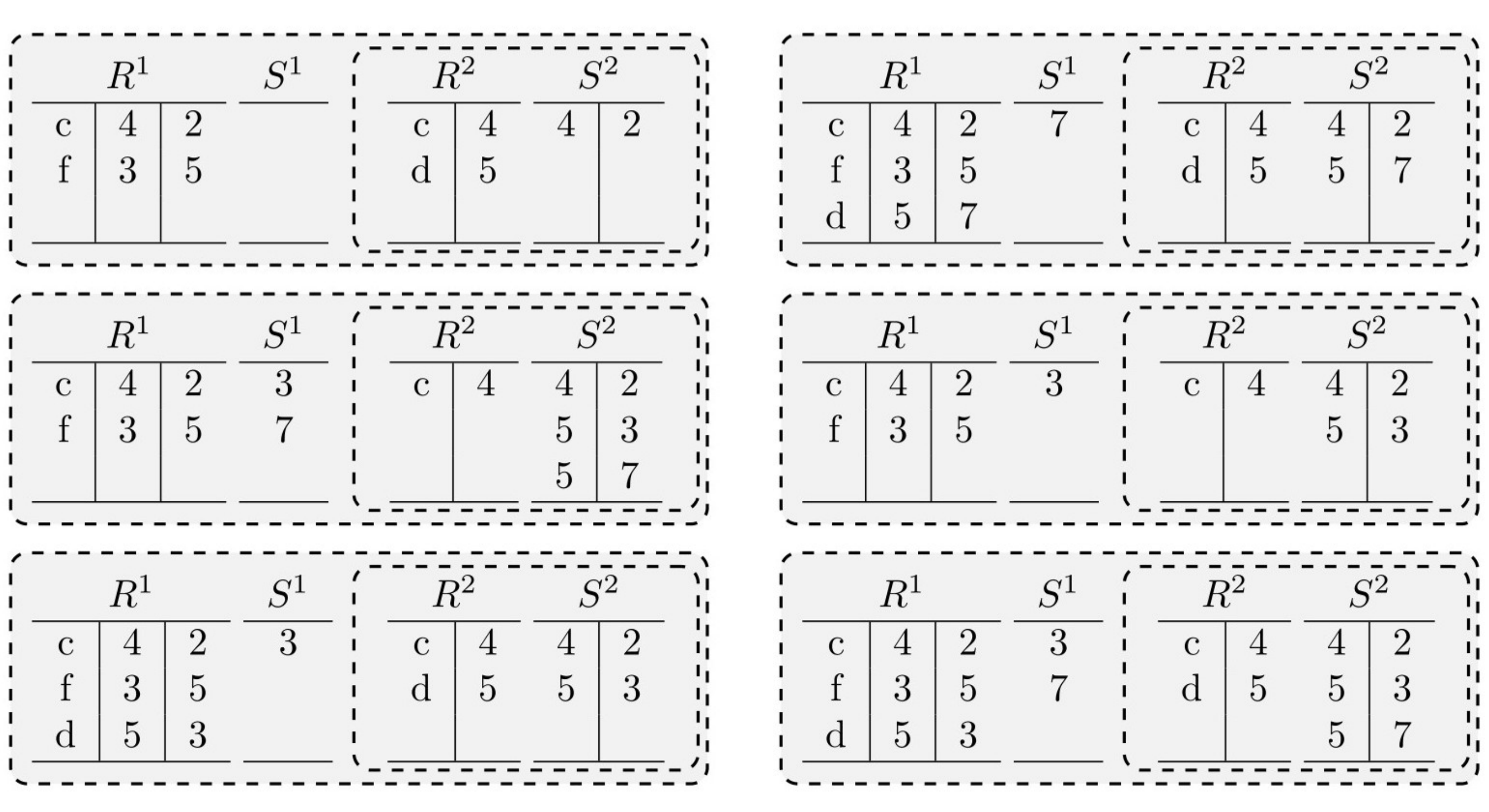}
\caption{Several neighborhood solution instances  for peer
\p{P1}\label{fig:varias}}
\end{figure}

%
%
%
%
%
%
%

\ignore{
\ni \red{In the previous example,} if \p{P1} has to answer a query ${\cal Q} \in {\cal L}(\p{P1})$, it will return -as will be formalized later in this section-
the answers that are simultaneously true in all its neighborhood solutions. For example, the expected answers to query $\mathcal{Q}(x): \exists y z
R^1(x,y,z)$  posed to \p{P1}  will be
$\langle c\rangle,  \langle f\rangle$, which are shared by all the six (restrictions to \p{P1}'s schema of the) neighborhood solutions for \p{P1}. We can see that the answers from a peer to a query are {\em certain answers}
\cite{imielinski}. The answers to queries ${\cal Q}_1$ and ${\cal Q}_2$ posed by $\p{P1}$ to $\p{P2}$ will be answered in the same way, taking into consideration the \p{P2}'s DECs.  This condition makes the data  moved
 from one peer to a neighbor always certain. In particular, this allows us to treat
external queries and inter-peer queries in a uniform manner. In particular, we will be in position to conceive and implement
the passage of data from one peer to a neighbor as a query answering process.
}



As the following example shows, trust relationships in combination with inconsistent DECs may cause that neighborhood solutions do not exist.

\begin{example} \label{ex:nosols} Consider
a PDES schema $\mathfrak{P} = \langle \mc{P}, \mf{S}, \Sigma,\trust\rangle$
with $\mc{P}=\{\p{P},\p{Q},\p{R}\}$, and
\begin{enumerate}[leftmargin=2em]
\item  $\mf{S}=\{\mathcal{S}(\p{P}),\mathcal{S}(\p{Q}),$ $\mathcal{S}(\p{R})\}$,
$\mathcal{S}(\p{P}) = \{P\}, \ \mathcal{S}(\p{Q})
= \{Q\}, \ \mathcal{S}(\p{R}) = \{R\}$.
\item $\Sigma=\{\Sigma(\p{P},\p{Q}), \Sigma(\p{P},\p{R}))\}$, \ \ $\Sigma(\p{P},\p{Q}) = \{\forall xy(Q(x,y) $
$\rightarrow P(x,y))\}$,\\ \hspace*{3.9cm}$\Sigma(\p{P},\p{R}) = \{\forall xy(P(x,y) \rightarrow
R(x,y))\}$.
\item $\trust=\{(\p{P},\nit{less},\p{Q}),$ $(\p{P},\nit{less},\p{R})\}$.
\item $\mathfrak{D}=\{D(\p{P}),D(\p{Q}) ,D(\p{R}) \}$, with $D(\p{P}) = D(\p{R})=\emptyset,$ and $D(\p{Q}) = \{Q(a,b)\}$.
\end{enumerate}
For peer $\p{P}$ and $D=D(\p{P}) \cup D(\p{Q}) \cup D(\p{R})= \{Q(a,b)\}$, there
is no neighborhood solution, i.e. $\NS(\p{P},D(\p{P})\cup D(\p{Q}) \cup D(\p{R}))$ is empty. This is because \p{P}'s DEC with $\p{Q}$ forces the insertion  of $P(a,b)$, but at the same time
the interaction of \p{P} with $\p{R}$ requires relation $P$ to be empty. Since peer $\p{P}$ trusts both peers more than itself, there is no neighborhood solution that satisfies the DECs and respect the trust relations.

\red{Notice that the DEC in  $\Sigma(\peer{P},\peer{R})$ acts
as a restriction on tuples for its owner, \p{P}, rather than as a tuple generator for it.}
\boxtheorem
\end{example}

\red{In the previous example, the trust relationships impose unsolvable conditions on neighborhood solutions. However, if all the  trust relationships are of the flexible form $(\p{P},\same,\p{Q})$, a peer always has solutions.} \ignore{The reason is that, with this kind of
trust relationships, we can always recreate the scenario of a single database whose consistency has to be restored in minimal way \wrt a set of ICs.
In this case, there are always repairs \cite{ABC99}.}

\begin{propositionS}\label{prop:Exists}
\red{Consider a PDES  schema $\mathfrak{P} = \langle
\mc{P}, \mf{S}, \Sigma,\trust\rangle$  and a peer $\p{P}\in\P$
whose trust relationships are all of the form  $(\p{P},\same,$ $\p{Q})$. Given an instance $\bar{D}$ for $\mc{S}(\mc{N}(\p{P}))$,  peer \p{P} always has a neighborhood
solution for $\bar{D}$.} \boxtheorem
\end{propositionS}

\ignore{
\dproof{\ref{prop:Exists}}{ Let $D_0$ be an empty instance
for the schema ${\cal S}({\cal N}(P))$. By being empty,  $D_0$ satisfies
$\bigcup_{\p{Q} \in {\cal N}(\p{P})} \Sigma(\peer{P},\peer{Q})$ (condition
(i) for a neighborhood solution). Also, since all the trust relationships are
of the
``same" kind, condition (ii) on neighborhood solutions is satisfied by $D_0$.
Then, either $D_0$ is a neighborhood solution, or there exists a neighborhood
solution $D''$ such that $D'' \preceq_{\bar{D}} D_0$.} }

\ignore{
\begin{remark} \label{rem:empty} As we see from the previous example, a peer may have no neighborhood solutions, i.e.
$\NS(\p{P}) = \emptyset$. In this case we will make the convention that
$\bigcap \NS(\p{P}) := \bigcap_{I \in \NS(\p{P})} I = \{\mathbf{inc}_\p{P}\}$, where $\mathbf{inc}_\p{P}$ is
a propositional built-in predicate in $\mathcal{S}(\p{P})$. An instance for the schema makes it true iff the instance contains the atom.
When $\NS(\p{P}) = \emptyset$, if another peer requests from \p{P} the intersection of its neighborhood
solutions, \p{P}  returns the instance $\{\mathbf{inc}_\p{P}\}$.

The idea is that when a peer becomes inconsistent, it makes another peer \p{Q} who request its data notice its
inconsistency by sending  this special instance. In this case, \p{Q} is expected to ignore its
mappings with \p{P}. This can be made precise and formally accommodated by modifying the DECS through the introduction of an extra
conjunct in the antecedent. For example, the DEC $\Sigma(\p{Q},\p{P}) = \{\forall x(P(x) \rightarrow Q(x))\}$ would
be replaced by   $\{\forall x(P(x) \wedge \neg \mathbf{inc}_\p{P} \rightarrow Q(x))\}$, which would
become trivially satisfied.  Alternatively, if we want to keep the built-ins in the consequent, we could
replace $\Sigma(\p{Q},\p{P})$ by $\{\forall x(P(x)  \rightarrow (Q(x) \vee \mathbf{inc}_\p{P}))\}$.

In order not to complicate the notation, we will refrain from explicitly introducing the $\mathbf{inc}_\p{P}$s in the  DECS. \boxtheorem
\end{remark} }

\vspace{2mm}
\ni
Having defined the notion of neighborhood solution for a peer, now we can define the notion of solution
for a peer, which is a local instance but takes all the accessible peers into account. The data distributed across different peers has to
be appropriately gathered  to build solution instances for the peer. This definition of solution instance is recursive,
and appeals to that of neighborhood solution. The intuition and idea are as follows: (a) Assume each neighbor \p{Q} of peer \p{P} has
the class $\nit{Sol}(\p{Q}\red{,\mathfrak{D}})$ of its solution instances (hence the recursion). Each of these \p{Q} passes to \p{P} the intersection
$\bigcap \nit{Sol}(\p{Q},\red{\mathfrak{D}})$ of its solution set.\footnote{When $\mc{C}$ is a class of sets, then we will usually denote with $\bigcap \mc{C}$ the intersection of
 its elements, i.e. $\bigcap \mc{C} := \bigcap_{C \in \mc{C}}C$.} With them and its own initial instance $D(\p{P})$, \p{P} builds an instance $D$ for its
neighborhood $\mc{N}(\p{P})$. The solutions for \p{P} become the restrictions to \p{P}'s schema of the neighborhood solutions for $\p{P}$
\wrt $D$, $\Sigma(\p{P})$, and \p{P}'s trust relationships.

\ignore{
 The
data that a peer \p{P} receives from a neighbor \p{Q} to build its
own solutions is the {\em intersection of the solutions} for \p{Q}.
After \p{P} collects this data, only $\p{P}$'s DECs.}

Under this recursive definition, the solutions for
the neighbors have to be determined, under the same semantics. Base
cases of the recursion are peers with no DECs or with only local constraints, that is, when either $\Sigma(\p{P})= \emptyset$ or $\Sigma(\p{P})= \{\Sigma(\p{P},\p{P})\}$ . These
peers are sinks in the accessibility graph of \p{P}. In this regard, we recall that we made the assumption that the
accessibility graphs are acyclic.

\ignore{
Now we define in precise terms the class, $\nit{Sol}(\p{P})$, of solution instances for  a peer \p{P}. This definition
now considers all the peers that are accessible from \p{P}, not only its neighbors. However, the
general definition uses the definition of neighborhood solution, to specify how its neighbors build
their own solutions in their turn. The solution instances in  $\nit{Sol}(\p{P})$ are recursively defined:  }

\begin{definition} \label{def:solutionSem1}  \red{Given an  instance $\mathfrak{D}$
for the PDES schema $\mathfrak{P} = \langle
\mc{P}, \mf{S}, \Sigma,\trust\rangle$, and   a peer
 \peer{P},} an instance $D$
for the schema ${\cal S}(\peer{P})$ is a {\em solution instance}~ (or simply {\em solution}) for
\peer{P}, denoted $D \in \red{\nit{Sol}^\mathfrak{P}\!(\p{P},\mathfrak{D})}$ \ (or simply, $\nit{Sol}(\p{P},\red{\mathfrak{D}})$), iff:
\begin{itemize}[leftmargin=2em]
\item[(a)] For $\Sigma(\p{P})=\emptyset$: \ $D = D(\p{P})$ \ (\p{P}'s initial instance, the projection of $\mf{D}$ on $\mc{S}(\p{P})$).
\item[(b)] For $\Sigma(\p{P})=\{\Sigma(\p{P,\p{P}})\}$: \ ${D} \in \nit{N\!S}^\mathfrak{P}(\p{P},
D(\p{P}))$.
\item[(c)] For $\emptyset \neq \Sigma(\p{P}) \neq \{\Sigma(\p{P,\p{P}})\}$: \ $D=\overline{D}\rs \mc{S}(\p{P})$, where $\overline{D} \in \nit{N\!S}^\mathfrak{P}(\p{P},
D(\p{P}) \cup \bigcup_{\p{Q} \in {\cal N}^\circ(\sp{P})} \nit{Core}(\p{Q},\red{\mathfrak{D}}))$.

Here, $\nit{Core}(\p{Q},\red{\mathfrak{D}}):= \bigcap \nit{Sol}^\mathfrak{P}\!(\p{Q},\red{\mathfrak{D}})$
when $\nit{Sol}^\mathfrak{P}\!(\p{Q},\red{\mathfrak{D}}) \neq \emptyset$; and $\nit{Core}(\p{Q},\red{\mathfrak{D}}):= \{\mathbf{inc}_\p{Q}\}$, otherwise. The propositional built-in predicate  $\mathbf{inc}_\p{Q}$
in $\mathcal{S}(\p{Q})$
is true of an instance iff  it is contained in the latter as an atom.
\boxtheorem
\end{itemize}
\end{definition}

\ignore{
\comlore{Reviwer 3 notes that case (a) is a special case of case (c) (since there are no neighbors in ${\cal N}^\circ(\sp{P})$). He thinks we should not include it as a separate case
but just explain it as a remark after the definition. I like it how it is now... I think it is clearer. What do you think? If we don't change it we would need to also explain below that both (a) and (b) are base cases of case (c)} }

The base cases of the recursion are (a) and (b), where the solutions of a peer can be computed without data
from other peers. Case (c) corresponds to the properly recursive case, where, before determining \p{P}'s solutions, there is an extended instance $\overline{D}$ around \p{P}
formed by its
local instance $D(\p{P})$ plus, for each neighbor \p{Q}, the intersection, $\nit{Core}(\p{Q},\red{\mathfrak{D}})$, of all its solutions.
The  combined database $\overline{D}$ is for the schema ${\cal S}({\cal N}(\p{P}))$, and starting from $\overline{D}$, neighborhood
solutions for \p{P}  can be determined; and
their restrictions to \p{P}'s schema become \p{P}'s solutions.

Although, in Definition \ref{def:solutionSem1}, cases (a) and (b) are special cases of (c), we include them explicitly, for clarity. Notice that case (b) amounts to obtaining the repairs of a local instance \wrt
a set of local integrity constraints.

The following is a immediate consequence of Proposition \ref{prop:Exists}.

\begin{corollaryS}\label{cor:Exists}
\red{Consider a PDES  schema $\mathfrak{P} = \langle
\mc{P}, \mf{S}, \Sigma,\trust\rangle$ with trust relationships all of the form  $(\p{R},\same,$ $\p{Q})$,  and a peer $\p{P}\in\P$.
Given an instance $\mathfrak{D}$ for $\mathfrak{P}$, peer \p{P} always has a
solution, that is, $\nit{Sol}^\mathfrak{P}\!(\p{P},\red{\mathfrak{D}})\neq \emptyset$}. \boxtheorem
\end{corollaryS}

\ignore{\dproof{\ref{cor:Exists}}{All we need is notice that the possibly inconsistent sink peers in the accessibility graph always have local repairs
under the kind of DECS considered (ICs in that case). A solution for a peer \p{P} can then be obtained by recursively propagating back neighborhood solutions (that always exist by Proposition
\ref{prop:Exists}) for peers along the paths that contain
\p{P}.} }

\begin{remark} \label{rem:empty} As we have seen, even in the absence of cycles in ${\cal G}({\cal
P})$, a peer may have no neighborhood solutions (cf. Example \ref{ex:nosols}), and then no solutions, i.e.
$\nit{Sol}(\p{P},\red{\mathfrak{D}}) = \emptyset$. That is why, in this case, we make the convention in Definition \ref{def:solutionSem1} that
$\nit{Core}(\p{P},\red{\mathfrak{D}}) =  \{\mathbf{inc}_\p{P}\}$.
When $\nit{Sol}(\p{P},\red{\mathfrak{D}}) = \emptyset$, if another peer requests from \p{P} the intersection of its
solutions, \p{P}  returns the instance $\{\mathbf{inc}_\p{P}\}$.

The idea is that when a peer becomes inconsistent, it makes another peer \p{Q} who requests its data notice its
inconsistency by sending  this special instance. In this case, \p{Q} is expected to ignore its
mappings with \p{P}. This can be made precise and formally accommodated by modifying the DECs through the introduction of an extra
conjunct in the antecedent. For example, the DEC $\Sigma(\p{Q},\p{P}) = \{\forall x(P(x) \rightarrow Q(x))\}$ would
be replaced by   $\{\forall x(P(x) \wedge \neg \mathbf{inc}_\p{P} \rightarrow Q(x))\}$, which would
become trivially satisfied.  Alternatively, if we want to keep the built-ins in the consequent, we could
replace $\Sigma(\p{Q},\p{P})$ by $\{\forall x(P(x)  \rightarrow (Q(x) \vee \mathbf{inc}_\p{P}))\}$.

In order not to complicate the notation, we will refrain from explicitly introducing the $\mathbf{inc}_\p{P}$s in the  DECs. \boxtheorem
\end{remark}

\red{According to this convention about the treatment of peers without solutions, a peer that becomes intrinsically inconsistent, ``irreparable",
is ignored by its neighbors when they receive the notification of inconsistency. However, this form of ignoring is put here on a solid logical
foot, that is compatible and uniform with the treatment of other peers. As long as a peer declares itself as inconsistent, there is no much
a neighbor can do. However, an inconsistent peer might decide to relax its own consistency requirements and send to other peers only ``partially
consistent" data, which would be transparent to those receiving peers.\footnote{\red{An alternative to this design choice could be, in the case a peer \p{P} trusts an inconsistent peer \p{Q} more than itself,
 that \p{P} becomes or declares itself inconsistent as well. This alternative may be worth exploring, but we do not pursue it here any further.}}}


\ignore{
\comlb{See footnote for reviewer 3.}
\comlore{Perfect}  }

\ignore{\comlore{Reviewer 3 thinks that if a peer \p{P1} trusts more a peer \p{P2} which is inconsistent then \p{P1} should be inconsistent too. He thinks it makes
sense when they trust each other the same but not when the peer trusts itself less... I'd like to keep it as is. It is common practice to drop the inconsistent peer, but I don't know if that
is sufficient justification. } }

\ignore{
\begin{remark} As in remark \ref{rem:empty}, when $\nit{Sol}(\p{P}) = \emptyset$, we will adopt the convention
that $\bigcap \nit{Sol}(\p{P}) = \{\mathbf{inc}_\p{P}\!\}$. \boxtheorem
\end{remark}  }

\vspace{2mm}\ni The peer consistent answers from a peer to a query are the semantically
correct answers, which means that when answering the query,  \red{the peer consistently
considers the data of its neighbors and the trust relationships with them.}

\begin{definition} \label{def:peercons}  \red{Consider an instance $\mathfrak{D}$
for the PDES  schema $\mathfrak{P} = \langle
\mc{P}, \mf{S}, \Sigma,\trust\rangle$, and   a peer
 $\peer{P} \in \mc{P}$.} Let  ${\cal Q}(\bar{x}) \in {\cal L}(\peer{P})$
be a query, with $\bar{x}$ a possibly empty list of free variables.
\begin{enumerate}[leftmargin=2em]
\item If $\nit{Sol}(\peer{P},\red{\mathfrak{D}}) \neq \emptyset$:
\begin{itemize}
\item[(a)]
If $\bar{x} \neq \emptyset$, a  finite sequence $\bar{c}$ of constants  in $\mc{U}$ of the same length as $\bar{x}$
is a {\em peer
consistent answer} (PCA) to ${\cal Q}$ from \peer{P} iff  $D
\models {\cal Q}[\bar{c}]$ for every $D \in \nit{Sol}(\peer{P},\red{\mathfrak{D}})$.
\item[(b)] If \red{$\mc{Q}$ is Boolean} and $D \models \mathcal{Q}$ for every $D \in \nit{Sol}(\peer{P},\red{\mathfrak{D}})$, then $\nit{yes}$
is the only PCA to $\mathcal{Q}$.
\end{itemize}
\item If $\nit{Sol}(\peer{P},\red{\mathfrak{D}}) = \emptyset$, then $\mathbf{inc}_\p{P}$ is the
only PCA to $\mathcal{Q}(\bar{x})$.
\boxtheorem
\end{enumerate}
\end{definition}

 \red{For illustration, in Example \ref{ex:intro2}, if \p{P1} has to answer a query ${\cal Q} \in {\cal L}(\p{P1})$,} it  returns
the answers that are simultaneously true in all its neighborhood solutions. For example, under this semantics, the  answers to query $\mathcal{Q}(x): \exists y z
R^1(x,y,z)$  posed to \p{P1}  will be
$\langle c\rangle,  \langle f\rangle$, which are shared by all the six neighborhood solutions for \p{P1} (or better, by their restrictions to \p{P1}'s schema). The answers to queries ${\cal Q}_1$ and ${\cal Q}_2$ posed by $\p{P1}$ to $\p{P2}$ are answered in the same way, taking into consideration the \p{P2}'s DECs.

We can see that the answers from a peer to a query are {\em certain answers}
\cite{imielinski}.
This condition makes the data  moved
 from one peer to a neighbor always certain. In particular, this allows us to treat
external queries and inter-peer queries in a uniform manner. In particular, we will be in position to conceive and implement
the passage of data from one peer to a neighbor as a query answering process.

\begin{example} \label{ex:main}
Consider the PDES schema $\mathfrak{P}$ and instance $\mathfrak{D}$ represented in graph $\G(\mathfrak{P})$ in Figure \ref{fig:ext+}.
Here, $\Sigma=\{\Sigma(\peer{P1},\peer{P2}),\Sigma(\peer{P2},\peer{P3}),\Sigma(\peer{P4},\peer{P3})\}$, and:
\begin{itemize}[leftmargin=2em]
\item[-] $\Sigma(\peer{P1},\peer{P2})\!=$ $\{\forall x y z$
$(R^2(x,y)$ $\wedge$ $S^2(y,z)$ $\rightarrow$ $R^1(x,y,z)),$
$\forall x$ $(S^1(x)$ $\rightarrow$ $S^2(5,x))\}$.

\item[-] $\Sigma(\peer{P2},$ $\peer{P3})\!=$ $\{\forall x  y $ $(
S^2(x,y)$ $\rightarrow$ $R^3(x,y))\}$.
\item[-] $\Sigma(\peer{P4},\peer{P3})\!=$ $\{\forall x y z$ $(
R^3(x,y)$ $\rightarrow$ $R^4(x,y,3))\}$.
\end{itemize}

\begin{figure}
\hspace*{1cm}\includegraphics[width=9cm]{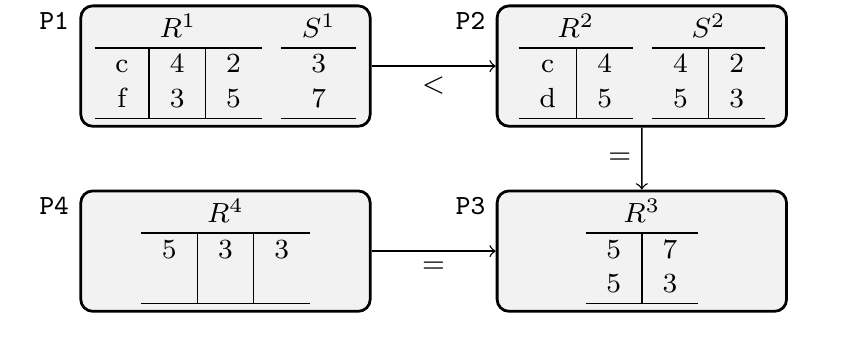}
\vspace{-3mm}\caption{PDES for Example \ref{ex:main}\label{fig:ext+}}
\end{figure}


\vspace{2mm}\ni In intuitive and procedural terms, if a query is posed to \p{P1}, it will send queries to
\p{P2}, to check the satisfaction of the DECs in
$\Sigma(\peer{P1},\peer{P2})$. But, in order for \p{P2} to
answer those queries, it will send queries to peer \p{P3} to
check the DECs in $\Sigma(\peer{P2},$ $\peer{P3})$. Since
\p{P3} is not connected to any other peer, it will answer
\p{P2}'s queries using its initial material instance $D(\p{P3})$.
Thus, the solutions for \p{P1} and its peer consistent answers will be affected by
the peers in ${\cal
AC}(\p{P1})=\{\p{P1},\p{P2},\p{P3}\}$.
More precisely, we can now apply the definitions of solution instance and peer consistent answer.

Since
\p{P3} has neither DECs
with other peers nor local integrity constraints, its only neighborhood solution is its
instance $D(\p{P3}) \in \mathfrak{D}$, which is sent back to \p{P2}. Now, \p{P2} has to
repair the extended instance $D = \{R^2(c,4),$ $R^2(d,5),$ $S^2(4,2),$ $S^2(5,3),$
$R^3(5,7),$ $R^3(5,3)\}$ \wrt $\Sigma(\peer{P2},\peer{P3})$, which is not satisfied due to the presence of tuple $S^2(4,2)$.
As
\p{P2} trusts \p{P3} the same as itself, it can modify its own
data or the data it got from \p{P3}.

Assuming  -for illustration purposes for the moment- that the preorder relation
on instances is given in terms of the symmetric set-difference, \p{P2} has two
neighborhood solutions: $\{R^2(c,4),$ $R^2(d,5),$ $S^2(5,3),$
$R^3(5,7),$ $R^3(5,3)\}$ and $\{R^2(c,4),$ $R^2(d,5),$ $S^2(4,2),$
$S^2(5,3),$ $R^3(5,7),$ $R^3(5,3),$ $R^3(4,2)\}$. They minimally depart from $D$, and their restrictions to \p{P2}'s schema lead to two
solutions for \p{P2}: $\nit{Sol}(\p{P2},\mathfrak{D}) = \{\{R^2(c,4), R^2(d,5),S^2(5,3)\},~
\{R^2(c,4),$ $R^2(d,5),$ $S^2(4,2),S^2(5,3)\}\}$.

Peer \p{P2} will send to \p{P1} the intersection of its solutions, namely~$\nit{Core}(\p{P2},\mathfrak{D})=\bigcap \nit{Sol}(\p{P2},\mathfrak{D})=\{R^2(c,4),$ $R^2(d,5),$ $S^2(5,3)\}$. Now, \p{P1} has to
repair the extended instance $\{R^1(c,4,2),$ $R^1(f,3,5),$ $S^1(3),$ $S^1(7),$
$R^2(c,4),$ $R^2(d,5),$ $S^2(5,3)\}$ \wrt
$\Sigma(\peer{P1},\peer{P2})$, which is not satisfied.

Since \p{P1} trusts \p{P2}
more, it will solve inconsistencies by modifying its own data, in this case inserting tuples into $R^1$ and deleting tuples
from $S^1$. This
produces only one neighborhood solution: $\{R^1(c,4,2),$
$R^1(f,3,5),$ $R^1(d,5,3),$ $S^1(3),$ $R^2(c,4),$ $R^2(d,5),$
$S^2(5,3)\}$. Thus, $\nit{Sol}(\p{P1},\mathfrak{D}) = \{\{R^1(c,4,2),$ $R^1(f,3,5),$
$R^1(d,5,3),$ $S^1(3)\}\}$.

If \p{P1} had received the query ${\cal Q}(x): \exists y z(R^1(x,y,z) \wedge S^1(y))$, the only peer consistent
answer would have been $\langle f\rangle$, obtained from its only solution.
 \boxtheorem
\end{example}

\red{Notice that when a peer \p{Q} passes its core, $\nit{Core}(\p{Q},\bar{D})$, with $\bar{D}$ a neighborhood instance, to a peer \p{P},
it is delivering the peers consistent answers to the atomic queries from \p{P} to \p{Q} of the form $\mc{Q}^R(\bar{x})\!: R(\bar{x})$, where $R \in \mc{S}(\p{Q})$.}
\capri{However, when \p{P} computes its local PCAs to a query, it does not use its own core, $\bigcap \nit{Sol}(\peer{P},\red{\mathfrak{D}})$, but the collection $\nit{Sol}(\peer{P},\red{\mathfrak{D}})$
as a whole. The reason is that this core is unnecessarily restrictive for peer consistent query answering. For example,
if the query is $\mc{Q}: \exists x P(x)$, and \p{P}'s solutions are $\{P(a)\}$ and $\{P(b)\}$, the PCA  to this Boolean query would be {\it no} if evaluated on the empty
core, but {\it yes} according to our definition.}

\ignore{\comlb{I added this paragraph, about instances vs. queries.}
\comlore{ok.}  }

\begin{remark} \label{rem:param}
\red{Notice that the definitions of solution instance for a peer and of peer consistent answer are parameterized by:}\\
  (a) The class of data exchange constraints. We have considered certain syntactic classes of FO-sentences, and we will also do so in the rest of this work. However, our presentation
  so far has been general enough to accommodate broader classes of FO sentences in the combined language of any two peers. The restriction on the classes of DECs
  has not been required or used yet.

  \vspace{1mm}\ni
  (b) A notion of satisfaction, $D \models \varphi$, where $D$ is a database instance and $\varphi$ is a FO sentence
  in the language of $D$'s schema. Most commonly, $D$ is a relational database considered as a FO structure, and classical logical satisfaction
  is used. However, if $D$ contains uncertain information, then we may have to depart from FO logic, as we will see in Section \ref{sec:nulls},
  for a particular PDES semantics.

   \vspace{1mm}\ni
   (c) The preorder relations $\preceq_D^{\Sigma(\sp{P})}$ between instances for the schema of an instance $D$. Different preorder relations
can be considered. In the examples above we have considered, just to fix ideas,  the common preorder based on the symmetric
  set difference between instances.

   \vspace{1mm}\ni
  (d) The repair semantics, i.e. by a characterization of the instances that minimally depart from a given one $D$ in order
  to satisfy local ICs or DECs between two peers. The repair semantics is based on the preorders $D'' \preceq_D^{\Sigma(\sp{P})}
D'$, and the associated minimality conditions (what we did
  in Definition \ref{def:localsolution}). \boxtheorem
\end{remark}

\section{Towards a Special  PDES Semantics  with NULL \ignore{\`a la SQL}}\label{sec:nulls}

Considering the rather abstract nature and flexibility of the semantic framework introduced in Section \ref{sec:semantics},
in this section, we  make  specific commitments about the semantic parameters
discussed in Remark \ref{rem:param}.

In the rest of this section, we consider a classical relational schema $\Sigma =(\mc{U}, \mc{R}, \mc{B})$, consisting of the data domain, a set of database predicates, and a set of built-in predicates.\footnote{In the
coming sections, where we will apply the semantics of this section, $\mc{R}$ will be the set of database predicates for a peer or a pair thereof.} Now, we will introduce in the schema some extra elements  related to the special constant, \nn, that we will use in the rest of this work.

\begin{remark}\label{rem:null}
We assume from now on that every attribute domain, $\nit{Adom}(A)$, contains the constant \nn. In particular, $\nn \in \mc{U}$.
Furthermore, among the built-in predicates in $\mc{B}$, we will also find
$\nit{I\!sNull}(\cdot)$, and  $\nit{I\!sNotNull}(\cdot)$. The first one is true only with constant $\nit{null}$, and  the second one,  on any constant $c$ other than $\nit{null}$.\footnote{Each is the negation of the other,
but we will keep  both in order to avoid using explicit negation.} Accordingly, \nn and these built-in predicates may appear in DECs as defined in (\ref{eq:UDEC}) and (\ref{eq:RDEC}), in particular, in integrity constraints,
and also in queries. They all become sentences or  formulas of the FO language $\mc{L}(\Sigma^\nulo)$ associated to $\Sigma^\nulo = (\mc{U}, \mc{R}, \mc{B}^\nulo)$, with $\mc{B}^\nulo
= \mc{B} \cup \{{\it IsNull}, {\it IsNotNull}\}$. \boxtheorem
\end{remark}

The  special constant $\nn$ in the data domain $\mc{U}$
is intended to behave and be used as the null value, \text{NULL}, in SQL relational databases.
The  new,
unary, built-in predicates  correspond to the SQL predicates
{\tt IS}\hspace{1mm}{\tt NULL} and {\tt IS}\hspace{1mm}{\tt NOT}\hspace{1mm}{\tt NULL}, used to check
null values.
Constant $\nn$ may appear in
database tuples, and will be used to restore consistency \wrt DECs and ICs.

 Using a single null, with its SQL semantics, is clearly different from using multiple
labeled null values, as is done in data exchange for enforcing the
satisfaction of existential quantifications \cite{kolaitis}. It is also different from using arbitrary elements of the underlying data domain
 for the same purpose \cite{CLR03b}. However, to a large extent, the  semantics of this section
could be developed without the
specific restrictions imposed on the representation and use of null values,  adopting other forms of handling incomplete and inconsistent data. \burg{Cf.  Section \ref{sec:related} and electronic \ref{app:disc} for additional
discussions of these issues.}


The semantics of \nn we  introduce next captures
the way nulls are
 handled by relational DBMSs that follow the SQL standard.
More precisely, our semantics
provides a partial logical reconstruction in first-order predicate
logic of the way nulls are handled in SQL databases.
It refines and extend previous work presented  in \cite{iidb06} on database repairs with and in the presence of ${\tt NULL}$.

The SQL Standard leaves many issues around ${\tt NULL}$
unspecified, and different DBMSs depart from the standard in different ways. As a consequence, it is not possible to provide a full
logical reconstruction of SQL databases with {\tt NULL}. For this reason, Accordingly, our semantics  concentrates on the notion
of  satisfaction of DECs and ICs, and query answering
for a broad classes of queries. Furthermore, the proposed semantics extends
the ``classical" notion of DEC and IC satisfaction, and query answering in databases without ${\tt NULL}$.\footnote{We reserve
the use of constant ${\tt NULL}$  to illustrate issues in relation its use  in SQL databases; and to emphasize this, also in subsection titles. Otherwise, we keep using only the constant \nn of the underlying data domain.}

\ignore{
In \cite{codd79}, Codd proposed three-valued logic with truth values \emph{true}, \emph{false}, and \emph{unknown} for
relational databases with {\tt NULL}. When a {\tt NULL} is involved in a comparison operation, the result is \emph{unknown}.
This logic has been adopted by the SQL standard, and partially implemented
in most common commercial DBMSs (with some variations). As a result, the semantics of
{\tt NULL} in both the SQL standard and the
commercial DBMSs is not quite clear; in particular, for IC satisfaction in the presence of {\tt NULL}.\\
The semantics for IC satisfaction with {\tt NULL}
 introduced in \cite{iidb06,Bravo07} presents a FO semantics of nulls in SQL databases. It is a reconstruction in classical logic of the treatment of {\tt NULL}  in
SQL DBs. More precisely, this semantics captures the notion of satisfaction of ICs, and also of query answering for a broad
class of queries in relational databases.  }

The rest of this section continues as follows. In Section \ref{sec:qa}, we illustrate  some of the elements of and issues around the notion of
query answer as used in SQL databases with ${\tt NULL}$.  It serves a motivation for Section \ref{app:nans}, where we formalize the semantics of query answering.
In Section \ref{sec:nullsDB}, we introduce a rewriting-based semantics for constraint satisfaction in the presence of ${\tt NULL}$, leading to a fully classical semantics. Finally, in Section \ref{sec:querRewr}, we apply the rewriting methodology to the semantics of query answering
under ${\tt NULL}$.

\subsection{Query answering under ${\tt NULL}$: motivation }\label{sec:qa}

A tuple $\bar{c}$ of elements of $\mc{U}$ is an answer to query $\mc{Q}(\bar{x})$, denoted $D \models_N \mc{Q}(\bar{c})$,
if the formula (that represents) $\mc{Q}$ is {\em classically true} when the
quantifiers on its {\em relevant variables} (or attributes) run over $({\cal U} \smallsetminus \{\nn\})$; and
those on  the non-relevant variables run over ${\cal U}$. The free relevant variables, i.e. relevant among those in $\bar{x}$, cannot  take the value $\nn$
either. (Relevance is made precise in Section \ref{app:nans}.)

\begin{example}
\label{ex:ansquery}
Consider  the instance $D_2$ and the query below:
\begin{multicols}{2}
\vspace{1mm}
{\small
\begin{tabular*}{4cm}{c|c|c|c|c c|c|}
  \cline{1-4} \cline{6-7}
  $R$ & $A$ & $B$ & $C$ &~~~~~& $S$ & $B$ \\ \cline{1-4}\cline{6-7}
   & $1$ & $1$ & $1$ & & &$\nn$ \\
   & $2$ &  $\nn$ & $\nn$ && & $1$\\
   & $\nn$ &  $3$ & $3$ & && $3$\\
  \cline{2-4} \cline{7-7}
 \end{tabular*}  }
\vspace*{-1.2cm}{\small \begin{equation}
\hspace*{5.2cm}
\mc{Q}_2(x)\!:
 \ \exists y \exists z(R(x,y,z) \ \wedge \ S(y) \ \wedge y>2). \ \ \ \ \ \ \label{eq:quer}
 \end{equation}}
 \end{multicols}

\vspace{4mm} A variable $v$ (quantified or not) in a conjunctive query is {\em relevant} if it appears (non-trivially) twice
 in the formula after the quantifier prefix. Occurrences of the form $v = \nn$ and $v \neq \nn$ do not count though.

In query (\ref{eq:quer}),
  the only relevant quantified variable is $y$, because it participates in a join and a built-in in the
 quantifier-free matrix of (\ref{eq:quer}).  So, there are two reasons for $y$ to be relevant. The only free variable is $x$, which is not relevant.
  As for query answers, the only candidate values for $x$ are: $\nn, 2, 1$. In this case,
 $\nn$ is a candidate value because $x$ is a non-relevant variable.

First, $x
= \nn$
is an answer to the query, because the formula $\exists y \exists z
(R(x,y,z)\wedge S(y) \wedge y > 2)$ is true in $D_2$, with a non-null witness value for $y$
and a witness value for $z$ that combined make the (non-quantified) formula true. Namely, $y=3, z=3$. So, it holds $D_2 \models_N \mc{Q}_2[\nn]$.

Next, $x = 2$ is not an answer. For this value of $x$, because the candidate value for $y$, namely $\nn$ that accompanies $2$ in $P$, makes the formula $(R(x,y,z)\wedge S(y)\wedge y > 2)$ false. Even if it were true, this value for $y$ would not be allowed.

Finally, $x = 1$ is not an answer, because the only candidate value for $y$, namely $1$, makes the formula false. In consequence, $\nn$ is the only answer.
\boxtheorem
\end{example}

 The next example with SQL queries and {\tt NULL} provides
additional intuition and  motivation for the formal semantics of Section
\ref{app:nans}. Notice the use in logical queries of the new unary predicates $\nit{I\!sNull}$ and $\nit{I\!sNotNull}$.

 \begin{example} \label{ex:nulls} Consider the schema $\mc{S} = \{R(A,B), S(B,C)\}$ and the instance in the table below. In it $\tt{NULL}$ is the SQL null. If this instance is stored in an SQL
 database, we can observe the behavior of the following queries when they are directly translated
 into SQL  and run on  an SQL DB:

\vspace{-1mm}
\begin{multicols}{2}
 \vspace{-2mm}
 {\small

 \begin{tabular*}{3cm}{c|c|c|c c|c|c|} \cline{1-3} \cline{5-7}
      $R$  &  $A$  &  $B$ & ~~~~~~~~&$S$  &  $B$  &  $C$ \\ \cline{1-3} \cline{5-7}
       &   a &   b &&&  b  &  h\\
         & a   &  c &&&{\tt NULL} & s\\
         & d &   $\tt{NULL}$&&&l &   m\\ \cline{6-7}
        &  d &   e\\
        &  u  &  u\\
        &  v  &  $\tt{NULL}$\\
     &     v  &   r \\
     & $\tt{NULL}$ & $\tt{NULL}$ \\ \cline{2-3}
     \end{tabular*} }

\ignore{
\begin{tabular}{c|c|c|} \cline{1-3}
      $R$  &  $A$  &  $B$ \\ \cline{1-3}
       &   a &   b \\
         & a   &  c\\
         & d &   $\tt{NULL}$\\
        &  d &   e\\
        &  u  &  u\\
        &  v  &  $\tt{NULL}$\\
     &     v  &   r \\
     & $\tt{NULL}$ & $\tt{NULL}$ \\ \hhline{~--}
     \end{tabular}

     \begin{tabular}{c|c|c|} \cline{1-3}
      $S$  &  $B$  &  $C$ \\ \cline{1-3}
    &b  &  h\\
    &{\tt NULL} & s\\
    &l &   m\\ \hhline{~--}
    \end{tabular}
}

\vspace{5mm}
\noindent (a)  $\mc{Q}_1(x,y)\!: R(x,y) \ \wedge \ y = \nn$

SQL:\hspace{1mm}\verb+Select * from R  where+
\verb+       B = NULL;+

\vspace{-1mm}Result: No tuple

\noindent (b) \
 $\mc{Q}_1'(x,y)\!: R(x,y) \ \wedge \ \nit{I\!sNull}(y)$

SQL: Now uses {\tt IS}\hspace{1mm}{\tt NULL}

Result:  $\langle \mbox{d}, {\tt NULL}\rangle$, $\langle \mbox{v}, {\tt NULL}\rangle$,\\ \hspace*{15mm} $\langle {\tt NULL}, {\tt NULL}\rangle$

\end{multicols}

\vspace{-5mm}
\begin{multicols}{2}
\noindent (c) \ $\mc{Q}_2(x,y): \ R(x,y)  \wedge  y \neq \nn$

SQL: \verb+Select * from R where     +\\
\verb+                 B <> NULL;+

Result: No tuple

\noindent (d) \ $\mc{Q}_2'(x,y): \ R(x,y)  \wedge  \nit{I\!sNotNull}(y)$

SQL: Now uses {\tt IS}\hspace{1mm}{\tt NOT}\hspace{1mm}{\tt NULL}

Answer: The five expected tuples

\noindent (e) \
$\mc{Q}_3(x,y): R(x,y) \wedge x = y$

SQL: \verb+Select * from R where A = B;+

Result: \ $\langle \mbox{u},\mbox{u}\rangle$

\noindent (f) \
$\mc{Q}_4(x,y): R(x,y) \wedge  x \neq y$

SQL: \verb+Select * from R where A <> B;+

Result:  \ $\langle \mbox{a},\mbox{b}\rangle, \langle \mbox{a},\mbox{c}\rangle, \langle \mbox{d},\mbox{e}\rangle, \langle \mbox{v},\mbox{r}\rangle$
\end{multicols}

\vspace{-3mm}
\noindent (g) \
$\mc{Q}_5(x,y,x,z):  R(x,y) \wedge R(x,z) \wedge y \neq z$

SQL: \verb+Select * from R r1, R r2 where+
\verb+r1.A = r2.A and r1.B <> r2.B;+

Result: \ $\langle \mbox{a},\mbox{b},\mbox{a},\mbox{c}\rangle, \langle \mbox{a},\mbox{c},\mbox{a},\mbox{b}\rangle$

\noindent (h) \ $\mc{Q}_6(x,y,z,t)\!: R(x,y) \wedge S(z,t) \wedge y=z$

SQL: \ \hspace{-1mm}\verb+Select * from R r1, S s1+
\verb+where r1.B = s1.B;+

Result: $\langle \mbox{a},\mbox{b},\mbox{b},\mbox{h}\rangle$

\noindent (i) As in (h), but now \ SQL: \ \verb+Select * from R r1 join S s1+
\verb+on r1.B = s1.B;+

Result:\footnote{The same result is obtained from
DBMSs that do not require an explicitly equality together with the join.} \ $\langle \mbox{a},\mbox{b},\mbox{b},\mbox{h}\rangle$

\noindent (j) $\mc{Q}_7(x,y,z,t): R(x,y) \wedge S(z,t) \wedge y \neq z$. \ \ \ In SQL:

\noindent \verb+Select R1.A, R1.B, S1.B, S1.C+
\verb+from R R1, S S1 where R1.B <> S1.B';+

\noindent Result: $\langle\mbox{a},\mbox{c},\mbox{b},\mbox{h}\rangle, \langle \mbox{d},\mbox{e},\mbox{b},\mbox{h}\rangle, \langle \mbox{u},\mbox{u},\mbox{b},\mbox{h}\rangle, \langle \mbox{v},\mbox{r},\mbox{b},\mbox{h}\rangle,$
$\langle \mbox{a},\mbox{b},\mbox{l},\mbox{m}\rangle, \langle \mbox{a},\mbox{c},\mbox{l},\mbox{m}\rangle, \langle \mbox{d},\mbox{e},\mbox{l},\mbox{m}\rangle,$ $\langle \mbox{u},\mbox{u},\mbox{l},\mbox{m}\rangle,\langle
\mbox{v},\mbox{r},\mbox{l},\mbox{m}\rangle$
\boxtheorem

\end{example}
We need to introduce predicates ${\it IsNull}$ and ${\it IsNotNull}$ in our formal treatment of nulls, because, as shown in Example \ref{ex:nulls}, in the presence of {\tt NULL}, SQL treats {\tt IS}\hspace{1mm}{\tt NULL} and
{\tt IS}\hspace{1mm}{\tt NOT}\hspace{1mm}{\tt NULL}  differently from classical $=$ and $\neq$, resp.  For example, the queries
\begin{equation}
\mc{Q}(x)\!:  \exists y(R(x,y)\wedge\nit{I\!sNull}(y)) \ \mbox{  and } \ \mc{Q}'(x)\!:$ $\exists y(R(x,y)\wedge y = \nn) \label{eq:que}
\end{equation}
 are both conjunctive queries\ignore{ of $L(\Sigma^\nulo)$}, but in
 SQL databases, they have different semantics.

\subsection{Query answering under ${\tt NULL}$: the semantics}\label{app:nans}
Here we  introduce the semantics of FO conjunctive query answering in relational databases with null values.
More precisely, in SQL relational databases with a single null value, $\nit{null}$, that is handled like the SQL {\tt NULL}.
We will exclude from the ``SQL-like" conjunctive queries those such as (a) and (c) in Example \ref{ex:nulls}. We will concentrate
only on conjunctive queries with built-ins.

\begin{definition} (a) $\nit{Conj}(\Sigma^\nulo)$ denotes the class of conjunctive queries in
$\mc{L}(\Sigma^\nulo)$ of the form
\begin{equation}
\mc{Q}(\bar{x}): \ \exists \bar{y}(A_1(\bar{x}_1) \wedge \cdots \wedge A_n(\bar{x}_n)), \label{eq:conjunctive}
\end{equation}
where $\bar{y} \subseteq \bigcup_i \bar{x}_i$, $\bar{x} = (\bigcup_i \bar{x}_i) \smallsetminus \bar{y}$, and
the $A_i$ are atoms containing any of the predicates in $\mc{R} \cup \mc{B}^\nulo$
plus terms, i.e. variables or constants in $\mc{U}$.\\
(b) $\nit{Conj}^\sql(\Sigma^\nulo)$  denotes the class of  conjunctive queries as in (a) whose
conjuncts are never
of the form  $t = \nn,  \ t \neq \nn$, with $t$ a term ($\nn$ or  not).
\boxtheorem
\end{definition}

For example, for the queries in (\ref{eq:que}), \ $\mc{Q}, \mc{Q'} \in \nit{Conj}(\Sigma^\nulo)$, \ $\mc{Q} \in \nit{Conj}^\sql(\Sigma^\nulo)$, but $\mc{Q'} \notin \nit{Conj}^\sql(\Sigma^\nulo)$.
Notice that $\nit{Conj}^\sql(\Sigma^\nulo), \nit{Conj}(\Sigma) \subseteq \nit{Conj}(\Sigma^\nulo)$.
The idea  is to force conjunctive queries {\em \`a la} SQL that
explicitly mention the null value in (in)equalities, to use the built-ins $\nit{InNull}$
or $\nit{I\!sNotNull}$.

\begin{definition} \label{def:relQ}
Given a query in $\nit{Conj}(\Sigma^\nulo)$  of the form
$\mathcal{Q}(\bar{x})\!\!: \exists \bar{y} \psi(\bar{x},\bar{y})$, with  $\psi$
quantifier-free,
a variable $v$ is  {\em relevant} for $\mathcal{Q}$  if it occurs at least twice in  $\psi$, without
considering the atoms ${\it IsNull}(v),$ ${\it IsNotNull}(v),$ $v~\theta~\nn,$ or $\nn~\theta~v,$ with
$\theta \in \mc{B}$. $\nit{RelV}(\mathcal{Q})$ denotes the set of relevant variables for $\mc{Q}$.
\boxtheorem
\end{definition}

For example, for the query $\mc{Q}(x): \exists y(P(x,y,z)\wedge Q(y)\wedge\nit{I\!sNull}(y))$, $\nit{RelV}(\mc{Q}(x)) = \{y \}$, because $y$ is used twice in the subformula $P(x,y,z)\wedge Q(y)$. \ignore{Actually, the notion of relevant variable can be applied to any first-order formula written in prenex normal form.}

As usual in FO logic, we consider assignments from the set, $\nit{Var}$, of variables to the underlying database domain $\mc{U}$ (that contains
constant $\nn$), i.e. $\sigma: \nit{Var} \rightarrow
\mc{U}$. Such an assignment can be extended to terms, mapping variables $x$ to $\sigma(x)$, and  $c \in
\mc{U}$ to $c$.  For an assignment $\sigma$, a variable $y$ and a constant $c$, $\sigma\frac{y}{c}$ denotes the
assignment that coincides with $\sigma$ everywhere, except possibly on $y$, that takes the value $c$. \ignore{We define similarly the extension $\bar{s}\frac{y}{c}$.} Given a formula $\psi$, $\psi[\sigma]$ denotes the formula obtained from $\psi$ by replacing its free variables by their values according to $\sigma$.

Now, given a formula (query) $\chi$ and an  assignment  $\sigma$, we verify if instance $D$ satisfies $\chi[\sigma]$ by assuming that the quantifiers on relevant variables range over $({\cal U} \smallsetminus \{\nn\})$, and
those on non-relevant variables range over ${\cal U}$.

\begin{definition} \label{df:nqs} Let $\chi \in \nit{Conj}(\Sigma^\nulo)$,
and $\sigma$ be an
assignment.  Instance $D$ with $\sigma$ {\em satisfies $\chi$ under the null-semantics},
denoted $D \models_{\!_N} \chi[\sigma]$, exactly  in
the following cases: \ (below $t,t_1, \ldots$ are terms; and $x, x_1,x_2$  variables)

\noindent     1. (a) $D \models_{\!_N} \nit{I\!sNull}(t)[\sigma]$, with $\sigma(t) =
\nn$. (b) $D \models_{\!_N} \nit{I\!sNotNull}(t)[\sigma]$, with\linebreak \hspace*{9mm} $\sigma(t) \neq
\nn$.

\noindent 2. \ $D \models_{\!_N} (t_1 < t_2)[\sigma]$, with $\sigma(t_1) \neq \nn \neq \sigma(t_2)$, and $\sigma(t_1) < \sigma(t_2)$.\footnote{Of course, when there is an order relation on $\mc{U}$. We could introduce ``$>$" similarly.}

     \noindent 3.
     \ (a) $D \models_{\!_N} (x = c)[\sigma]$, with $\sigma(x) = c \in
(\mc{U}\smallsetminus \{\nn\})$. \ \  (or
symmetrically)\footnote{Notice the
use of the symbols $=$ and $\neq$ both at the object and the meta levels.}

     \ (b) $D \models_{\!_N} (x_1 = x_2)[\sigma]$, with $\sigma(x_1) = \sigma(x_2)
\neq \nn$.

 ~(c) $D \models_{\!_N} (c = c)[\sigma]$, with $c \in (\mc{U}\smallsetminus
\{\nn\})$.

\noindent 4.  \ (a) $D \models_{\!_N} (x \neq c)[\sigma]$, with $\nn \neq \sigma(x) \neq c \in (\mc{U}\smallsetminus
\{\nn\})$. \ \
(or symmetrically)

\ (b) $D \models_{\!_N} (c_1 \neq c_2)[\sigma]$, with $c_1 \neq c_2$, and $c_1, c _2 \in (\mc{U}\smallsetminus
\{\nn\})$.

\noindent
5. \ $D \models_{\!_N} R(t_1,\dots,t_n)[\sigma]$, with $R \in {\cal R}$, and $R(\sigma(t_1),$ $\dots,$ $\sigma(t_n)) \in D$.

\noindent  6. \ $D \models_{\!_N} (\alpha \wedge \beta)[\sigma]$, with $\alpha$, $\beta$ quantifier-free, $\sigma(y) \neq \nn$ for
every $y \in \nit{RelV}(\alpha \wedge \beta)$,\linebreak \hspace*{8mm} and
  $D \models_{\!_N} \alpha[\sigma]$ and $D \models_{\!_N} \beta[\sigma]$.

\noindent  7. $D \models_{\!_N} (\exists y \ \alpha)[\sigma]$ when: (a) if $y \in \nit{RelV}(\alpha)$, there is $c$ in $({\cal U} \smallsetminus
\{\nn\})$ with \linebreak \hspace*{4mm} $D \models_{\!_N} \alpha[\sigma \frac{y}{c}]$; or (b) if $y \not \in \nit{RelV}(\alpha)$, there is $c$ in ${\cal
    U}$ with $D \models_{\!_N} \alpha[\sigma \frac{y}{c}]$.\boxtheorem
\end{definition}

This semantics also applies to $\nit{Conj}^\sql(\Sigma^\nulo)$.

\begin{definition} \label{def:nvs}  Let $\mathcal{Q}(\bar{x}): \exists \bar{y} \psi(\bar{x},\bar{y})$ be
 in $\nit{Conj}(\Sigma^\nulo)$, with $\bar{x} = x_1, \ldots, x_n$, and $\psi$ quantifier-free.

\noindent (a)  A tuple $\langle c_1,\dots,c_n\rangle \in \mathcal{U}^n$
is an {\em $N$-answer from $D$ to $\mc{Q}$}, denoted $D \models_{\!_N} \mathcal{Q}[c_1,\ldots,c_n]$, iff there is an assignment $\sigma$ such
that  $\sigma(x_i)=c_i$, for $i=1,\dots,n$; and $D
\models_{\!_N} (\exists \bar{y} \psi)[\sigma]$.

\noindent (b) If $\mc{Q}$ is a sentence (a Boolean query), the $N$-answer  is {\it yes} iff $D \models_{\!_N} \mc{Q}$,
and $\nit{no}$, otherwise.

\noindent (c) $\mc{Q}^\nit{N\!sem}\!(D)$ denotes the set of
$N$-answers to $\mc{Q}$ from instance $D$. \ignore{\burg{Similarly,  $V^\nit{N\!sem}\!(D)$ denotes a view extension
according to the $N$-answer semantics: $V^\nit{N\!sem}\!(D)$ $=$ $(\mc{Q}^V\!)^\nit{N\!sem}\!(D)$.}}
\boxtheorem
\end{definition}

Notice that $D
\models_{\!_N} (\exists \bar{y} \psi)[\sigma]$ in (a) above requires, according to Definition \ref{df:nqs}, that the relevant
variables in the existential prefix $\exists \bar{y}$  do not take
the value $\nn$. The free variables, i.e. in $\bar{x}$, may take the value $\nn$ only when they are not
 relevant in the query. For illustration, in Example \ref{ex:ansquery}, since the free variable $x$ is not relevant,
 $\mc{Q}_2^\nit{N\!sem}\!(D_2) =\{\langle \nn\rangle\}$.

 \begin{example}\label{ex:first}
Consider the instance $D$ and the conjunctive query below.

\begin{multicols}{2}
{\small \hspace*{3mm}\begin{tabular*}{3cm}{c|c|c|}\cline{1-3}
  $R$ & $A$ & $B$ \\ \cline{1-3}
   & $a$ & $b$\\
   & $c$ &  $d$\\
   & $e$ &  $\nn$\\
  \cline{2-3}
 \end{tabular*}
\quad
 \begin{tabular*}{3cm}{c|c|c|}\cline{1-3}
  $S$ & $B$ & $C$ \\ \cline{1-3}
   & $b$ & $f$ \\
   & $d$ & $g$\\
   & $\nn$ & $j$ \\
  \cline{2-3}
 \end{tabular*} }

 $\mc{Q}(x,z)\!: \ \exists y (R(x,y) \wedge S(y,z))$.
\end{multicols}

Here, under the classical semantics,
$\mc{Q}(D) = \{\langle a, f\rangle,$ $\langle c, g \rangle, \langle e, j\rangle\}$,
treating $\nn$ as any other constant. However,
 $\mc{Q}^\nit{N\!sem}\!(D) = \{\langle a, f\rangle,$ $\langle c, g \rangle\} \subseteq \mc{Q}(D)$.
\boxtheorem
\end{example}

It is easy to prove that, for queries in $\nit{Conj}(\Sigma^\nulo)$: \
 $\mc{Q}^\nit{N\!sem}\!(D) \subseteq \mc{Q}(D)$.
 Furthermore, the $N$-query answering semantics coincides with classical FO query answering semantics in databases without \nn\ignore{ \cite{Bravo07,iidb06}}. More precisely, if $\nn \notin \mc{U}$ (and then it does not appear in $D$ or $\mc{Q}$ either): \ $D \models_{\!_N} \mc{Q}[\bar{t}]$ \  iff  \ $D \models \mc{Q}[\bar{t}]$.

\subsection{Constraint satisfaction under ${\tt NULL}$ via FO rewriting\ignore{: motivation}} \label{sec:nullsDB}

The notions of relevant attributes (or variables) and formula satisfaction under the null-semantics can be both extended to more complex
formulas. In particular, they can be applied to constraint satisfaction under SQL ${\tt NULL}$ \cite{iidb06,Bravo07}.
As expected, the
 satisfaction of a constraint $\psi$ by a database
that may contain $\nn$ depends upon the presence of
$\nn$ in the {\em
relevant attributes} of $\psi$.
The following  is a generalization of  Definition \ref{def:relQ} to a larger class of formulas.

\begin{definition}  \label{def:rel} For $\psi \in \mc{L}(\Sigma^\nulo)$
 in prenex normal form,\footnote{That is, of the form $\bar{Q}\chi$, where $\bar{Q}$ is a prefix of quantifiers, and
$\chi$ is a quantifier-free formula. Actually, in this work all the formulas  have a quantifier prefix of the form
$\bar{\forall}\bar{\exists}$.} a variable $v$ is {\em relevant}
 if it occurs at least twice in  $\psi$, without
considering occurrences in quantifiers or atoms of the forms ${\it IsNull}(v),$ ${\it IsNotNull}(v),$ $v~\theta~\nn,$ or $\nn~\theta~v,$ where
$\theta$ a built-in comparison predicate.  $\nit{RelV}(\psi)$ denotes the set of relevant variables of $\psi$. 
\boxtheorem
\end{definition}

The constraints we are considering in this work, particularly those of the form (\ref{eq:RDEC}),  may not be in prenex normal form, but can be
easily transformed while keeping the same variables and their occurrences, so that relevant variables can be determined.
\ignore{However, in all the cases of sentences that commonly appear as mappings in  data exchange or  as integrity constraints in databases, relevance can be easily detected without having to do this transformation.
\ignore{Notice that also conjunctive queries  (which we privilege in this work) are in prenex normal form, \burg{and the notion of relevant variable can be applied to
them  \ (cf. \ref{app:queryAns})}.}}

\begin{example} \label{ex:gen} Consider the {\em referential integrity constraint} (RIC) on schema $\mc{R} = \{P(A,B,C),$ $ R(A,B,E)\}$:\ignore{\footnote{This is a case of
an RDEC on a single peer, i.e. belonging to a set of DECs of the form $\Sigma(\p{P},\p{P})$.}} \
$\psi\!: \ \forall x \forall y \forall z (P(x,y,z) \rightarrow \exists v R(x,y,v))$; and
\ignore{ It is
sometimes denoted as an inclusion dependency,
 $P[A,B] \subseteq R[A,B]$.} the instance $D$:

\begin{center}
\begin{tabular*}{5cm}{c|c|c|c| c c c c c c|c|c|c|}\cline{1-4} \cline{10-13}
  $P$ &$A$ & $B$ & $C$ &&&&&& $R$ & $A$ & $B$ & $E$  \\
\cline{1-4} \cline{10-13}  
   & $a$ & $5$ & $d$ &&&&&& &$a$ & $5$ & $3$ \\
     &$b$ & $\nn$ & $a$ &&&&&& & $a$ & $3$ & $7$ \\
       \cline{2-4} \cline{11-13}
\end{tabular*}\end{center}


\vspace{3mm}
DBMSs implement the so-called ``simple semantics" of the SQL Standard for
satisfaction of ICs. According to it, the database $D$ above satisfies the RIC.
This is because, for
every tuple $t$ in $P$, if $t[A]$ and $t[B]$ are different
from $\nn$, there is a tuple $t'$ in $R$ with $t[A,B]=
t'[A,B]$.

In this case, and informally, the attributes that are relevant  for checking
the satisfaction of the RIC (i.e. those we attempt
to capture through the relevant variables) are $A$ and $B$, in both $P$ and $R$. If we try
to insert tuple $P(c,d,\nn)$ into $P$, the DBMS will reject the
insertion, because none of the attributes that are relevant for checking the
constraint are $\nn$, and there is no tuple $R(c,d)$.

More precisely,  the set of relevant variables for $\psi$  is
$\nit{RelV}(\psi)=\{x,y\}$, because $x$ and $y$ appear twice in $\psi$. Accordingly, the values for
attributes $C$ and $E$ are not relevant when checking the satisfaction of $\psi$, which makes sense.
\boxtheorem
\end{example}
\ignore{Relevant variables in a formula, e.g. an IC or a query, are relevant in the
context of commercial DBMSs
that follow (closely enough) the SQL standard, at least \wrt the treatment of the SQL constant {\tt NULL}. }

\ignore{
\burg{Basically, a constraint is satisfied when any of the
relevant variables takes the value ${\tt NULL}$ or when the constraint is satisfied
according to the classical notion of first-order logic satisfaction, with ${\tt NULL}$
treated as any other constant.?????}} \ignore{In the rest of this work we  keep using the constant \nit{null} (as opposed to ${\tt NULL}$), but our semantics treats
it as the SQL ${\tt NULL}$.}


We will now formalize  constraint satisfaction under the null-semantics. In this direction, \ignore{order to introduce the semantics of $\nn$ in our PDES context, at least the part of it that we need, we  start
by generalizing the classes of DECs introduced in Section \ref{sec:semantics},} we consider a single class of constraints that includes those in (\ref{eq:UDEC}) and (\ref{eq:RDEC}), and can be handled in a
uniform manner. It also includes all the common constraints used in data management.
More precisely, we consider constraints that are sentences in $\mc{L}(\Sigma^\nulo)$ of the form:

\begin{equation}\label{eq:GenDEC} \forall \bar{x}(\bigwedge_{i = 1}^{n}
R_i(\bar{x}_i) ~\longrightarrow~ \bigvee_{j=1}^m C_j), \vspace{-2mm}
\end{equation}
where $m,n \neq 0$,  $R_i$ is a predicate in $\mc{R}$, \ignore{${\cal S}(\p{P}) \cup {\cal
S}(\p{Q})$,} $C_j$ is a conjunctive formula of the form \ $\exists
\bar{y}_j\bigwedge_{k=1}^l Q_{jk}(\bar{x}_{jk}',\bar{y}_{jk})$, where each
$Q_{jk}$ is a predicate in $\mc{R}$  or a built-in,\footnote{Occurrences of variables in built-ins have to be {\em safe}, i.e.
they also appear in $\bar{x}$ or in database predicate in the same conjunction.}  $\bar{x}=\bigcup_{i=1}^n \bar{x}_i$, $\bar{x}_{jk}'
\subseteq \bar{x}$, and $\bar{y}_j=\bigcup_{k=1}^l \bar{y}_{jk}$.

Without loss of generality, we  assume that the existentially
quantified variables (the $\bar{y}_j$) do  not appear in $\bar{x}$ and are different for each conjunctive literal
$C_j$. Notice that (\ref{eq:GenDEC}) allows formulas with only built-in atoms in the consequent.
We  also assume  that they do not have any {\em explicit} occurrence of the constant $\nit{null}$.\footnote{This is not
an essential requirement, but will simplify the presentation. Furthermore, all reasonable
DECs and local ICs do not require the explicit use of \nit{null} as a constant. If we need to express a relational NOT-NULL-constraint,
we can say, e.g. $\forall x \forall y(R(x,y) \rightarrow \nit{I\!sNotNull}(x))$.} Those formulas
may contain the  $\nit{I\!sNull}$ or $\nit{I\!sNotNull}$, but \ignore{their occurrences are all  in formula
$\varphi$ that involves the} as built-in predicates.

In order to define $N$-satisfaction of a constraint $\psi \in {\mc L}(\Sigma^\nulo)$ of the form (\ref{eq:GenDEC}), we first
rewrite it  into
a new FO formula $\psi^N$, which makes explicit the role played by the relevant attributes in $\psi$ (as in Definition \ref{def:rel}) and the way nulls
are handled in them. Next, satisfaction is defined in terms of the rewriting.

\ignore{
The next definition formalizes the rewriting methodology
for the class of formulas of the form
(\ref{eq:GenDEC}), that includes our UDECs and RDECs.   The definition  introduces at the same time the notion of formula satisfaction
in the presence of \nn,  indirectly via the rewriting.\footnote{As mentioned above, a direct, semantic definition that uses the relevant variables \burg{is
also given in \ref{app:queryAns}, as an extension of the query answering semantics in the presence of
nulls}.} }

\begin{definition}  \label{def:pseudosat}
Let $\psi
\in \mc{L}(\Sigma^\nulo)$ be a  constraint
of the form (\ref{eq:GenDEC}), i.e. $\forall \bar{x}(\bigwedge_{i = 1}^{n}
R_i(\bar{x}_i) ~\longrightarrow~ \bigvee_{j=1}^m C_j)$. \\ (a) \ {\em The $N$-rewriting of $\psi$} is the $\mc{L}(\Sigma^\nulo)$-sentence:
\begin{equation}\label{eq:Satrule2} \psi^N\!: \ \ \forall
\bar{x}(\bigwedge_{i = 1}^{n} R_i(\bar{x}_i)
   ~\rightarrow~ (\!\!\bigvee_{v \in \nit{RelV}(\psi) \cap \bar{x}} \!\!\!\!
   \nit{I\!sNull}(v) ~\vee~ \bigvee_{j=1}^{m} C_j^N)),
\end{equation}
with $\bar{x}= \cup_{i=1}^n \bar{x}_i$ and
\begin{equation}\label{eq:CjN}C_j^N= \exists \bar{y}_j (\bigwedge_{k=1}^l
Q_{jk}(\bar{x}_{jk}',\bar{y}_{jk}) \ \ \wedge \bigwedge_{\red{w
\in ((\nit{RelV}(\psi) \sm \bar{x})\cap \bar{y}_j)}} \!\!\!\! \nit{I\!sNotNull}(w)).
\end{equation}
(b) For an instance $D$, possibly containing \nn, $\psi$
 is {\em $N$-satisfied by
$D$}, denoted ~$D \models_{_N} \psi$, ~iff~ $D
\models \psi^N$.
Here, $D\models \psi^N$ refers
to classical first-order satisfaction, with $\nn$ treated as any
other constant of the domain.\boxtheorem
\end{definition}

We can see from Definition \ref{def:pseudosat} that there are
basically two cases when a ground instantiation of $\psi$ (obtained by assigning constants to the variables in $\bar{x}$) is immediately satisfied due to the presence of \nn: (a) When $\nn$
appears in any of the relevant attributes in the antecedent. (b) At least one of the conjunctive formulas $C_j$
is true, considering that when they are  checked according to equation
(\ref{eq:CjN}),  $\nn$ is treated as any other constant, but
the variables in existential joins do not take the value $\nn$ \ (hence the condition based on $w
\in \ (\nit{RelV}(\psi) \sm \bar{x})\cap \bar{y}_j$ in (\ref{eq:CjN})). \ The rewriting
can be applied in particular to our UDECs and RDECs, as follows.

\ignore{\comlb{This is what we have for UDECS, which is not what we have below:
\begin{eqnarray*}
 {\forall}\bar{x}(\bigwedge_{i = 1}^{n} R_i(\bar{x}_i) ~\longrightarrow~
\bigwedge_{k=1}^l(\bigvee_{j=1}^{m} Q_{kj}(\bar{y}_{kj}))),
\end{eqnarray*}
where the $R_i$ are predicates in ${\cal S}(\p{P}) \cup {\cal
S}(\p{Q})$, the $Q_{kj}$ are  atomic formulas with predicates  in ${\cal S}(\p{P}) \cup {\cal
S}(\p{Q})$ or atoms with  built-in predicates  in $\mc{B}$
(e.g. $=, \neq, <,$ $\mathbf{false}, \nit{I\!sNull}, \nit{I\!sNotNull}, ...$), and $\bar{x}=\bigcup_{i=1}^n \bar{x}_i$
and $\bar{y}_{kj} \subseteq \bar{x}$.
}  }

For a UDEC
$\psi$ of the form (\ref{eq:UDEC}), \ $\psi^N$ is:
\begin{equation}
\forall \bar{x}(\bigwedge_{i = 1}^{n} R_i(\bar{x}_i)
   ~\rightarrow~ (\!\!\!\!\bigvee_{\red{v \in \nit{RelV}(\psi)}} \!\!\!\!\!\!
   \nit{I\!sNull}(v) ~\vee~ \bigwedge_{k=1}^l(\bigvee_{j=1}^m Q_{kj}(\bar{y}_j)))). \label{eq:rewUDEC}
    \end{equation}

For  an RDEC $\psi$ of the form (\ref{eq:RDEC}), \
$\psi^N$ is:
\begin{eqnarray}
\forall \bar{x}(\bigwedge_{i = 1}^{n} R_i(\bar{x}_i)
   \ \rightarrow \ (&&\!\!\!\!\!\!\!\!\!\!\!\!\!\bigvee_{v \in (\nit{RelV}(\psi) \cap \bar{x})} \!\!\!\!\!\!\!\!
   \nit{I\!sNull}(v) ~\vee \exists \bar{y}(\bigwedge_{k=1}^l
(Q_{k}(\bar{x}_{k},\bar{y}_k) \wedge  \varphi_k(\bar{x}_{k}',\bar{y}_k')) \wedge\nonumber\\
   &&~~~~~~~~~~~~~~~~~~~~~~~~~~~~~~~~~~\bigwedge_{\red{w \in
(\nit{RelV}(\psi) \sm \bar{x})}}\!\!\!\!\!\!\nit{I\!sNotNull}(w))\ \ )\ \ \ ). \label{eq:rewRDEC}
\end{eqnarray}
\vspace{-5mm}\begin{example} (example \ref{ex:gen} cont.)
The rewriting $\psi^N$ of the RIC $\psi$ is, according to
(\ref{eq:rewRDEC}):
$$\psi^N\!: \ \forall x \forall y \forall z (P(x,y,z) \ \rightarrow \ \nit{I\!sNull}(x) \vee \nit{I\!sNull}(y) \vee \exists wR(x,y,w)).$$
$D$ classically satisfies $\psi^N$, treating $\nn$ as any other constant. Then, $D \models_N \psi$.
\boxtheorem
\end{example}

\begin{example}\label{ex:newEx1}
For \ $\psi\!: \ \forall x(R(x) \rightarrow \exists y(T(x,y) \wedge S(y))$, \ $\nit{RelV}(\psi)=\{x,y\}$. From (\ref{eq:rewRDEC}):
$$ \psi^N\!: \   \forall x(R(x) \ \ \rightarrow \ \  \nit{IsNull}(x) \vee \exists y(T(x,y) \wedge S(y) \wedge \nit{IsNotNull}(y)).$$
\begin{itemize}[leftmargin=2em]
\item[(a)] For $D = \{R(a)\}$, $D \not \models  \psi^N$; and then,  $D \not \models_N  \psi$.

\item[(b)] For $D = \{R(a), T(a,\nn), S(\nn)\}$, $D \not \models  \psi^N$; and then,  $D \not \models_N  \psi$.

\item[(c)] For $D = \{R(a), T(a,b), S(b)\}$, $D \models  \psi^N$; and  then,  $D \models_N  \psi$.

\item[(d)] For $D = \{R(\nn)\}$, $D \models  \psi^N$; and then,  $D \models_N  \psi$.

\item[(e)] For $D = \emptyset$, $D \models  \psi^N$; and then,  $D \models_N  \psi$. \boxtheorem
\end{itemize}
\end{example}

\begin{example} \label{ex:gen0} \ignore{(example \ref{ex:gen} continued)} Consider the schema $\mc{S} = \{R(A,B,E)\}$
with the \red{primary} {\em key constraint} (KC) $R\!: A B \rightarrow E$, \red{expressing that attributes $A, B$, together, functionally determine attribute $E$}.\footnote{\red{A key constraint
is a particular kind of functional dependency, where a set of attributes of a relational predicate functionally determines all the attributes of the predicate. Declaring a
  key constraint as {\em primary} in an SQL-based relational DBMS has in particular the effect that {\tt NULL} is not accepted in key attributes.}} \ignore{, that, in combination with the  referential integrity constraint  \ $\psi\!:\forall x y z (P(x,y,z) \rightarrow \exists v R(x,y,v))$, produces a foreign-key constraint.}
An SQL database with the instance $D = \{R(a,5,3), R(a,3,7)\}$, which contains no nulls, would satisfy the
KC.

The
insertion of $R(\nn,4,5)$  into $D$ would be rejected since the KC would be violated:
\ $\nn$ is not allowed in a key attribute.
The insertion of $R(a,5,\nn)$ would also be rejected, but now due to the potential occurrence of two $R$-tuples $t_1$, $t_2$ with
$t_1[A]=t_2[A]=a$ and $t_1[B]=t_2[B]=5$, but being unknown whether
$t_1[E]=t_2[E]$. 

In order to fully capture this KC in predicate logic and the intended semantics
we just described,
we need the following sentences to represent the KC:
\begin{eqnarray}
&&\!\!\!\!\!\!\psi_1\!: \ \forall x \forall y \forall z_1 \forall z_2 (R(x,y,z_1) \wedge
R(x,y,z_2) \rightarrow z_1=z_2), \label{eq:fd}\\
&&\!\!\!\!\!\!\psi_2\!: \  \forall x \forall y\forall z_1 \forall z_2 (R(x,y,z_1) \wedge
R(x,y,z_2) \wedge \IsN(z_1) \rightarrow \IsN(z_2)), \label{eq:withnulls}\\
&&\!\!\!\!\!\!\psi_3\!: \ \forall x \forall y\forall z (R(x,y,z)
 \wedge \IsN(x) \rightarrow \mathbf{false}), \label{eq:null}\\
&&\!\!\!\!\!\!\psi_4\!: \  \forall x \forall y \forall z (R(x,y,z)
 \wedge \IsN(y) \rightarrow \mathbf{false}). \label{eq:null2}
 \end{eqnarray}
The first two constraints ensure
 that, if two tuples coincide in attributes $A$ and $B$, then they have the
 same value in $z$ or they are both $\nn$.
  The last two constraints ensure
 that the values in attributes $A$ or $B$ in relation $R$ cannot be
 $\nn$.

Actually, (\ref{eq:withnulls}) can be written as
 \begin{eqnarray*}
\forall x \forall y \forall z_1 \forall z_2 (R(x,y,z_1) \wedge
R(x,y,z_2)  \rightarrow \neg \IsN(z_1) \vee \IsN(z_2)),
\end{eqnarray*}
or equivalently, as a UDEC of the form (\ref{eq:UDEC}):
 \begin{eqnarray*}
\psi_2'\!: \ \forall x \forall y \forall z_1 \forall z_2 (R(x,y,z_1) \wedge
R(x,y,z_2)  \rightarrow \nit{I\!sNotNull}(z_1) \vee \IsN(z_2)),
\end{eqnarray*}
\ignore{
which can be further combined with (\ref{eq:fd}) into a single formula, obtaining an UDEC,
 \begin{eqnarray*}
\psi\!: \  \forall x y z_1 z_2 (R(x,y,z_1) \wedge
R(x,y,z_2)  \rightarrow \ (\red{\nit{I\!sNotNull}(z_1)} \vee \IsN(z_2)) \ \red{\wedge \ z_1 = z_2}
\end{eqnarray*}   }
Similarly, (\ref{eq:null}) and (\ref{eq:null2})
can be written  as
  UDECs of the form (\ref{eq:UDEC}):
 \begin{eqnarray*}
\psi_3'\!: \ \forall x \forall y \forall z (R(x,y,z) \
 \rightarrow \ \nit{I\!sNotNull}(x)),\\
\psi_4'\!: \ \forall x \forall y \forall z (R(x,y,z) \
 \rightarrow \ \nit{I\!sNotNull}(y)).
\end{eqnarray*}
 So, finally, the KC is represented by the UDECs \ $\psi_1, \psi_2', \psi_3', \psi_4'$.

For illustration, considering  that $\nit{RelV}(\psi_1) = \{x,y,z_1,z_2\}$, and $\nit{RelV}(\psi_3') = \emptyset$, we obtain from (\ref{eq:rewUDEC}):
\begin{eqnarray}
&&\!\!\!\!\!\!\!\!\!\!\psi_1^N\!: \ \forall x \forall y \forall z_1 \forall z_2 (R(x,y,z_1) \wedge
R(x,y,z_2) \ \rightarrow \ \nit{I\!sNull}(x) \vee \nit{I\!sNull}(y) \vee \nonumber\\
&& \hspace*{4.8cm}\nit{I\!sNull}(z_1) \vee \nit{I\!sNull}(z_2) \vee z_1=z_2),\\
&&\!\!\!\!\!\!\!\!\!\!(\psi_3')^N\!: \ \forall x \forall y \forall z (R(x,y,z) \
 \rightarrow \ \nit{I\!sNotNull}(x)).
 \end{eqnarray}
 Finally, observe that when the
 atom $z_1 = z_2$ in $\psi_1^N$ is evaluated on the domain, the constant $\nit{null}$ is treated as any other constant,
 i.e. $\nit{null} = c$ is true only when constant $c$ is $\nit{null}$\ignore{ (cf. Appendix \ref{app:queryAns} for more details)}.
\boxtheorem
\end{example}

We have managed to reduce
constraint satisfaction under ${\tt NULL}$ as handled by SQL databases to formula satisfaction  in FO predicate logic,
without assigning any special status to \nn, which is treated as an ordinary constant.
We considered only constraints of the
form (\ref{eq:GenDEC}), which include all our DECs and ICs for peers, in which case, the generic set of database predicates $\mc{R}$
will contain those for a single peer or a pair thereof. However, the semantics can be extended to more general FO formulas \cite{Bravo07}.

\subsection{Query answering under ${\tt NULL}$ via FO rewriting}\label{sec:querRewr}

The notion of query answer under ${\tt NULL}$, that of $N$-answer given in Section \ref{app:nans}, being interesting {\em per se},  allowed us to motivate the treatment of constraints
via relevant variables. However, query answering in presence of \nit{null} can also be treated via the rewriting  in Definition \ref{def:pseudosat}. This is achieved
by considering conjunctive queries as a special case of (\ref{eq:GenDEC}), without the  $\forall \bar{x}$ and empty antecedents; and obtaining
their  relevant variables  through
Definition \ref{def:rel}.

More precisely, a conjunctive query $\mc{Q}(\bar{x}) \in \nit{Conj}(\Sigma^\nulo)$, i.e. of the form (\ref{eq:conjunctive}),  is
rewritten into a new conjunctive query as follows:
\begin{equation}
\mc{Q}^N(\bar{x})\!: \ \exists \bar y(A_1(\bar{x}_1) \wedge \cdots \wedge A_n(\bar{x}_n) \ \wedge
\! \! \! \!\bigwedge_{v\in \nit{RelV}(\mc{Q})} \!\!\! v \neq \nn). \label{eq:rew}
\end{equation}
 It holds: \ $D \models_{\!_N} \mc{Q}[\bar{c}]  \ \ \mbox{ iff } \  \ D \models \mc{Q}^N[\bar{c}]$, where the latter is classic FO satisfaction, with $\nn$ treated as an ordinary constant in the domain.

 This transformation ensures that  relevant variables range over $({\cal U} \smallsetminus \{\nn\})$. \ $\mc{Q}^N(\bar{x}) \in \nit{Conj}(\Sigma^\nulo)$, and may contain
atoms of the form  $\nit{I\!sNull}(t)$ or $\nit{I\!sNotNull}(t)$. However, they can be replaced by $t = \nn$ or $t \neq \nn$, resp.,
leading to a query in $\nit{Conj}(\Sigma)$, but with the same answers as (\ref{eq:rew}).

\begin{example}\label{ex:QASQL} Consider  the instance $D$ below,\ignore{$D=\{P(f,7),$ $
P(f,5),$ $P(\nn,8),$ $P(b,\nn)\}$,} and the query $\mc{Q}(x)\!: \ \exists y
(P(x,y) \wedge y>5)$, for which $\nit{RelV}(\mc{Q})=\{y\}$. In this case, the bound variable is the only relevant variable, and then
ranges over non-null values when checking query satisfaction.

\begin{multicols}{2}
Accordingly,
 $D \models_N \mc{Q}[f]$ holds, because $\exists y
(P(f,y) \wedge y>5)$ is true in $D$, with $7$ as a non-null value for $y$
that makes the formula true. This result is confirmed by the rewriting of $\mc{Q}(x)$.

\begin{center}\begin{tabular*}{3cm}{c|c|c|}\cline{1-3}
  $P$ & $A$ & $B$\\ \cline{1-3}
   & $f$ & $7$ \\
  & $f$ & $5$ \\
   & $\nn$ & $8$ \\
   & $b$ & $\nn$ \\ \cline{2-3}
\end{tabular*}\end{center}
\end{multicols}
\vspace{-3mm}For   \ $\mc{Q}^N(x)\!: \ \exists y
(P(x,y) \wedge y>5 \ \wedge \ \nit{I\!sNotNull}(y))$, \  $D \models \mc{Q}^N[f]$ holds,
under classical satisfaction, with $\nit{null}$ treated as any other constant. Similarly, $D \models_N \mc{Q}[\nn]$.
\boxtheorem
\end{example}
\begin{example} (example \ref{ex:ansquery} continued) The query in (\ref{eq:quer}) can be rewritten as

\vspace{1mm}
\centerline{$\mc{Q}_2^N(x)\!: \ \exists y\exists z
(R(x,y,z)\wedge S(y)\wedge y > 2\wedge y\neq\nn).$}

\vspace{1mm}\noindent
  We had $D \not \models_{\!_N} \mc{Q}_2[1]$. Now, $D \not \models \exists y\exists z(R(1,y,z)\wedge S(y)\wedge y>2\wedge y\neq\nn)$,
 classically,  with $\nn$ treated as an ordinary constant.
 \ As expected, $D \not \models \mc{Q}_2^N[2]$ due to the new
 conjunct $y\neq\nn$.

Finally, $D \models \mc{Q}_2^N[\nn]$ because $D \models (R(\nn,3,3)\wedge S(3)\wedge 3>2 \wedge 3 \neq \nn)$. Since $\nn$ is treated as any other constant, we can compare it with $3$. By the {\em unique names assumption},  $3 \neq \nn$ holds.
\boxtheorem
\end{example}

Our query answering $N$-semantics for $\nit{Conj}(\Sigma^\nulo)$ can be applied in particular any SQL-like conjunctive query, but first expressing it as a query
$\mc{Q}$
 in  $\nit{Conj}^\sql(\Sigma^\nulo)$, and then computing and classically evaluating $\mc{Q}^N$.\\

The  notions of constraint satisfaction and query answer in the presence of
${\tt NULL}$ \`a la SQL we introduced in this section coincide with the classic notions in databases without
$\nit{null}$.

In the rest of this work,  the  notion of formula satisfaction that we have denoted with $\models_N$
will be simply denoted with $\models$. When constraints are not satisfied in this sense,  we will apply a
particular repair semantics  that captures the
special  role of \nit{null}.  It is introduced  in the next section.

\section{Solution Semantics with NULL}\label{sec:nullsRep}


In the preceding section, we introduced  a notion of formula satisfaction with database instances that
may contain \nn. We now use it for defining the semantics of a peer system where the data movement process
relies on consistency restoration of peer instances \wrt  DECs. The value \nn \ will be used to replace existentially
quantified variables in consequents of referential DECs and local
referential  ICs as a possibility to consider for DEC and IC enforcement. Another possibility for the same task is tuple deletion.

The repair semantics that supports consistency restoration will have to capture and be sensitive to the (possibly multiple) presence of \nn in the database and its
use for consistency enforcement. In particular, it has to give an account of the facts  that: \begin{enumerate}\item When atoms are inserted into the database, an existential variable that appears in a join  or in a built-in atom
in an RDEC's consequent (a so-called {\em problematic existential
variable}) is never replaced by (or takes the value)
\nn. \item Arbitrary non-null constants
from the domain are not used for these problematic variables either. Actually, rather than introducing arbitrary values for existential variables of this kind, tuple deletions
from antecedents in DECs will be privileged.
\end{enumerate}
\subsection{A restricted chase}\label{sec:chase}

In order to achieve the just stated goals, we first introduce \ignore{ (in Section \ref{sec:chase})}  a restricted, {\em ad hoc} form
of the  chase \cite{AHV1995}, as an auxiliary construct. It is applied only with the DECs and ICs at hand that do not have problematic existential variables
in their consequents. The enforcement of this subset of the constraints introduces tuples with \nn \ for existential  variables  (that do not appear in joins though).

In the end, this chase will return a finite set of atoms that will be used 
as a possibly generous upper-bound for the admissible tuple insertions that create repairs and solutions on the basis of the whole sets of DECs at hand.

As any other form of chase, our restricted chase enforces the satisfaction
of constraints, but in our case the notion of satisfaction corresponds to that introduced in Section \ref{sec:nullsDB}.

In the rest of this subsection, we consider a database instance $D$ and a set of DECs $\Sigma = \Sigma_1 \cup \Sigma_2$, with $\Sigma_1$ a set of UDECs of the form (\ref{eq:UDEC})
and $\Sigma_2$ a set of RDECs of the form (\ref{eq:RDEC}). We consider the subset, $\Sigma_2^{-}$ of $\Sigma_2$ that contains all the RDECs except for those involving existentially quantified variables in joins  or built-in atoms. Accordingly, we consider $\Sigma^{-} := \Sigma_1 \cup \Sigma_2^{-}$.

\begin{definition} \label{def:Rchase} The {\em restricted chase}, $r\mbox{-}\nit{Chase}^{\nn}(D,\Sigma^{-})$, with an instance $D$ and $\Sigma^{-} = \Sigma_1 \cup \Sigma_2^{-}$,
\capri{is the instance $D'$ obtained as a fix-point of the following iterative procedure:} 
\begin{enumerate}[leftmargin=2em]
\item $D'_0 := D$.
\item Given  instance $D'_s$, if $D'_s \models \Sigma^{-}$ (relative to $\models_N$), then $D'_{s+1} := D'_s$.

\item Given  instance $D'_s$,  if a ground instantiation, $\varphi\!\!\downarrow$, of a constraint $\varphi \in \Sigma_1$ (with constants or \nn)
is not satisfied by $D'_s$ (relative to $\models_N$), but its antecedent is,
then $D'_{s+1}$ is obtained from $D'_s$ by adding the ground database atoms appearing in every disjunct of every conjunct in the consequent of $\varphi\!\!\downarrow$ (but only when the built-ins in the conjunct are satisfied).

Notice that this excludes the generation of atoms with any kind of built-ins, including the atom $\mathbf{false}$.

\item Given  instance $D'_s$, if a ground instantiation (obtained by replacing universal variables with constants or \nn, but keeping the existential quantifiers) of a constraint in $\Sigma_2^{-}$ is not satisfied
by $D'_s$ , but its antecedent is,
then replace the existentially quantified variables by \nn, and build $D'_{s+1}$ by adding to $D'_s$ the corresponding ground atoms in the conjunct in the consequent (but only when the built-ins in it are satisfied). \boxtheorem
\end{enumerate}
\end{definition}

\ignore{
\comlb{Above in blue: we should say something about the order of application of DECS, to make it deterministic.}\\
\comlore{I don't see a problem with the order. Can you think of an example?}\\
\comlb{Along the same line: we are generous making everything true in the consequente *when* the DEC is not satisfied. However, when it is,
which can be due to the truth of one of the disjuncts,
we do not do anything, but we could still insert tuples to make the other disjuncts true; a form of saturation process.}\\
\comlore{But this could result in repairs that are less interesting... When the DEC is not satisfied we cannot avoid adding all possible disjuncts as true.
But in the second case we can so I'd rather not add the rest}
}

\capri{This procedure finitely terminates as we show below, but first some intuitions, explanations and examples.}
The chase procedure propagates  and invents values due to the enforcement of
 RECs with non-problematic existential quantifiers and UDECs. As a result, the constants in the chase instance will be those already appearing in the initial instance $D$,
those that appear explicitly
in consequents of DECs, or \nn \ as a value invented for (non-problematic) existentially quantified variables.

\begin{example} \label{ex:chase} Consider the following set of constraints, classified according to Definition \ref{def:Rchase}:
\begin{eqnarray}
\Sigma_1&=& \{\forall x \forall y(T(x,y) \rightarrow
R(x,y)), \label{eq:unoC}\\
&&\ \ \forall x \forall y  \forall z(R(x,y) \wedge
S(y,z) \rightarrow Q(x,y,z) \vee T(x,z)), \label{eq:tresC}\\
&&\ \ \forall x \forall y  \forall z(Q(x,y,z) \rightarrow S(x,y) \wedge R(y,z)),\label{eq:and}\\
&&\ \ \forall x \forall y \forall z(T(x,y) \wedge T(x,z) \rightarrow y = z),\label{eq:egd}\\
&& \ \ \forall x \forall y ( T(x,y) \wedge S(x,y)
\rightarrow {\bf false}),\label{eq:dosC} \\
\Sigma_2^{-} &=& \{\forall x \forall y ( R(x,y) \rightarrow
\exists z (Q(x,y,z) \wedge x \neq y),\label{eq:cuatroCC}\\
&& \ \ \forall x \forall y \forall z( Q(x,y,z) \rightarrow
\exists w(R(x,z) \wedge S(x,w))\}.\label{eq:cuatroC}
\end{eqnarray}
Independently from the instance at hand (initial or not) at   a chase step, constraints (\ref{eq:egd}) and (\ref{eq:dosC}), will never be applied (or enforced).

If $D$ contains $T(a,\nn)$, then the ground instantiation $T(a,\nn) \rightarrow
R(a,\nn)$  of (\ref{eq:unoC}) is satisfied under $\models_N$, even if $D$ does not contain $R(a,\nn)$. This is because both variables $x,y$
in (\ref{eq:unoC})
are relevant (cf. Definition \ref{def:rel}). So, $R(a,\nn)$ is not generated. However, if $T(a,b) \in D$ and $R(a,b) \notin D$, the
latter atom is generated.

Now, the ground instantiation  $R(a,a) \rightarrow
\exists z (Q(a,a,z) \wedge a \neq a)$  of (\ref{eq:cuatroCC}) will not create new tuples
because of the inequality $a \neq a$. However, if $R(a,b) \in D$, and the  instantiation $R(a,b) \rightarrow
\exists z (Q(a,b,z) \wedge a \neq b)$
is not satisfied, the tuple $Q(a,b,\nn)$ will be created.

The ground instantiation  $R(a,b) \wedge
S(b,c) \rightarrow \ Q(a,b,c) \vee T(a,c)$  of (\ref{eq:tresC}), with $R(a,b),$
 $S(b,c) \in D$, will generate both $Q(a,b,c)$ and $T(a,c)$ (if it was not already satisfied).

 Notice that creation of $Q(a,b,\nn)$ above, due to (\ref{eq:cuatroCC}), feeds the antecedent of (\ref{eq:and}). However,
 its instantiation $Q(a,b,\nn) \rightarrow S(a,b) \wedge R(b,\nn))$ is satisfied, because \nn appear for variable $z$ that is relevant.  So, $R(b,\nn)$ is not generated.
 \boxtheorem \end{example}

 \begin{remark}  \label{rem:rem} (a) Notice that in the DEC (\ref{eq:and}) all variables, $x,y,z$, are relevant. As a consequence, any ground atom with predicate $Q$ with \nn in it will
  not trigger the rule.  We can see that, every time the chase introduces \nn in a predicate position that turns out to be relevant in another DEC, this latter DEC will not be triggered
  (because it will be automatically satisfied). This possibly very common situation will contribute to a fast termination of the chase.

   \vspace{1mm}\ni
(b) We can assume that the chase applies in parallel all possible instantiations of DECs in $\Sigma^{-}$. It is clear that this restricted \nn-based chase always terminates, because only the elements of the
initial active domain plus possibly \nn are used to fill arguments in database predicates, which leads to a saturation point in polynomial-time in the size of
 instance $D$ and the schema. In particular, the resulting instance is of polynomial size in the size of $D$.

  \vspace{1mm}\ni
 \capri{(c) Instead of using the chase procedure (in case we wanted a more model-theoretic or declarative approach to the chase instance), we could replace
 each sentence in $\Sigma_1$ by a collection of Datalog rules, one for each disjunct in the consequent; and each sentence in $\Sigma_2^{-}$ by a Datalog rule obtained
 replacing existential variables by \nn. The chase instance would coincide with the minimal model of the resulting Datalog program \cite{AHV1995}.}
  \boxtheorem
\end{remark}

\vspace{1mm}
Notice that $r\mbox{-}\nit{Chase}^{\nn}(D,\Sigma^{-})$ may not satisfy $\Sigma$, for the trivial reason that not all DECs in $\Sigma$ are considered
in the chase. Actually,  $r\mbox{-}\nit{Chase}^{\nn}(D,\Sigma^{-})$ may not even satisfy $\Sigma^{-}$, because it ignores DECs with only built-ins in
consequents. For illustration, in Example \ref{ex:chase}, the chase of $D=\{T(a,b),T(a,c)\}$ will not satisfy (\ref{eq:egd}), and the chase of $D=\{T(a,b),S(a,b)\}$ will not satisfy (\ref{eq:dosC}).

\ignore{\comlb{What else should we say about the chase? Properties? }\\
\comlore{That the resulting instance does not necessarily satisfy the constraints.}\\
\comlb{I guess only because of the trivial reason that we may not have the full set $\Sigma$, but only $\Sigma'$. But this chase
should satisfy $\Sigma'$. Am I missing something?}\\
\comlore{When for a constraint the antecedent is satisfied by the consequent has built-ins that are not satisfied the chase will not satisfy that constraint.
For example the chase of $D=\{T(a,b),T(a,c)\}$ will not satisfy (26) and the chase of $D=\{T(a,b),S(a,b)\}$ will not satisfy (27)}
}

The next step will be  imposing $r\mbox{-}\nit{Chase}^{\nn}(D,\Sigma^{-})$ as an upper bound on
the extensions of possible repairs. With this we will discard candidate repairs that introduce arbitrary, non-mandatory, non-null constants. In other words,
we will use the result of this chase to test candidate repairs (or solutions), demanding their inserted atoms
to be contained in the chase applied with the subset $\Sigma^{-}$ of the set $\Sigma$ of DECs at hand.

\subsection{Repair semantics with NULL and solutions}\label{sec:nullrep}

A repair semantics for single relational databases and sets of ICs that may include referential ICs is proposed in \cite{iidb06,Bravo07}.
It introduces null values of the kind described in Section \ref{sec:nullsDB}. We will
adopt and adapt this repair semantics in the context of PDESs, by taking into account the restrictions imposed by the
chase we just introduced, and
also the trust relationships.

So as repairs of an inconsistent instance \wrt a set of ICs,  solutions for a peer in a PDES are
expected to
stay ``close" to the original peer's
instance, while satisfying its DECs. In order to capture ``closeness" in our null-based semantics, we need to compare
instances and their tuples, which  may contain  constant $\nn$.

\begin{definition}   \cite{LL94} For constants $c, d \in {\cal U}$, $c$ provides {\em less
or equal information} than  $d$, denoted $c \sqsubseteq d$, iff $c =
\nn$ or $c=d$. For sequences $\bar{s}_1=\langle c_1, \dots, c_n\rangle$ and
 $\bar{s}_2=\langle d_1,
\dots, d_n\rangle$, with $c_i, d_i \in \mc{U}$, $\bar{s}_1$ provides less or equal information than $\bar{s}_2$, denoted $\bar{s}_1 \sqsubseteq \bar{s}_2$, iff
$c_i\sqsubseteq d_i$ for every $i=1,\dots,n$. Finally, $\bar{s}_1
\sqsubset \bar{s}_2$ means $\bar{s}_1 \sqsubseteq \bar{s}_2$ and
$\bar{s}_1 \not = \bar{s}_2$. \boxtheorem
\end{definition}

\ignore{
\comlb{I used ``sequence" in definition. We should reserve tuple for ground atoms. There are some changes of this kind to make in ``your" earlier sections.}
\comlore{Why isn't this a ground atom? You consider a tuple with null not to be ground? But we defined
database instances as a set of ground atoms even if it includes \nn. In some places were we were talking about trust relations as tuples I changed it. But in this particular case (of the previous definition) I would use tuple.
Also, we used both  $\langle,\rangle$ and $(,)$ to refer to tuples in a database. I changed it to the former.}  }

\begin{definition} \label{def:minD}  Consider $D,D',D''$
be database instances for the same schema, and a set $\Sigma$ of DECs of the form (\ref{eq:UDEC}) or (\ref{eq:RDEC}) in the language of the schema. Let $\Sigma^{-}$ be the subset of $\Sigma$
that excludes the DECs with problematic existential variables (cf. Section \ref{sec:chase}).
\begin{itemize}[leftmargin=2em]
\item[(a)] \ $D'$ {\em is at least as close to} $D$ {\em as} $D''$ (is to $D$), denoted $D' \leq_{D}^\Sigma
D''$, ~iff~ one of the following holds:
\begin{enumerate}[leftmargin=2em]
\item $D'' \ \not \subseteq \ \rch(D,\Sigma^{-})$.

 \item For every $P\oa \in \Delta(D,D')$, there exists
$P(\bar{a}') \in
    \Delta(D,D'')$, such that: \\ i. \ $\bar{a} \sqsubseteq \bar{a}'$, \ and \ ii. \ if
$\bar{a} \sqsubset \bar{a}'$, then $P(\bar{a}') \not \in
    \Delta(D,D')$.
\ignore{\begin{itemize}
\item[(a)] $\bar{a} \sqsubseteq \bar{a}'$; and
 \item[(b)] if
$\bar{a} \sqsubset \bar{a}'$, then $P(\bar{a}') \not \in
    \Delta(D,D')$.
    \end{itemize} }
 \end{enumerate}
\item[(b)] \ $D'$ {\em is closer
to} $D$ {\em than} $D''$ (is to $D$), denoted $D' <_{D}^\Sigma D''$, iff $D' \leq_{D}^\Sigma D''$, but not $D''
\leq_D^\Sigma D'$.  \boxtheorem
\end{itemize}
\end{definition}

\ignore{
The next definition characterizes a class of null-based repairs. It is the one to be used on neighborhood instances, which will allow us to capture the notion of
neighborhood solution for a peer and its neighborhood, on the basis of our special  semantics.}

\begin{definition}\label{def:repair}
Given an instance $D$ and a set $\Sigma$ of DECs of the forms (\ref{eq:UDEC}) or (\ref{eq:RDEC}), a {\em null-based repair}
of $D$ \wrt  $\Sigma$ is an instance $D'$ such that $D' \models_N \Sigma$, and there is no $D''$, such that
$D'' \models_N \Sigma$ and $D'' <_D^\Sigma D'$. \boxtheorem
\end{definition}

This definition  ensures, in particular, that
a database that, due to the enforcement of an RDEC with a non-problematic existential variable, inserts a tuple with $\nn$ for that variable, is closer to $D$ than
another that adds some other, arbitrary constant. At the same time, the definition makes us prefer repairs obtained
via tuple deletions instead of tuple insertions (with values null or not), when enforcing RDECs with problematic
existential variables.

\ignore{\capri{It is possible to give a purely declarative definition of {\em null-based repair}. This can be
done along the lines of the model-theoretic, universal-solution-based semantics of data exchange systems, which provides a justification for the use of the (procedural) chase  \cite{kolaitis} as a proper representative of that semantics.
In order to simplify the presentation, we prefer to directly use the chase.}}

\red{The null-based repair semantics, in particular Definition \ref{def:repair},  is also of independent interest in the context of single inconsistent databases
\wrt a set of ICs that are like the DECs we are considering in the PDES setting. Actually, this is the official and detailed formalization of the
repair semantics first sketched and used in \cite{iidb06} for databases with referential ICs. The idea in that case is that inconsistencies are preferably repaired through the
insertion of tuples with the special constant \nn, that is used as the null in SQL. When this is not possible, tuple deletions are preferred. This happens  either because a constraint cannot be solved via tuple insertions
at all (e.g. functional dependencies) or an inclusion dependency would have to be satisfied through the use of \nn for variables in a join. The next two examples show how this repair semantics would be used for
 repairing a single database.}

 \begin{example}\label{ex:new} (example \ref{ex:newEx1} cont.)
Consider $\Sigma = \{\forall x(R(x) \rightarrow \exists y(T(x,y) \wedge S(y))\}$, and the instance $D = \{R(a)\}$.

Here, \ $\Sigma^{-} = \emptyset$, and $\rch(D,\Sigma^{-}) = \{R(a)\}$. Furthermore, $D \not \models \Sigma$ \ (this is case (a) in Example \ref{ex:newEx1}).
\begin{itemize}[leftmargin=2em]
 \item[i.] $D'_1 = \{R(a), T(a,\nn), S(\nn)\}\not \models  \Sigma$ \ (this is case (b) in Example \ref{ex:newEx1}). So, $D'_1$ is not a repair.

 \item[ii.] $D'_2 = \emptyset \models \Sigma$ \ (this is case (e) in Example \ref{ex:newEx1}). \ $D'_2$ is a repair.

 Indeed, for any instance $D''$ that is not contained in $\rch(D,\Sigma^{-})$, \ $D'_2  \leq_D^\Sigma D''$ holds.
 So, the only possible improvement on $D_2'$ could be $\{R(a)\}$, which does not satisfy $\Sigma$.

 \item[iii.] $D'_3 = \{R(a), T(a,a), S(a)\} \models \Sigma$ \ (as case (c) in Example \ref{ex:newEx1}), but it is not a repair.

 Indeed, first notice that  $D'_2  \leq_D^\Sigma D'_3$, because $D'_3 \not \subseteq \rch(D,\Sigma^{-})$.

 Now, $D'_3  \not \leq_D^\Sigma D'_2$, because condition 2. of Definition \ref{def:minD} does not hold for $S(a)$ (or $T(a,a)$).
 We have obtained that  $D'_2  <_D^\Sigma D'_3$, i.e. the former, which is a repair, is strictly closer to $D$ than the latter.
 The same holds for any instance of the form $D'_4 = \{R(a), T(a,b), S(b)\}$ (and its supersets).
 \boxtheorem
\end{itemize}
\end{example}

\begin{example}\label{ex:new+}
Consider $\Sigma = \{\forall x \forall y \forall z(T(x,y) \wedge T(x,z) \rightarrow y = z), \
 \forall x \forall y ( T(x,y) \wedge S(x,y)
\rightarrow {\bf false})\}$, containing a functional dependency and a denial constraint. The instance $D = \{T(a,b), T(a,c), S(a,c)\}$
is inconsistent \wrt $\Sigma$.

Here, $\Sigma^{-} = \Sigma$. However, the restricted chase does not apply any of the ICs, because they contain only built-ins
in the consequent. So,   $\rch(D,\Sigma^{-}) = D$. As a consequence, only repairs based on tuple deletions are acceptable.
In this case, $D' = \{T(a,b), S(a,c)\}$ is the only repair. \boxtheorem
\end{example}

Since the $\leq_D^\Sigma$ relations can take the place of the generic relation $\preceq_D^{\Sigma(\sp{P})}$ used in Section \ref{sec:semantics},
  $\leq_D$ determines, through the instantiation of Definition \ref{def:localsolution}, a concrete repair semantics that determines null-based neighborhood solutions
  for a peer \p{P}, in which case the definition  has to be applied with $\Sigma(\p{P})$.
However, Definition \ref{def:repair} does not take into account the trust relationships in a neighborhood that are imposed in Definition \ref{def:localsolution}.
As a consequence, to have a null-based notion of neighborhood solution, we have to impose the additional
requirement that certain database relations
in a neighborhood instance have to stay untouched (they belong to a peer that is most trusted).

Having a specific notion of neighborhood solution, it is possible to obtain, following the developments in Section \ref{sec:semantics}, also specific definitions of
solution and core for a peer, and the notion of peer-consistent answer. In the rest of this section we provide some additional examples that show aspects of this specific semantics, and in the following subsections we investigate
computational problems related to this semantics.

\begin{example}\label{ex:newSol} (example \ref{ex:new} continued)
Consider a PDES instance $\mathfrak{D}$
for the schema $\mathfrak{P}$ with  peers \peer{P1} and \peer{P2}, with $\mc{S}(\peer{P1}) = \{R\}$, $\mc{S}(\peer{P2}) = \{T,S\}$,
and $\trust=\{(\peer{P1}, \nit{same}, \peer{P2})\}$. Assume the only DECs are those in
$\Sigma(\peer{P1},\p{P2}) = \{\forall x(R(x) \rightarrow \exists y(T(x,y) \wedge S(y))\}$. So,   $\Sigma^{-}(\peer{P1},\p{P2}) = \emptyset$.

 Assume that the instance of the neighborhood of \peer{P1} is $D = \{R(a)\}$.
 Here, $D \not \models \Sigma(\peer{P1},\p{P2})$.
  So, in order to obtain a neighborhood solution, we have to
 repair  $D$ with respect to $\Sigma(\peer{P1},\p{P2})$.

 Since \p{P1} trusts \p{P2} the same as itself, in principle
 we could insert tuples into \p{P2}'s relations or delete tuples from \p{P1}'s relation. However, in this case (cf. Example \ref{ex:new}),
since $\rch(D,\Sigma^{-}(\peer{P1},\p{P2})) = \{R(a)\}$,
 the only neighborhood solution is the empty instance: \ $\emptyset \models_N  \Sigma(\peer{P1},\p{P2})$, and
 $\emptyset <_D^{\Sigma(\sp{P1})} D'$,  for any $\mc{S}(\p{P1},\p{P2})$-instance $D'$ that inserts tuples.
 \boxtheorem
\end{example}

Similarly, the generic Definition \ref{def:solutionSem1} of solution and core for peers, can be instantiated with the null-based solution semantics, i.e. on the basis of the
 $\leq_D^\Sigma$ relations, obtaining specific versions of those definitions. Accordingly, we recall that, given an instance $\mathfrak{D} = \{D(\p{Q}) \ | \ \p{Q} \in \mc{P}\}$ for the schema $\mathfrak{P}$,
the {\em core} of peer \p{P} is the intersection of its solutions: \ $\mathit{Core}(\p{P},\red{\mathfrak{D}}):= \bigcap \mathit{Sol}(\p{P},\mathfrak{D})$.

\begin{example}\label{ex:trans4} (example
\ref{ex:trans0} continued) ~ Consider the following
peers' instances:
$D(\p{P1})$=$\{R^1(a,2)\}$, $D(\p{P2})=\{R^2(c,4),R^2(d,5)\}$,
$D(\p{P3})=\{R^3(c,4)\}$, and $D(\p{P4})=\{R^4(d,5,1)\}$. So, the PDES instance is $\mathfrak{D} = \{D(\p{P1}),D(\p{P2}),D(\p{P3}),D(\p{P4})\}$.

If we want the solutions for  \p{P4},  we first need  the solutions for \p{P3} and
\p{P2}, who also needs the solution for \p{P3}.  Since \p{P3} has no DECs
with other peers, its only neighborhood solution is its local instance
$D(\p{P3})$, which is sent back to \p{P2} or \p{P4} if request.

Peer \p{P2} needs to find the neighborhood solutions for
$\{R^2(c,4),R^2(d,5),R^3(c,$ $4)\}$ \wrt
$\Sigma(\peer{P2},\peer{P3})$. Since \p{P2} trusts \p{P3} the same as
itself, it can modify its own data or the data it got from \p{P3}.
There are two neighborhood solutions for \p{P2}:
$\{R^2(c,4),R^2(d,5)\}$ and $\{R^2(d,5),R^3(c,4)\}$, that restricted to \p{P2}'s schema lead to two
solutions for \p{P2}: $\nit{Sol}(\p{P2},\mathfrak{D})$=$\{\{R^2(c,4),R^2(d,5)\},$ $\{R^2(d,5)\}\}$.
 Peer \p{P2} sends to \p{P4} the intersection of its solutions:
$\nit{Core}(\p{P2},\mathfrak{D})=\{R^2(d,5)\}$.

Neighborhood solutions for
\p{P4} are obtained by repairing $\{R^4(d,5,$ $1),$
$R^2(d,5),$ $R^3(c,4)\}$ \wrt $\Sigma(\peer{P4},\peer{P2})$, and
$\Sigma(\peer{P4},\peer{P3})$. The DECs in
$\Sigma(\peer{P4},\peer{P2})$ are already satisfied, but not those
in $\Sigma(\p{P4},\p{P3})$. Since \p{P4} trusts the data in \p{P3}
more than its own, the only neighborhood solution for \p{P4} is obtained by inserting a tuple with $\nn$ into \p{P4}'s:
\ $\{R^4(d,5,1),R^2(d,5),R^3(c,4),$ $R^4(c,4,\nn)\}$. Consequently,
$\nit{Sol}(\p{P4},\mathfrak{D})=\{\{R^4(d,5,1),$ $R^4(c,4,\nn)\}\}$.
\boxtheorem
\end{example}

\red{Due to a  particular combination of DECs, trust relationships, and
local instances, it is possible that a peer has not solution, which may happen with or without the null-based
repair semantics. For illustration,  Example \ref{ex:nosols}, now with the null-based solution semantics,  still does not have solutions.}

\subsection{Complexity of neighborhood solutions}\label{sec:neiNull1}

We now investigate the complexity of decision problems related to the general case of DECs of the form (\ref{eq:GenDEC}) (cf. Section
\ref{sec:nulls}), in combination with the null-based repair semantics we introduced in this section.

We concentrate
mostly on the case that is directly relevant to our forthcoming answer-set programming (ASP) approach to specifying solutions for individual peers when they have already gathered their neighbors solutions.
That is, we consider specifications and reasoning at the neighborhood level (cf. Section \ref{sec:P2Ptransit}). Accordingly, we start by analyzing the complexity of deciding if an instance is a
neighborhood solution (cf. Definition \ref{def:localsolution}).

\begin{definition}
 Consider a PDES  schema $\mathfrak{P} = \langle
\mc{P}, \mf{S}, \Sigma,\trust\rangle$  
 and a peer $\p{P}\in\mc{P}$. \ 
\red{Given an instance $\bar{D}$ for the neighborhood schema $\S(\N(\p{P}))$ around \p{P}, }
\ignore{Given instances
$J_\p{Q}$ for every  peers $\p{Q} \in \N(\p{P})$,} and an instance $J$ for the schema $\S(\N(\p{P}))$, the ${\sf NeighborhoodSol}$
decision problem is about determining if $J$ is a neighborhood solution for \p{P} and $\bar{D}$, i.e. about membership of
the set:

\vspace{1mm}
${\sf NeighborhoodSol}(\Ps,\p{P})=\ignore{$\hspace*{-3mm}&$}\{(J,\ignore{(J_\p{Q})_{\p{Q}\in
\N(\p{P})}}\bar{D})\,|\, J \in
\mathit{\NS}(\p{P}, \red{\bar{D}}\ignore{\bigcup_{\p{Q}\in\N(\p{P})}J_\p{Q}})\ignore{$ when each\newline \hspace*{3.8cm} $J_{\p{Q}}$ is an instance for
$\S(\p{Q})} \}$.   \boxtheorem \hspace{-3mm}
\end{definition}

This decision problem is parameterized by peer schemas and selected peers. The inputs  are database instances. Then, we are considering data complexity.

\begin{propositionS}\label{prop:SolutionGivenCore}
    ${\sf NeighborhoodSol}$  is
   $\nit{coN\!P}$-complete in  the size of
$J\cup \bar{D}$. \boxtheorem 
\end{propositionS}

 Notice that Proposition \ref{prop:SolutionGivenCore} has to be interpreted as follows  (and similarly those that follow in this section): For every peer schema $\Ps$, ${\sf NeighborhoodSol}(\Ps,\p{P})$ is in \nit{coNP}; and
there is a peer schema $\Ps_0$ and a peer $\p{P}_0$ in it, such that
the associated problem ${\sf NeighborhoodSol}(\Ps_0,\p{P}_0)$ is \nit{coNP}-hard.

\ignore{

\red{\comlb{This is a proposition we need later on, and still needs a proof.}}

\red{\begin{propositionS}\label{prop:thenewone}
Given a PDES $\mathfrak{P}$, and a peer $\p{P} \in \mc{P}$ with neighborhood instance $D$ for schema $\mc{S}(\mc{N}(\p{P}))$. The problems of deciding if
 an arbitrary instance $D'$ for   $\mc{S}(\mc{N}(\p{P}))$ is a neighborhood solution for $D$
is \nit{coNP}-complete in data (i.e., $|D| + |D'|$.). \boxtheorem
\end{propositionS}}

\red{\comlb{I left the following below, that I removed from the proof in the next section. However, since there is no (published) proof in \cite{Bravo07} of the exact result
we need, we should have a real proof for the proposition above. And the chase should be involved in it.\\
Mas aun, la membership deberia ser esencialmente como aquella en la dem. de la prop. \ref{prop:SolutionGivenCore}, pero hecho de menos la
referencia al chase en todo esto.}}

}

\ignore{
\comlb{Check the next claim after giving the proof. It would be good to have this.}

\comlore{Even though the following claim is true in the sense that the results holds already for a single peer with local ICs, the proof
uses a more general case. A single peer can be used in the proof but it becomes more complex to follow since there can be both insertions and deletions.
}
}

Proposition \ref{prop:SolutionGivenCore} holds already for PDES with a single peer
 with
local integrity constraints; actually with (cyclic) sets of RDECs of the simple form $\forall \bar{x}(R(\bar{x})
\longrightarrow \exists \bar{y}~Q(\bar{x}',\bar{y}))$. The proof in this case is a bit more involved since tuple deletions and insertions are in principle
possible, but the latter were blocked with the trust relationships. A proof can be found
in \cite{Bravo07}.\footnote{It is also possible to modify the proof just given by introducing auxiliary, dummy joins in the
consequent of the RDEC in $\Sigma(\p{P1},\p{P1})$, to make attributes relevant and avoid so the insertion  of
tuples with \nn, which has the effect of enforcing deletions.}  \
The proposition also holds in the case where $\bar{D} = \bigcup_{\p{Q} \in \mc{N}(\p{P})} J_\p{Q}$, but
every $J_\p{Q}$ with $\p{Q}\in\N^\circ(\p{P})$ is fixed, i.e. only  $J$ and
$J_\p{P}$ vary.

\ignore{
\blue{This result can be obtained by reduction from repair checking for single databases \wrt local ICs of the UDEC and RDEC forms. Actually, hardness already holds
with (cyclic) sets of RDECS of the simple form $\forall \bar{x}(R(\bar{x})
\longrightarrow \exists \bar{y}~Q(\bar{x}',\bar{y}))$ \cite{Bravo07,iidb06}.\footnote{For a survey of complexity result for repairs and CQA, see \cite{bertossi11}. Specific references are given in the proofs.} In fact, to obtain the minimal
repairs of an instance $D_0$ over a schema $S_0$ \wrt a set $\IC_0$ of integrity
constraints, we construct a PDES schema $\mathfrak{P} = \langle
\mc{P}, \mc{S}, \Sigma,\trust\rangle$ with $\p{P}\in \mc{P}$, $S(\p{P})=S_0 \in \mc{S}$, $\Sigma=\{\Sigma(\p{P},\p{P})\}$, $\Sigma(\p{P},\p{P})=\IC_0$, $(\p{P},\same,\p{P})\in \trust$ and an instance $\mathfrak{D}=\{D(\p{P})\}$ over  $\mathfrak{P}$ with $D(\p{P})=D_0$. The neighborhood solutions for peer $\p{P}$
coincide with the minimal repairs of $D$ \wrt $\IC$.}\\
}

\ignore{
\red{\comlb{Do we need the problem of deciding if {\em there exists}
a neighborhood solution for an instance $D$? What complexity you think it has?} }\\
\comlore{There are  syntactic conditions over the constraints that ensure the existence of a NS, but not the opposite. I think the problem in general is NP-hard... but I'm not sure}
}

\subsection{Complexity of the core and peer-consistent answers}\label{sec:neiNull2}

\begin{definition} \label{def:core} Consider a PDES  schema $\mathfrak{P} = \langle
\mc{P}, \mf{S}, \Sigma,\trust\rangle$  and a peer \p{P} in it.
 Given a neighborhood instance
$\bar{D}= \bigcup_{\p{Q} \in \N(\p{P})} J_\p{Q}$ for $\mc{S}(\N(\p{P}))$, the {\em  local core} of \p{P} is the intersection of the
neighborhood solutions for \p{P} and $\bar{D}$, but restricted to \p{P}'s schema $\S(\p{P})$:
\ $\mathit{localCore}(\p{P},\bar{D}):= (\bigcap \NS(\p{P},\bar{D}))\rs \S(\p{P})$. \boxtheorem
\end{definition}

\red{In this definition, $J_\p{Q}$ denotes an arbitrary instance for peer \p{Q}, which may be different from what we have called the initial instance $D(\p{Q})$ for \p{Q} (cf. Definition \ref{def:solutionSem1}). Actually, $J_\p{Q}$ could be $D(\p{Q})$, but also a neighborhood
solution for \p{Q} restricted to its schema, or the intersection of the latter, etc.}
In this regard, we recall that a neighborhood solution for \p{P} is defined in terms of its neighborhood schema, its local instance $D(\p{P})$ in $\mathfrak{D}$, and instances $J_{\p{Q}}$ for other peers \p{Q} in its neighborhood
(cf. Definition \ref{def:localsolution}). Each
$J_\p{Q}$ for a neighboring peer \p{Q} is typically the restriction to $\mc{S}(\p{Q})$ of the intersection of \p{Q}'s local neighborhood solutions, which may not necessarily be the same as the
initial
instance
$D(\p{Q}) \in \mathfrak{D}$.

We can see that the problems of defining and computing neighborhood solutions
for a peer can be formulated with arbitrary instances for the neighbors, and do not require taking into consideration the recursive relations between
peers. However,  the computation of the solutions of a peer is a global problem that does include recursion. In this case, we fix all the instances to be the initial ones for each peer.


\vspace{2mm}
Notice from Definition \ref{def:solutionSem1}, that when $\bar{D}=D(\p{P}) \cup
\bigcup_{\p{Q} \in \N^\circ(\p{P})} \mathit{Core}(\p{Q},\red{\mathfrak{D}})$ in Definition \ref{def:core}, it holds
$\mathit{localCore}(\p{P},\bar{D}) = \mathit{Core}(\p{P},\red{\mathfrak{D}})$. In particular, Proposition \ref{prop:SolutionGivenCore} remains true if
$J_\p{Q}=\mathit{Core}(\p{Q},\red{\mathfrak{D}})$, for $\p{Q} \in \N^\circ(\p{P})$, and $J_\p{P}=D(\p{P})$.

\ignore{Now we study the problem
of checking if a tuple $R(\bar{t})$, with $R \in \mc{S}(\p{P})$, belongs to the local core of  \p{P} with
an instance $D$ for $\S(\N(\p{P}))$.}

\begin{definition}
Consider a PDES  schema $\mathfrak{P} = \langle
\mc{P}, \mf{S}, \Sigma,\trust\rangle$  and a peer $\p{P}\in\P$, the {\sf
InLocalCore} decision problem is about membership of the set:

\vspace{1mm}
${\sf InLocalCore}(\Ps,\p{P})=\{(R(\bar{t}),\red{\bar{D}})\,|\, R(\bar{t}) \in
\mathit{localCore}(\p{P},\red{\bar{D}})$ and $\bar{D}$ is an instance \linebreak \hspace*{5.5cm}for $\mc{S}(\mc{N}(\p{P}))\ignore{ $D$ $=$ $\bigcup_{\p{Q} \in \N(\p{P})}J_\p{Q}}\}$.  \boxtheorem
\end{definition}
\begin{propositionS}\label{lemma:PCAGivenCore}
For a PDES  schema $\mathfrak{P} = \langle
\mc{P}, \mf{S}, \Sigma,\trust\rangle$  and a peer $\p{P}\in\P$, \
 ${\sf InLocalCore}(\mathfrak{P},\p{P})$ is $\Pi^P_2$-complete
 in\ignore{ \ $\sum_{\p{Q} \in \N(\p{P})}|J_\p{Q}|$} the size of \red{$\bar{D}$}, the input neighborhood instance. \boxtheorem
%
\end{propositionS}
\ignore{This result can be
proven by reduction from the problem of consistent
query answering (CQA), which is $\Pi^P_2$-complete for referential integrity constraints \cite{iidb06}.
However, we\\
\red{We establish
hardness by reduction
from the satisfiability problem for quantified boolean formulas (QBF) $\beta$ of
the form \ $\forall p_1\cdots \forall p_k \exists q_1\cdots \exists q_l
\psi$, where $\psi$ is a propositional formula in CNF, which is $\Pi^P_2$-complete \cite{comp,papa}.}
}

\red{Despite the assumption  that the peer graphs $\G(\mathfrak{P})$ are acyclic (cf. Remark \ref{rem:cycles}),
sets $\Sigma$ of DECs
may contain cycles through inclusion dependencies, in particular
 with existential quantifiers, i.e. RDECs.}

 \red{ More precisely, assume $\Sigma$ is (or contains) a set of sentences $\varphi$ of the form $\forall \bar{x}(\psi(\bar{x})
\rightarrow \exists\bar{y}\chi(\bar{x},\bar{y}))$, of the form (\ref{eq:GenDEC}). When $\bar{y}$ is not empty, $\varphi$ is called
{\em existential}. A directed graph can be associated to $\Sigma$. The nodes are the database predicates, and there is an edge from predicate $P$
to predicate $Q$ when $P$ appears in the antecedent, and $Q$ in the consequent  of a $\varphi \in \Sigma$. The edge is {\em marked} if $\varphi$ is existential.
$\Sigma$ is {\em ref-acyclic} if there no cycles with marked edges in the graph.\footnote{The condition of ref-acyclicity was already used in \cite{iidb06}, as {\em RIC-acyclicity},
for sets of ICs on a single database schema.} }

\red{For example,
$\IC_1 = \{\forall x(S(x)$ $\rightarrow$ $Q(x)),$ $\forall x(Q(x)$ $\rightarrow$
$\exists y$ $T(x,y))\}$ and $\IC_2 = \{\forall x(S(x)$ $\rightarrow$ $Q(x)),$ $\forall x(Q(x)$ $\rightarrow$
 $S(x))\}$ are  ref-acyclic sets,  whereas $\IC_3$ $=$
$\IC_1$ $\cup$ $\{\forall xy$ $(T(x,y)$ $ \rightarrow$ $Q(y))\}$ is not. This notion can be
applied without changes to the sets $\Sigma(\p{P})$ in PDESs.}


\begin{example} \label{ex:cycle} Given a PDES $\mathfrak{P} = \langle
\mc{P}, \mf{S}, \Sigma,\trust\rangle$ with $\Pe = \{\p{P1}, \p{P2}\}$, $\mf{S} = \{\mc{S}(\p{P1}), \mc{S}(\p{P1})\}$, $\mc{S}(\p{P1})=\{R^1(\cdot,\cdot)\}$, $\mc{S}(\p{P2})=\{R^2(\cdot,\cdot)\}$,
 $\Sigma(\p{P1},\p{P2}) = \{\forall x \forall z(R^1(x,z) \rightarrow \exists y R^2(x,y)),$ $ \forall x \forall z(R^2(x,z)
\rightarrow \exists y R^1(x,y)\}$, $\Sigma(\p{P2},\p{P1}) = \emptyset$; and
$\trust $ $= \{(\p{P1},\nit{less},$ $\p{P2})\}$.
 In this case, the peer graph $\G(\mathfrak{P})$ is acyclic. However, schema $\mathfrak{P}$ is not ref-acyclic,
 because $\Sigma(\p{P1}) = \Sigma(\p{P1},\p{P2})$ has a cycle through RDECs. \boxtheorem
\end{example}

\red{In some cases we will make the assumption that
the sets of DECs at hand are {\em ref-acyclic}. For this reason, we make notice that the proof of Proposition \ref{lemma:PCAGivenCore}
uses a
ref-acyclic set of DECs. So, the proposition  still holds for this class of DECs, which will become
relevant in Section \ref{sec:P2Ptransit}.  \burg{(Cf. electronic Appendix \ref{sec:cycles} for an additional discussion.)}}

\begin{corollaryS}\label{cor:PCAGivenCore} For a PDES  schema $\mathfrak{P} = \langle
\mc{P}, \mf{S}, \Sigma,\trust\rangle$  and a peer $\p{P}\in\P$, with
  ref-acyclic sets of DECs, the \,${\sf InLocalCore}(\mathfrak{P},\p{P})$ decision problem is $\Pi^P_2$-complete
 \red{in the size of\, $\bar{D}$\ignore{=$ $\bigcup_{\p{Q} \in \N(\p{P})}J_\p{Q}$}, the  input neighborhood instance}. \boxtheorem
\end{corollaryS}

We can also apply the null-based solution semantics to Definition \ref{def:peercons}, obtaining a specific definition of {\em peer consistent answer}.
More precisely, given a PDES schema $\mathfrak{P}$, an instance $\mathfrak{D}$ for it, a peer $\p{P}$ and a query $\mc{Q}(\bar{x}) \in \mc{L}(\peer{P})$, the set of {\em peer consistent answers}
to $\mc{Q}$ from $\peer{P}$ is
\begin{equation}
 \red{\nit{PCA}_{\mathfrak{P},\p{P}}^\mathfrak{D}(\mc{Q})} = \{ {t} ~~~~|~~~~ D \models_N \mc{Q}[{t}], \mbox{ for all } D \in \nit{Sol}(\peer{P},\mathfrak{D}) \ \},
 \end{equation}
 with $\nit{Sol}(\peer{P},\mathfrak{D})$ as defined in this section, on the basis of the $\leq_D^{\Sigma(\p{P})}$ relations.

Proposition \ignore{Lemma} \ref{lemma:PCAGivenCore} and its Corollary \ref{cor:PCAGivenCore}
still hold when,  for each $\p{Q} \in \N^\circ(\p{P})$, $J_\p{Q}=\mathit{Core}(\p{Q},\red{\mathfrak{D}})$, and $J_\p{P}=D(\p{P})$. Therefore,  the local computation of peer consistent answers
to a conjunctive query -for which the
cores of the neighboring peers are used- is also $\Pi^P_2$-complete.

\begin{corollaryS}\label{cor:PCA} \red{For a PDES  schema $\mathfrak{P} = \langle
\mc{P}, \mf{S}, \Sigma,\trust\rangle$, an instance $\mathfrak{D}$ for $\mathfrak{P}$,  and a peer $\p{P}\in\P$, \
deciding answers to a conjunctive query posed to  \p{P} that are true in $\bigcap \nit{NS}(\p{P},\bar{D})$, where
$\bar{D} = D(\p{P}) \cup \bigcup_{\p{Q} \in \N^\circ(\p{P})}\nit{Core}(\p{Q},\mathfrak{D})$,
\ignore{all the neighborhood solutions for \p{P} after it has gathered its neighbors'  cores} is $\Pi^P_2$-complete in the size
of $\bar{D}$. This is also true for $\mathfrak{P}$ with ref-acyclic sets of DECs. \boxtheorem}
\end{corollaryS}

\red{ The complexity of the decision and computational problems at the neighborhood
level we have obtained so far are the most interesting, due to the modular and local manner solutions and peer-consistent answers are computed.
These results will be useful in  Section \ref{sec:P2Ptransit}, where specification and computation of neighborhood solutions and PCAs are addressed.}

We  make only some final remarks in relation to global solutions and PCAs on their basis.
Corollary \ref{cor:PCA}  already tells us that deciding peer-consistent answers (which involves global solutions as opposed to neighborhood solutions) will be at least $\Pi^P_2$-hard. Actually,
using Proposition \ref{lemma:PCAGivenCore} it is possible to provide a
non-deterministic polynomial time algorithm (in data) with a $\Pi^P_2$-oracle to decide if an instance for a peer is a (global) solution for
the peer. \ignore{
{\sf Solution} problem, which makes it a member of $\Delta_3^{\!P}$, in data.} On this basis, one can obtain 
that \nit{PCA}, as a decision problem, belongs to $\Pi^P_3$, in data (that includes those of all peers in the system).

\subsection{The  import case}\label{sec:uic}


In  this section we consider a common situation, namely {\em the import case}, where we find \trust \  relationships are only of the form $(\p{P},\less, \p{Q})$ when $\p{P} \neq \p{Q}$;
and
the DECs are  used for importing data from other peers. \ignore{More
precisely, for every peer \p{P}, its DECs $\Sigma(\p{P},\p{Q})$ to any other peer
\p{Q} are  {\em import DECs}, i.e., of the
forms (\ref{eq:UDEC}) or (\ref{eq:RDEC}), but the consequents contain only
one database atom, with predicate in ${\cal S}(\p{P})$ (plus
possibly built-ins); and  all predicates in the antecedents belong to \p{Q}'s
schema. More formally:}

\begin{definition} \label{def:importDECs} Consider a PDES  schema $\mathfrak{P} = \langle
\mc{P}, \mf{S}, \Sigma,\trust\rangle$, and two different peers \p{P} and \p{Q}:

(a) An {\em import
UDEC} (IUDEC) {\em to} \p{P} {\em from} \p{Q}
is a UDEC in ${\cal L}(\p{P},\p{Q})$ of the form:
\begin{equation}\label{eq:importUDEC}
 {\forall}\bar{x}(\bigwedge_{i = 1}^{n} R_i(\bar{x}_i) ~\longrightarrow~
(Q(\bar{x}') ~\vee~ \varphi(\bar{x}'')),
\end{equation}
where the $R_i$ are predicates in ${\cal S}(\p{Q})$, $Q$ is a predicate in
${\cal S}(\p{P})$, $\bar{x}', \bar{x}'' \subseteq \cup \bar{x}_i \ \capri{= \bar{x}}$, and $\varphi(\bar{x}'')$ is a
disjunction of built-ins \capri{(representing a conjunction of built-ins on the variables in the antecedent)}.\\
\vspace{1mm}
(b) An {\em import RDEC} (IRDEC) {\em to}  \p{P} {\em from} \p{Q} is an RDEC in ${\cal
L}(\p{P},\p{Q})$ of the form:
\begin{equation}\label{eq:importRDEC} \forall \bar{x}(\bigwedge_{i = 1}^{n} R_i(\bar{x}_i)
~\longrightarrow~ \exists \bar{z} (Q(\bar{y},\bar{z}) ~\wedge~ \varphi_1(\bar{x}',\bar{z}')) \vee \varphi_2(\bar{x}'')),
\end{equation}where the $R_i$ are predicates in ${\cal S}(\p{Q})$, $Q$ is a predicate in
${\cal S}(\p{P})$, $\bar{x}', \bar{x}'', \bar{y}  \subseteq   \cup_i\bar{x}_i = \bar{x}$, $\bar{z}' \subseteq \bar{z}$,
and
$\varphi_1$ ($\varphi_2$) is a conjunction (disjunction) of built-ins. \capri{Notice that $\varphi_1$ basically imposes conditions with built-ins on the values for
$\bar{z}$ that, under the \nn-based semantics, will be all \nn. \ ($\varphi_2(\bar{x}'')$ plays the same role on $\varphi$ in (\ref{eq:importUDEC}).)} \\
\vspace{1mm}
(c) A PDES is of the {\em import kind} if, for every peer \p{P}, the DECs in $\Sigma(\p{P},\p{Q})$ with any peer \p{Q} different from $\p{P}$, are
import DECs to \p{P} from \p{Q}. Furthermore, if this set of DECs is non-empty, $(\p{P},\nit{less},\p{Q}) \in \trust$. \boxtheorem
\end{definition}


Notice that this definition does not make any assumptions on the possible sets of local constraints, $\Sigma(\p{P},\p{P})$.


\begin{example} \label{ex:newEx2} Consider a PDES with $\mc{P} = \{\p{P1}, \p{P2}, \p{P3}, \p{P4}\}$, $\mc{S}(\p{P1}) = \{R^1(\cdot,\cdot)\},  \mc{S}(\p{P2}) = \{R^2(\cdot,\cdot),$ $ S^2(\cdot,\cdot)\},$
$\mc{S}(\p{P3}) = \{R^3(\cdot,\cdot)\}, \mc{S}(\p{P4}) = \{R^4(\cdot,\cdot,\cdot)\}$, and the following sets of DECs:
\begin{eqnarray*}
\Sigma(\peer{P1},\peer{P2}) &=& \{\forall x \forall y(R^2(x,y) \rightarrow
(R^1(x,y) \vee x \leq y))\}, \label{eq:uno}\\
\Sigma(\peer{P4},\peer{P2}) &=&\{\forall x \forall y  \forall z(R^2(x,y) \wedge
S^2(y,z) \ \rightarrow \ R^4(x,y,z))\}, \label{eq:tres}\\
\Sigma(\peer{P4},\peer{P3}) &=& \{\forall x \forall y ( R^3(x,y) \rightarrow
\exists z R^4(x,y,z))\}. \label{eq:cuatro}\\
\Sigma(\peer{P3},\peer{P2}) &=& \{\forall x \forall y ( R^3(x,y) \rightarrow
R^2(x,y))\}.
\end{eqnarray*}
The first three sets of DECs are formed by import DECs, but not the last one. \boxtheorem
\end{example}

\subsubsection{The unrestricted case}\label{sec:ui}

\red{The
{\em  unrestricted import case} of PDES $\mathfrak{P}$ as in Definition \ref{def:importDECs} occurs when, for every peer \p{P} $\in \mc{P}$, $\Sigma(\p{P},\p{P}) = \emptyset$. That is, in this case, a peer may have DECs of the forms (\ref{eq:importUDEC}) and (\ref{eq:importRDEC}) to neighboring peers, in
whom it has more trust than in itself, but no local ICs.}

As a consequence of  an unrestricted local repair process,
 a peer will get data from its neighbors and will integrate them, at least virtually, into its
own neighborhood instance. The data from a neighboring peer will be obtained by just posing it a conjunctive query, the one corresponding to
the antecedent of the DEC.

For illustration,
in Example \ref{ex:newEx2},
if \p{P4} uses its import DEC from \p{P2} (that in $\Sigma(\peer{P4},\peer{P2})$) to retrieve data from  \p{P2}, then \p{P4} sends to \p{P2} the query: ${\cal Q}(x,y)\!: \ R^2(x,y) \wedge
S^2(y,z)$.

\red{In the import case we can define a Datalog program for which its minimal model is the neighborhood solution of the peer.}

\begin{definition} \label{def:impProg}
 \red{Given an unrestricted import PDES  schema $\mathfrak{P} = \langle
\mc{P}, \mf{S}, \Sigma,\trust\rangle$, a peer $\p{P}\in\mc{P}$ and an instance $\bar{D}$ for the neighborhood schema $\S(\N(\p{P}))$ around \p{P},
{\em the import program} $\mathfrak{I}(\p{P},\bar{D})$ is a Datalog program containing:}
\begin{enumerate}[leftmargin=2em]
\item The facts: $R(\bar{a})$, for every atom  $R(\bar{a}) \in \bar{D}$.

\item For every IUDEC in $\Sigma(\p{P})$ of the form
      (\ref{eq:importUDEC}), the rule: \ignore{ (cf. with (\ref{eq:importUDEC}))}
      $$Q(\bar{y}) \leftarrow
      R_1(\bar{x}_1), \dots R_n(\bar{x}_n), \neg \varphi(\bar{x}'').$$
\item For every import RDEC in $\Sigma(\p{P})$ of the form
      (\ref{eq:importRDEC}), the rule: $$Q(\bar{y},\overline{\nn})
      \leftarrow R_1(\bar{x}_1), \dots R_n(\bar{x}_n), \ \red{\neg \varphi_2(\bar{x}'')}.$$
\end{enumerate}
\red{Here, $\neg \varphi(\bar{x}'')$ and $\neg \varphi_2(\bar{x}'')$ become conjunctions of  built-in literals, i.e. atomic formulas with a built-in or negations thereof; and
$\overline{\nn}$ is a sequence of nulls for variables $\bar{z}$. According to the discussion of (\ref{eq:importRDEC}), conditions associated to $\varphi_1$ become conditions
on $\bar{x}'$, which can all be made part of $\bar{x}''$.}  \boxtheorem
\end{definition}



\ignore{\comlb{A la proposicion le falta contexto (la queja del reviewer 1): El programa habla de "neighborhood solution", pero la proposicion
y su demostracion habla de todo el grafo asociado al peer. Puedes arreglar esto? Hay que partir con una instancia para G(P).}
\comlore{fixed} }


\begin{propositionS}\label{prop:Import} \red{Given an unrestricted import PDES  schema $\mathfrak{P} = \langle
\mc{P}, \mf{S}, \Sigma,\trust\rangle$ and an instance $\mathfrak{D}$ over it, every peer $\p{P}\in\mc{P}$ }
 has a unique solution instance. Furthermore, there is an algorithm, \red{that uses the import program $\mathfrak{I}(\p{P},D')$, and computes the solution for a peer \p{P} in polynomial time in the size of
 $\mathfrak{D} \rs \mc{S}(\AC(\p{P}))$.}
  \boxtheorem
\end{propositionS}

This uniqueness result relies on the fact that we are repairing IRDECs through the insertion of null values, as sanctioned by the official
repair semantics. Otherwise, uniqueness would hold in general only for IUDECs.

\subsubsection{The restricted case}\label{sec:ri}

This case  appears when, under all the import assumptions above, peers are allowed to have local constraints, that is, it may be that $\Sigma(\p{P},\p{P})\neq \emptyset$. In
this case, a peer will import data without restrictions from its neighbors, but when building neighborhood solutions, also the local
ICs will be taken into account. In this case,
 \p{P} may have none or several solutions.

\begin{example} \label{ex:NoSol2} Consider the PDES  schema $\mathfrak{P} = \langle
\mc{P}, \mf{S}, \Sigma,\trust\rangle$ corresponding to Figure \ref{fig:exNoSol}. Here,  $\mc{P} =\{\p{P1}, \p{P2},
\p{P3}\}$, $\mf{S} = \{{\cal S}(\peer{P1}),$ $ {\cal S}(\peer{P2}), {\cal S}(\peer{P3})\}$,
${\cal S}(\peer{P1}) =
\{R^1(\cdot,\cdot)\}$,~ ${\cal S}(\peer{P2})$ $ =
\{R^2(\cdot,\cdot)\}$, ${\cal S}(\peer{P3}) =
\{R^3(\cdot,\cdot)\}$.

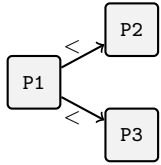
\begin{figure}
  \centering
  \begin{footnotesize}
\begin{tikzpicture}
 \tikzstyle{peer}=[rectangle,rounded corners=.5ex,draw,fill=black!5!,thick, inner
  sep=2pt,minimum width=7mm,minimum size=7mm]
 \tikzstyle{trust}=[thick,draw]
  \node[peer] (P1) at (1.3,0)
        {\p{P1}};
  \node[peer] (P2) at (2.6,0.7)
        {\p{P2}}
    edge [<-,trust] node[near end,above] {$<$} (P1);
  \node[peer] (P3) at (2.6,-0.7) {\p{P3}}
    edge [<-,trust] node[near end,below] {$<$} (P1);
\end{tikzpicture}\end{footnotesize}
\caption{Accessibility graph for Example
\ref{ex:NoSol2}\label{fig:exNoSol}}
\end{figure}

The local  instances are: \ $D(\peer{P1}) = \{\}$, $D(\peer{P2}) = \{R^2(a,b)\}$,
$D(\peer{P3}) = \{ R^3(a,c)\}$.  The DECs are:
\begin{itemize}[leftmargin=2em]
\item[1.] $\Sigma(\peer{P1},\peer{P2})\!= ~\{\forall x \forall y(R^2(x,y)
\rightarrow R^1(x,y))\}$,
\item[2.]
$\Sigma(\peer{P1},\peer{P3}) =$ $\{\forall x \forall y(R^3(x,y)
\rightarrow R^1(x,y))\}$, and
\item[3.] $\Sigma(\p{P1},\p{P1}) = \{\forall x \forall y \forall z(R^1(x,y) \wedge R^1(x,z)
\rightarrow y = z)\}$.
\end{itemize}
 $\Sigma(\peer{P1},\peer{P2})$ and $\Sigma(\peer{P1},\peer{P3})$ contain import DECs. However, due to the local constraints in $\Sigma(\p{P1},\p{P1})$, there are no solutions for $\p{P1}$. \boxtheorem

\end{example}
\begin{example} \label{ex:sevSols}
Consider $D(\p{P}) = \{P(a,b), P(a,c)\}, D(\p{Q}) = \{Q(a,d)\}$, and $\trust=\{(\p{P},\less,$ $\p{Q})\}$.
$\Sigma(\p{P},\p{Q}) = \{\forall x \forall y(Q(x,y) \rightarrow P(x,y)\}$, and
$\Sigma(\p{P},\p{P}) = \{\forall x \forall y \forall z \forall v(P(x,y) \wedge P(x,z) \wedge P(x,v)
 \rightarrow  y = z \vee z = v \vee v = y )\}$. The local constraints in  $\Sigma(\p{P},\p{P})$ ensures that there are at most two tuples in $P$ with the same first attribute.
In this case, $\Sigma(\p{P},\p{Q})$ is of the import kind and therefore $P(a,d)$ will belong to all solutions. The local constraint in $\Sigma(\p{P},\p{P})$ will force the removal of either $P(a,b)$ or $P(a,c)$, and therefore \p{P}  has two solutions: $\{P(a,d), P(a,b)\}$ and
$\{P(a,d), P(a,c)\}$. \boxtheorem
\end{example}

\ignore{
\comlb{Para que pusimos este ejemplo con esta IC local: $\forall x y z v(P(x,y) \wedge P(x,z) \wedge P(x,v)
 \rightarrow  y = z \vee z = v \vee v = y \ \vee \ \mathbf{false})$ tan rara? Cual era la idea?}

\comlore{Creo que debe haber sido incialmente:

$\{\forall x y z v(P(x,y) \wedge P(x,z) \wedge P(x,v) \wedge y \neq z \wedge z \neq v \wedge v \neq y
 \rightarrow \mathbf{false})\}$

Como al final movimos los built in a la derecha nos quedo con a disyuncion y el $\mathbf{false}$ esta de mas. Basta con dejarla como:

$\{\forall x y z v(P(x,y) \wedge P(x,z) \wedge P(x,v)
 \rightarrow  y = z \vee z = v \vee v = y \ \}$

 Modifique el texto para dejarlo mas claro
}

\comlb{No he revisado nada de la seccion siguiente. Solo la reemplace por tu nueva version.}
}

%

\section{Answer Set Programs and the Solutions for a
Peer} \label{sec:P2Ptransit}

Continuing with the special PDES semantics based on the use of \nn introduced in Section
\ref{sec:nulls}, in this section we show how to specify neighborhood  solutions for a
peer (which is the central notion in this paper, upon which the recursive notion
of global solution is defined) by means of disjunctive logic programs with stable model semantics  \cite{GL91,Eiter97}.
As above, we assume that  DECs and ICs have the forms (\ref{eq:UDEC}) or (\ref{eq:RDEC}).

\red{Our aim in this section is to show the gist of the approach, by considering in precise terms a particular but still
interesting case, one where the RDECs have a limited syntactic form, without joins in the consequents. The general case, i.e. with arbitrary UDECs and RDECs as considered in this
work so far, is possible and not difficult from the conceptual or technical point of view, but more difficult or lengthy to present.}

\ignore{
\comlore{Reviwer 3 wants us to CHANGE the logic programming specification for one using "FO formulas and completion".  He has his own idea of
how it should be specify and provides details. He defines distance in terms of number of insertions and deletions. ``Divino lo encuentro", pero eso
no es lo que estamos haciendo aca. How do you answer politely to a suggestion like this one????}
\comlb{Yo diria algo asi: ``The idea of using FO formulas and completion is clearly interesting, and worth exploring, but possibly in a separate paper. We cannot see, in our case, the advantage of doing so.
Furthermore, for good or bad, ASP seems to be better known and more used -at least in data management- that the proposed alternative. Choosing the latter could probably limit the impact of our research."} }

\begin{remark} \label{rem:refacyclic}
In this section we  make the additional assumption that given a schema $\mathfrak{P} = \langle
\mc{P}, \mf{S}, \Sigma,\trust\rangle$, for each \p{P} $\in \Pe$, its set of DECs $\Sigma(\p{P})$
is {\em ref-acyclic}. In this case, we say that $\mathfrak{P}$ is ref-acyclic. \red{The formulation
of the programs is independent from this assumption, but -as we will see later in this section- it ensures desirable properties of the program.}\boxtheorem
\end{remark}

Now we  show how to specify the neighborhood solutions for a
peer \p{P} as the stable models of a
disjunctive logic program $\Pi(\p{P})$. {\em The program can be used for this purpose under the assumption that \p{P} has already
gathered, for each of its neighbors} \p{Q},{\em \ the  intersection of its solutions}, i.e. $\nit{Core}(\p{Q},\mathfrak{D})$ \ (cf. Definition \ref{def:core}).
However, the program works with arbitrary instances for the neighbors. Accordingly, the  facts of the program
come from the union $\mc{D}$ of the given instances for the peers
in ${\cal N}(\p{P})$, the neighborhood centered around  peer \p{P}. As a consequence, $\Pi(\p{P})$  can be used to compute the restrictions to \p{P}'schema of
\p{P}'s neighborhood solutions, on the basis of a neighborhood instance $\mc{D}$.
\ignore{The idea is that each neighbor
of \p{P} contributes to this union with the intersection of its solutions.}

In $\Pi(\p{P})$, we  find each predicate $R \in {\cal S}({\cal N}(\p{P}))$ and
also its copy  $R\_$  that has an extra argument (an augmented {\em nickname}). This argument is the last one and is used to place an annotation constant.
Thus, if $R$ has arity $n$, then $R\_$ has arity $n+1$. The possible annotation constants are: $\ta, \fa, \mathbf{t^\star}, \fs, \tss$.
Their occurrences in database tuples have the following intended semantics:

\begin{center}\begin{tabular}{|l|l|}\cline{1-2}
~Annotation & ~The tuple $R(\bar{a})$ ...\\ \cline{1-2}
~$R\_(\bar{a},\ta)$ & ~is made true in (inserted into) the database\\
~$R\_(\bar{a},\fa)$ & ~is made false in (deleted from) the database\\
~$R\_(\bar{a},\mathbf{t^\star})$ & ~was true or is made true\\
~$R\_(\bar{a},\fs)$ & ~was false or is made false\\
~$R\_(\bar{a},\tss)$ & ~was true or made true, and is not deleted\\ \cline{1-2}
\end{tabular}
\end{center}

 This intended semantics is formally captured in the program by means
of appropriate rules. The  (possibly virtual)
insertions and  deletions are made in order to  satisfy \p{P}'s DECs.
Actually, for each DEC $\psi$, a rule captures through its
disjunctive head the alternative virtual updates that can be
performed to satisfy $\psi$ (cf. rules 2. and 3. in Definition
\ref{def:solProgIII} for UDECs, and 4. and 5. for RDECs). That is why we use
annotations $\ta$ or $\fa$ in rule heads.

The annotations $\trs, \fs$  are used  to  execute  sequences
of virtual updates, which may be necessary when there are interacting DECs.
Finally, atoms annotated with $\tss$ are those that become true in
a solution. They are the relevant atoms, and are used to
read off the database atoms in the solutions (rules 8. below).

Before presenting the logic program, let us recall that $\nit{RelV}(\psi)$ denotes the set of relevant variables of a constraint $\psi$,
those where the occurrence of $\nn$ is
relevant for its satisfaction (see Definition \ref{def:rel}). Those
attributes/variables  receive a special treatment in the program.

{\em In the
following, for a DEC $\psi$, $A(\psi)$ and $C(\psi)$ denote the set of
database atoms (without built-ins) in the antecedent of, resp. the
consequent, of $\psi$.}

A consequence of repairing using $\nn$, is that any RDEC with joins
between existential quantifiers can only be repaired by deleting
tuples and not by inserting tuples with $\nn$. Thus, the most
relevant RDECs in this setting are of the form:
\begin{equation}\label{eq:RDECsimple} \forall \bar{x}(R(\bar{x})
\longrightarrow \exists \bar{y}~Q(\bar{x}',\bar{y})), \vspace{-2mm}
\end{equation}
\capri{For this reason, and to simplify the presentation, the logic program that follows considers RDECs of this form only.} \ignore{It is rather straightforward to generalize it to
the more general case with RDECs of the form (\ref{eq:RDEC}).} \ignore{The
solution program for the more general setting can be found in
 \ref{app:SolutionP}.}


\ignore{
\comlb{Loreto: Please check carefully the program below; I changed many things.}
\comlore{Done}
}

\begin{definition} \label{def:solProgIII}  Consider a
PDES schema $\mathfrak{P} = \langle
\mc{P}, \mf{S}, \Sigma,\trust\rangle$ and a peer $\p{P} \in {\cal P}$ with local instance $D(\p{P})$ (for the schema $\mc{S}(\p{P})$). \red{Let $\bar{D}$ be an instance for the schema $\mc{S}({\cal
N}(\p{P}))$, i.e. for \p{P}'s neighborhood
${\cal N}(\p{P})$, with $\bar{D}\rs \mc{S}(\p{P}) = D(\p{P})$.}
\ignore{and a collection of instances ${\cal D}= (D^{\p{Q}})_{\p{Q} \in {\cal
N}(\p{P})}$ over  the schemas ${\cal S}(\p{Q})$, such that $D^{\p{P}} = D(\p{P})$.} The {\em
solution program} $\Pi(\p{P};\bar{D})$ for $\p{P}$ contains:

\vspace{1mm}\noindent 1.~ The following facts: \red{$dom(a)$, for every
$a \in  \nit{Adom}(\bar{D})$; \nit{dom}(\nn), and $R(\bar{a})$, for every atom
 $R(\bar{a}) \in \bar{D}$}.\footnote{We can also add
$\nit{I\!sNull(null)}$, and $\nit{I\!sNotNull}(c)$, for every non-null constant in the active domain, but this is not necessary if
these two predicates do not appear in the program.}


\vspace{1mm}\noindent 2.~ For every UDEC $\psi \in
\Sigma(\p{P},\p{Q})$ of the form (\ref{eq:UDEC}), with $\p{Q} \in
{\cal N}(\p{P})$ and $(\p{P}, \less, \p{Q} ) \in \trust$, the rule:

\vspace{1mm} $\bigvee \{R\!\_ (\bar{x}_{i},\fa) \ | \ R(\bar{x}_i) \in
A(\psi), R \in {\cal S}(\p{P})\} \vee
\bigvee\{Q\!\_ (\bar{y}_{j}, \mathbf{t}) \ | \ Q(\bar{y}_j) \in C(\psi), Q \in {\cal S}(\p{P})\}$\\
\hspace*{2.7cm}$\leftarrow~\bigwedge_{i=1}^{n} R_{i}\_
(\bar{x}_{i},\trs),~ \bigwedge_{j=1}^m \!\! Q_{j}\_
(\bar{y}_{j},\fs),~\bigwedge_{x_l \in \nit{RelV}(\psi)} x_l\neq \nn.$

\vspace{1mm}\noindent Here, when $Q_{j} \in \mc{B}$ (i.e. a built-in), $Q_{j}\!\_
(\bar{y}_{j},\fs)$ simply denotes the negated atom  $Q_{j}\!\_
(\bar{y}_{j})$, e.g. $=$ becomes $\neq$, $\nit{I\!sNull}$ becomes, $\nit{I\!sNotNull}$, etc., without annotations.
Notice that the condition $Q \in {\cal S}(\p{P})$ in the head of the rule discards built-in atoms (cf. Example
\ref{ex:builts} below).

\ignore{$\bar{\varphi}$ is a conjunction of
atoms that is equivalent to the negation of $\varphi$.}

\vspace{1mm}\noindent 3.~
 For every UDEC $\psi \in
\Sigma(\p{P},\p{Q})$ of the form (\ref{eq:UDEC}), with $\p{Q} \in
{\cal N}(\p{P})$ and $(\p{P}, \same, \p{Q} )$ $\in \trust$, the
rule:

\vspace{1mm} $\bigvee \{R\_(\bar{x}_{i},\fa) \ | \ R(\bar{x}_i) \in
A(\psi)\}  \vee \bigvee \{Q\!\_ (\bar{y}_{j},
\mathbf{t}) \ | \ Q(\bar{y}_j) \in C(\psi)\}
~\leftarrow~\bigwedge_{i=1}^{n} R_{i}\_ (\bar{x}_{i},\trs),$\\
\hspace*{7cm}$\bigwedge_{j=1}^m \!\! Q_{j}\_ (\bar{y}_{j},\fs),
\bigwedge_{x_l \in \nit{RelV}(\psi)} x_l\neq \nn$.

\vspace{1mm}\noindent We apply the same convention as in 2. about the rule body. This case also covers UDECs in $\Sigma(\p{P},\p{P})$, i.e.
local universal ICs for \p{P}.

\vspace{1mm}\noindent 4.~ For every RDEC $\psi \in
\Sigma(\p{P},\p{Q})$ of the form (\ref{eq:RDECsimple}), with $\p{Q}
\in {\cal N}(\p{P})$ and  $(\p{P},\same,\p{Q})$ $\in \trust$, the
rule:

\vspace{1mm} $R\_(\bar{x},\mathbf{f}) \vee
Q\!\_(\bar{x}',\overline{\nn},\mathbf{t}) \leftarrow
R\_(\bar{x},\mathbf{t^{\star}}),  \n \ \nit{aux}_\psi(\bar{x}'),~
 \bar{x}'\neq \nn.$

\vspace{1mm} \noindent \red{Here, $\nit{aux}_\psi$ is an auxiliary predicate used to obtain a {\em safe} rule (i.e. its variables in negated atoms also appear in a positive atom), whose  body captures the violations of
 (\ref{eq:RDECsimple}). In FO logic this would be done through: \ $R(\bar{x}) \wedge \neg \exists \bar{y} Q(\bar{x}',\bar{y})$, which basically leads
to the following rules that define the auxiliary predicate (taking into account
the special semantics of \nn):}

For every $y_i \in \bar{y}$: 
\begin{equation}\label{eq:aux2}
\nit{aux}_\psi(\bar{x}') \leftarrow Q\!\_(\bar{x}',\bar{y},\trs), \n
Q\!\_(\bar{x}',\bar{y},\fa), \bar{x}'\neq \nn, y_i \neq \nn.
\end{equation}
\begin{equation}\label{eq:aux1}
\nit{aux}_\psi(\bar{x}') \leftarrow Q(\bar{x}',\overline{\nn}), \n
Q\!\_(\bar{x}',\overline{\nn},\fa),\bar{x}'\neq \nn. 
\end{equation}
This case covers the RDECs in $\Sigma(\p{P},\p{P})$, i.e. local referential ICs
for \p{P}. \ (Cf. Example \ref{ex:builts} below.) \ \red{Here, and in the rest of this work, atoms
of the form $Q\!\_(\bar{x}',\overline{\nn}, \ldots)$ associated to an atom in a DEC, have all the
existential variables appearing in the latter and predicate $Q$ replaced by \nn. Furthermore, for $\bar{x} = x_1, \ldots, x_n$,~ $\bar{x} \neq \nn$ abbreviates
$x_1 \neq \nn, \ldots, x_n \neq \nn$.}

\vspace{1mm}\noindent
 5.~ For every RDEC $\psi \in
\Sigma(\p{P},\p{Q})$ of the form (\ref{eq:RDECsimple}), with $\p{Q}
\in {\cal N}(\p{P})$ and  $(\p{P},\less,\p{Q}) \in \trust$, the
rules:

\vspace{1mm}\noindent (a) If  $R \in {\cal S}(\p{P})$:~~
$R\_(\bar{x},\mathbf{f})  \leftarrow
R\_(\bar{x},\mathbf{t^{\star}}),  \n \ \nit{aux}_\psi(\bar{x}'),~
 \bar{x}'\neq \nn.$

 \noindent
(b)   If  $Q \in {\cal S}(\p{P})$:~~
$Q\!\_(\bar{x}',\overline{\nn},\mathbf{t}) \leftarrow
R\_(\bar{x},\mathbf{t^{\star}}),  \n \ \nit{aux}_\psi(\bar{x}'),~
 \bar{x}'\neq \nn.$

 \vspace{1mm}\noindent
Plus  rules (\ref{eq:aux1}), (\ref{eq:aux2}).

\ignore{
\item \label{it:uicP2PICS} For every UIC $\psi  \in \IC(\p{P})$
    of the form (\ref{eq:UDEC}), the rule: \vspace{-3mm}{\small
$$\bigvee_{i=1}^{n} P_{i}\_ (\bar{x}_{i},\fa) \vee
\bigvee_{j=1}^{m} Q_{j}\_ (\bar{y}_{j}, \mathbf{t}) ~\leftarrow~
\bigwedge_{i=1}^{n} P_{i}\_ (\bar{x}_{i},\trs), \bigwedge_{j=1}^m
\!\!Q_{j}\_ (\bar{y}_{j},\fs), \!\!\!\!\bigwedge_{x_l \in {\cal
A}(\psi)}\!\!\!\! x_l\neq \nn, ~\bar{\varphi}.\vspace{-4mm}$$}
\comlb{This one should follow from \ref{it:UDECS}.}
\item \label{it:ricP2PS} For every RIC $\psi \in \IC(\p{P})$  of
    the
form (\ref{eq:RDEC}), the rules:\\
{ $P\!\_(\bar{x},\mathbf{f}) \vee
Q\!\_(\bar{x}',\overline{\nn},\mathbf{t}) \leftarrow
P\!\_(\bar{x},\mathbf{t^{\star}}),  \n ~\nit{aux}_\psi(\bar{x}'),~
 \bar{x}'\neq \nn.$\\}
 { $\nit{aux}_\psi(\bar{x}') \leftarrow
Q(\bar{x}',\overline{\nn}),$ $\n
Q\!\_(\bar{x}',\overline{\nn},\fa),$ $\bar{x}'\neq \nn.$}\\
 For every { $y_i \in \bar{y}$}:\\
{ $\nit{aux}_\psi(\bar{x}') \leftarrow
Q\!\_(\bar{x}',\bar{y},\trs),$ $\n Q\!\_(\bar{x}',\bar{y},\fa),$
$\bar{x}'\neq \nn,$ $y_i \neq \nn$.}}

\vspace{1mm}\noindent 6.~ For each predicate $R \in {\cal S}({\cal
N}(\p{P}))$, the  annotation rules:

 \vspace{1mm}$R\_(\bar{x},\fs) \leftarrow dom(\bar{x}), \n R(\bar{x}).$ \hspace{2cm} $R\_(\bar{x},\fs)
\leftarrow R\_(\bar{x},\fa).$

 $R\_(\bar{x},\trs) \leftarrow R(\bar{x}).$ \hspace{3.85cm} $R\_(\bar{x},\trs)
\leftarrow R\_(\bar{x},\ta).$

\vspace{1mm}\noindent 7.~ For each predicate $R \in {\cal S}({\cal
N}(\p{P}))$,  the program  constraint:

$\leftarrow~R\_(\bar{x},\ta),$ $R\_(\bar{x},\fa).$

\vspace{1mm}\noindent 8.~ For each predicate $R \in {\cal
S}(\p{P})$, the interpretation rule:

$R\_(\bar{x},\tss)$ $\leftarrow R\_(\bar{x},\trs),$ $\n
R\_(\bar{x},\fa).$ \boxtheorem
\end{definition}

\ignore{\comlb{Loreto: Necesitamos en segundo literal en el cuerpo de (\ref{eq:aux1})? De hecho, necesitamos (\ref{eq:aux1})?
Alguna sutileza con los nulos ahi que haya que explicar?}
\comlore{Si, se necesita. Esa regla caputra si es que la IC ya se
satisface con un atomo con nulos que eran parte de la base de datos
$D$ }
}

\begin{example} \label{ex:builts}If the UDEC $\psi$ is $\forall x \forall y (R(x) \ \rightarrow \ S(x,y) \vee \nit{I\!sNull}(y))$, then the
corresponding rule according to (2) above is: \ $R(x,\fa) \leftarrow R(x,\trs), S(x,y,\fs),  \nit{I\!sNotNull}(y), x \neq \nit{null}$. \

Notice that $\nit{RelV}(\psi) = \{x\}$.
Since the program interprets $\nit{null}$ as any other constant in the domain, this rule can be replaced by
\ $R(x,\fa) \leftarrow R(x,\trs), S(x,y,\fs),  y \neq \nit{null}, x \neq \nit{null}$.

If the RDEC $\psi$ is $\forall x(R(x) \rightarrow \exists y S(x,y))$, then the
corresponding rules according to 4. above are:
\begin{eqnarray*}
R\_(x,\mathbf{f}) \vee
S\!\_(x,\nn,\mathbf{t}) &\leftarrow& R(x,\trs), \n \ \nit{aux}(x),  x \neq \nit{null}.\\
\nit{aux}(x) &\leftarrow& S(x,\nn), \n
S\!\_(x,\nn,\fa),x \neq \nn.\\
\nit{aux}(x) &\leftarrow& S\!\_(x,y,\trs), \n
S\!\_(x,y,\fa), x \neq \nn, y \neq \nn.
\end{eqnarray*}
In this case, the RDEC is ``repaired", as usual with referential ICs in databases, by either deleting the tuple
from table $R$ or inserting a tuple with a null value in table $S$, both cases equally acceptable. The two cases for item 5. in the program correspond
each to only one of these possible choices.
\boxtheorem
\end{example}

\red{Since  instance $\bar{D}$ and its active domain are finite, the program has a finite number of facts in item (1)}
The most relevant part of the program corresponds to
the rules (2)-(5). They capture through their bodies the violations of DECs; and through their heads, the (possibly alternative) virtual updates that are necessary on the peers'
instances to restore
the satisfaction of DECs and
local ICs. In the bodies of these  rules associated to DECs or ICs $\psi$,
the conditions of the form $x \neq \nn$, with $x$ a variable
appearing in a relevant attribute of $\psi$,  are used to capture the
special semantics of null values introduced in Section \ref{sec:nulls}\ignore{ \burg{and  \ref{app:queryAns}}}. \ignore{An atom of the form $P(\bar{x},\overline{\nn},...)$ in the
program represents an atom with possibly several occurrences of \nn,
not necessarily in its last arguments, e.g. $P(x,\nn,y,\nn, ...)$.
For $\bar{x} = x_1, \ldots, x_n$,~ $\bar{x} \neq \nn$ abbreviates
$x_1 \neq \nn, \ldots, x_n \neq \nn$.}

The fixed semantics of annotations is captured by the
rules (6)-(8). They appear in every solution program.
The program constraints in (7)
 discard
models where an atom is both inserted and deleted.

The
instances  $\bar{D}\rs \mc{S}(\p{Q}))$ used in the program for \p{P}'s neighbors may not coincide with the initial, physical
instances $D(\p{Q})$ in an instance $\mathfrak{D}$ for the PDES $\mathfrak{P}$ (except when \p{Q} is \p{P}). However, the idea is that \p{P},
given a global instance $\mathfrak{D}$,
uses its program with $D(\p{P})$ and $\bar{D}\rs \mc{S}(\p{Q})) = \nit{Core}(\p{Q},\mathfrak{D})$
for each
$\p{Q} \in {\cal N}^\circ(\p{P})$. In this way, the program  computes through its stable models the neighborhood solutions for $\p{P}$, whose
restrictions to \p{P}'s schema will be the solutions for \p{P}.
Notice that  atoms annotated with $\tss$ in a stable model of \p{P}'s
program have predicates in ${\cal S}(\p{P})$ only; and they define a database
instance for \p{P}.

\ignore{The idea is to apply the solution program $\Pi(\p{P}; D(\p{P}), (D^\p{Q})_\p{Q})$ with $D^\p{Q}
= \bigcap \nit{Sol}(\p{Q},\mathfrak{D}) = \nit{Core}(\p{Q},\mathfrak{D})$,\footnote{As usual, for a class of sets ${\cal C}$, $\bigcap {\cal C}$ denotes the intersection of the elements of ${\cal C}$.}
for $\p{Q} \in {\cal N}^\circ(\p{P})$.
Here, each of the classes $\nit{Sol}(\p{Q},\mathfrak{D})$ is obtained
using a corresponding program for \p{Q}.  }

The adoption of the stable model semantics for
the solution programs \cite{GL91}, plus the appropriate use of conditions with \nn \ in rules' bodies,  guarantee
the minimal discrepancy between the generated and the original instances according to Definitions \ref{def:localsolution} and
\ref{def:minD}.

\begin{example} \label{ex:trans01}  Consider  $\Pe = \{\p{P1},\p{P2}\}$, with $D(\p{P1})=\{R^1(a,2)\}$, $D(\p{P2})=\{R^2(d,5)\}$, and neighborhood instance \red{$\bar{D} = D(\p{P1})
\cup D(\p{P2})$}. Assume $(\peer{P1},\less,\peer{P2}) \in \trust$, and the only set of DECs in the system is $\Sigma(\peer{P1},\peer{P2}) = \{\forall x y(R^2(x,y) \rightarrow
R^1(x,y))\}$.

The solution program $\Pi(\p{P1};\red{\bar{D}})$ contains: ~(we itemize in correspondence to Definition \ref{def:solProgIII})

\ignore{
\comlb{Loreto: Originalmente no estaba \nn \ en el dom. Alguna sutileza?}
\comlore{Para efectos practicos, en el programa no afecta si es que
$\nn$ esta o no en $dom$. Antes ocupabamos $dom(x)$ para excluir al
null en algunas reglas, pero ahora ponemos directamente $x\neq null$
(lo que queda mucho mas claro!). Como $\nn$ esta en $\cal U$ preferi
dejarlo como parte de dom.}
}

\vspace{1mm}
 1.~ The facts
$\nit{dom}(a), \nit{dom}(d), \nit{dom}(2), \nit{dom}(5), \nit{dom}(\nn),
    R^1(a,2), R^2(d,5).$

\vspace{1mm} 2.~ $R^1\!\_(x,y,\ta) \leftarrow R^2\!\_(x,y,\trs),$
$R^1\!\_(x,y,\fs), x \not= \nn, y \not = \nn.$

\vspace{1mm} 6.~
    $R^1\!\_(x,y,\fs)   \leftarrow R^1\!\_(x,y,\fa).$~~~~~~~~~~~
    $ R^1\!\_(x,y,\fs)   \leftarrow dom(x), dom(y), \n R^1(x,y).$

    \hspace*{5mm}$R^1\!\_(x,y,\trs)   \leftarrow R^1\!\_(x,y,\ta).$~~~~~~~~
    $ R^1\!\_(x,y,\trs)   \leftarrow R^1(x,y).$

\vspace{1mm} 7.~  $\leftarrow R^1\!\_(x,y,\ta), R^1\!\_(x,y,\fa).$

\vspace{1mm} 8.~ $R^1\!\_(x,y,\tss)   \leftarrow R^1\!\_(x,y,\trs), \n
R^1\!\_(x,y,\fa).$

\vspace{1mm}
\ni  The rule in 2. makes
sure that a \nn-free $R^2$-tuple that is not in $R^1$, is also virtually
inserted into $R^1$ (if it had \nn, the DEC would be satisfied since all attributes are relevant). Since \p{P1} trusts \p{P2} more
than itself, virtual
 changes affect only peer $\p{P1}$. According to Definition \ref{def:solProgIII}, we should also include rules similar to those
 in 6. and 7. for $R^2$.
 However, they are not necessary because $R^2$ does not change. If we decide to omit them, we have to replace the atom
 $R^2\!\_(x,y,\trs)$ in the body of  rule
 in 2. by $R^2(x,y)$.

 Notice that since \p{P2}, which is a sink peer, does not have local ICs, its instance $D(\p{P2})$ is also its only solution, and the
 one passed over to \p{P1}, who uses it to build the facts of its program.
\boxtheorem
\end{example}

\begin{example} \label{ex:RDEC} Consider  $\Pe = \{\p{P1},\p{P2}\}$, with $D(\peer{P1}) =
\{R^1(s,t),$ $R^1(a,\nn)\}$, $D(\peer{P2})= \{R^2(c,d),$ $ R^2(a,e)\}$, and $\trust = \{(\peer{P1},\same,
\peer{P2})\}$.

Assume the only set of DECs are \ $\Sigma(\p{P1},\p{P1})$ $=$ $
\{\forall x y z (R^1(x,y) \wedge R^1(x,z) \rightarrow
y = z)\}$, and
$\Sigma(\peer{P1},$ $\peer{P2})$ $=$ $\{\forall xy(R^2(x,y)$
$\rightarrow \exists z R^1(x,z))\}$. The neighborhood instance is $\bar{D} = D(\peer{P1}) \cup D(\p{P2})$ (the second disjunct is also \p{P2}'s
only solution).

The solution program $\Pi(\p{P1};\bar{D})$ for \p{P1}
is as follows:  (omitting  rules 6. and 7.):

\vspace{1mm} 1.~ $\nit{dom}(a), \ldots, \nit{dom}(\nn),
    R^1(a,\nn), R^1(s,t), R^2(c,d),~ R^2(a,e).$

\vspace{1mm} 3.~ $R^1(x,y,\fa) \vee R^1(x,z,\fa) \leftarrow
R^1(x,y,\trs), R^1(x,z,\trs), ~x \neq \nn, ~y \neq \nn,$ \\
\hspace*{5cm}$ ~z \neq \nn,$ $~y \neq z.$

\vspace{1mm} 4.~ $R^2\!\_(x,y,\fa) \vee R^1\!\_(x,\nn,\ta) \leftarrow R^2\!\_(x,y,\trs), \n \ \aux(x), x \not= \nn.$

\hspace*{5mm}$\nit{aux}(x) \leftarrow R^1(x,\nn), \n R^1\!\_(x,\nn,\fa).$

\hspace*{5mm}$\nit{aux}(x) \leftarrow R^1(x,y,\trs), \n R^1\!\_(x,y,\fa), x \not=
\nn, y \not = \nn.$

\vspace{1mm} 8.~ $R^1\!\_(x,y,\tss)   \leftarrow R^1\!\_(x,y,\trs), \n
R^1\!\_(x,y,\fa).$

\vspace{1mm}\ni Rule  3. takes care
of  \p{P1}'s the local functional dependency. In case of a violation, one of the two tuples in conflict has to be deleted.
Rule 4. has the role of satisfying the
RDEC. Since \p{P1} trusts \p{P2} the same as itself, consistency is restored  by either deleting the tuple from $R^2$ or introducing
a \nn ~into $R^1$.

Here, only
$R^1\!\_$~-atoms become annotated with $\tss$. These annotations are used to build solutions for \p{P}.
In this example, we have two stable models, whose restrictions to $\tss$-annotated atoms are: $\{R^1\!\_(a,\nn,\tss),
R^1\!\_(s,t,\tss), R^1\!\_(c,\nn,\tss)\}$, corresponding to the insertion of tuple $R^1(c,\nn)$; and $\{R^1\!\_(a,\nn,\tss),$
$R^1\!\_(s,t,$ $\tss)\}$, corresponding to the deletion of tuple $R^2(c,d)$.
They correspond to the two solutions for \p{P}:
$\{R^1(a,\nn),
R^1(s,t), R^1(c,\nn)\}$ and $\{R^1(a,\nn),
R^1(s,t)\}$.
\boxtheorem
\end{example}

Notice that each peer \p{P} in a PDES has a solution program that, except for the facts in it, is fixed. It depends
only on its DECS in $\Sigma(\p{P})$ and the trust relationships to its neighbors. The same program can be used
with different initial neighborhood instances $\bar{D}$. On their basis, the program will compute stable models,
whose double-starred atoms will determine a local instance.

\begin{definition} \label{def:corresp} Consider a  PDES $\mathfrak{P} = \langle
\mc{P}, \mf{S}, \Sigma,\trust\rangle$, a peer $\p{P} \in \mc{P}$, and a neighborhood instance $\bar{D}$ for
$\mc{S}(\N(\p{P}))$. Let $M$ be a stable model of $\Pi(\p{P};\bar{D})$. The database
instance for peer \p{P} (and schema $\mc{S}(\p{P})$) associated to $M$ is $D_M =\{ R(\bar{a})$ $|$ $
R \in \mc{S}(\p{P}) \mbox{ and } R\_(\bar{a},\tss) \in M \}$. \boxtheorem
\end{definition}

The following result tells us that the instances for a peer obtained via the stable models of its program are all and only the solutions for the peer, under the assumption
that its neighbors contribute with their own cores.

\ignore{\begin{definition}
\blue{Consider} a  PDES $\mathfrak{P} = \langle
\mc{P}, \mc{S}, \Sigma,\trust\rangle$, with peer instance $\mathfrak{D} = (D(\p{Q}))_{\p{Q} \in \mc{P}}$, and $\p{P} \in {\cal P}$. \
The solution program $\Pi(\p{P}; D(\p{P}), (\nit{Core}(\p{Q},\mathfrak{D}))_{\p{Q} \in \N^\circ(\p{P})})$ (c.f. Definition \ref{def:core})
 is denoted with $\Pi(\p{P})$. \ Program $\Pi(\p{P})$ without the facts in 1. in Definition \ref{def:solProgIII} is denoted with $\Pi^{\!-}\!(\p{P})$.
\boxtheorem
\end{definition}

\ni We can see that each of the peers has a fixed, facts-free program $\Pi^{\!-}\!(\p{P})$. After \p{P} receives each of the $\bigcap \nit{Sol}(\p{Q},\mathfrak{D})$ from its neighbors \p{Q}, it  forms its
program $\Pi(\p{P})$ by adding the facts to $\Pi^{\!-}\!(\p{P})$. \ignore{ \p{P} will change its program $\Pi^{\!-}\!(\p{P})$  only
if local ICs or DECS or trust relationships with its neighbors change.}
}

\ignore{\comlb{\red{The editor (Thomas Eiter), based on the reviews, wants a proof of the proposition below:} {\em ``In particular, for Proposition 6.1.,
more details on the proof should be provided, where it may be
sufficient to include the proof from the other work and thoroughly
discuss the necessary adaptations and changes."}} }

\begin{propositionS}  \label{theo:corresp}
\red{ Consider a PDES $\mathfrak{P} = \langle
\mc{P}, \mf{S}, \Sigma,\trust\rangle$, with an instance $\mathfrak{D}$ for $\mathfrak{P}$, and $\p{P} \in {\cal P}$ whose set of DECs $\Sigma(\p{P})$ is
ref-acyclic. Given the neighborhood  instance for $\mc{S}(\mc{N}(\p{P}))$: $\bar{D} :=
D(\p{P}) \cup \ \bigcup_{\p{Q} \in {\cal N}^\circ(\p{P})} \nit{Core}(\p{Q},\mathfrak{D})$, it holds:}

\vspace{2mm}\hspace*{1cm} $\nit{Sol}(\p{P},\mathfrak{D}) = \{D_M \ | \
M \mbox{ is a stable model of program } \Pi(\p{P},\bar{D})\}$.  \boxtheorem
\ignore{
 the stable models of the
 solution program $\Pi(\p{P})$
 are in one-to-one correspondence with the solutions for \p{P}. More precisely,
   the instances of the form $D_\mm$, where $\mm$ is a stable
 model of the program, are  all
 and the only solution instances for \p{P}.}
\end{propositionS}

This result generalizes  one about  the correctness of similar programs for null-based repairs of the kind considered
in this work, for single databases \wrt denial constraints and referential constraints and null values \cite{iidb06}. The hypothesis of ref-acyclicity is necessary
for the soundness of the program (every instance $D_M$ is a solution). Otherwise, only the completeness of the program can be guaranteed (every solution
is a $D_M$) \cite{iidb06}.
\ignore{\burg{We do not give the lengthy proof of this result here; it follows the same lines as those of Theorem 5.4 in \cite{Bravo07}.}\footnote{\red{Accessible from: \ {\tt http://people.scs.carleton.ca/$^\sim$bertossi/papers/Thesis36.pdf}}} }

Under the assumption that we have already computed the (intersection
of the) solution instances for \p{P}'s neighbors, the program
for \p{P} allows us to compute its solution instances. This generates
a  recursive process that can be applied because $\G(\p{P})$ is
acyclic.  Terminal peers \p{P'} in $\G(\p{P})$, i.e. without outgoing
edges,  become the base cases for the recursion. \red{If their local instances $D(\p{P'})$ are (locally) consistent, they pass those instances to their
neighbors. Otherwise, they first, using program $\Pi(\p{P'};D(\p{P'}))$ compute local repairs for $D(\p{P'})$, and pass the intersection of them to their neighbors.}

Several optimizations can be applied to solution programs \cite{CB10}. An important one has to do with the materialization of the
{\em closed-world-assumption}, which results through the rule  $ R^1\!\_(x,y,\fs)   \leftarrow dom(x), dom(y),$ \linebreak $\nit{not} \ R^1(x,y)$
in 6. in Example \ref{ex:trans01}. This is clearly undesirable and can be avoided. We do not provide optimized versions here, because they
are more difficult to read.

Proposition \ref{theo:corresp} still holds if
\p{P}, instead of collecting the intersection of the
solutions of a neighbor \p{Q}, uses the intersection of the
solutions for \p{Q} restricted to the subschema of \p{Q} that
contains \p{Q}'s relations that appear in $\Sigma(\p{P},\p{Q})$,
which are those \p{P} needs to run its program.

\red{As we expressed at the beginning of this section, our solution programs can be much more general. In particular, they can be modified
to include rules for REDCs with existential quantifiers and joins in the consequents. For example, if the DEC is, say
$\forall x y(R(x,y) \wedge S(y,z) \rightarrow \exists w(P(x,w) \wedge Q(w,z)))$, and the antecedent is true with \nn-free tuples,\footnote{If it is true via tuples
with nulls, then due to the relevance of attributes, the DEC is immediately satisfied.} but
not the consequent, one of the (satisfying) tuples in the antecedent has to be deleted. All we need to handle this DEC is a disjunctive, ``deleting" rule.
Built-in comparisons (with \nn) are used by the rule as well. }

\begin{remark}
For simplicity, the programs we introduced above do not consider inconsistent peers, i.e. a peer \p{P}
whose instance is $D^\p{P} = \{\mathbf{inc}_\p{P}\}$. As the following example shows, it is easy to modify the rules
to capture this situation. This change is compatible with the corresponding general DECs (cf. Remark \ref{rem:empty}). \boxtheorem
\end{remark}

\begin{example} (example \ref{ex:trans01} continued) The system is exactly as before, except that now  $D^\p{P2} =
\bigcap \nit{Sol}(\p{P2},\mathfrak{D}) = \{\mathbf{inc}_\p{P2}\}$, reflecting the fact that \p{P2} has no solutions (due to
its mappings and trust relationships to other peers, which we are not showing here). The general program
that allows for inconsistent peers
would now have, instead of 1. and 2.:

 $1'.~ \nit{dom}(a),  \nit{dom}(2), \ignore{\nit{dom}(\nn),}
    R^1(a,2), \mathbf{inc}_\p{P2}.$

\vspace{1mm} 2'.~ $R^1\!\_(x,y,\ta) \leftarrow R^2\!\_(x,y,\trs),$
$R^1\!\_(x,y,\fs), \n \ \mathbf{inc}_\p{P2}, \ x \not= \nn, y \not = \nn.$

\vspace{1mm}\ni In this case, and as expected, the rule  has no effect on \p{P1}. \boxtheorem
\end{example}

\subsection{ASPs and PCAs}

With a solution program for \p{P}, PCAs to a query ${\cal Q}$
posed to \p{P} can be obtained by running  a query
 program  in combination
 with the solution program. First a query program
 $\Pi({\cal Q})$ has to be produced, which is rather standard, and next, $\Pi(\p{P}) \cup \Pi({\cal Q})$
 is run under the {\em cautious stable model semantics}, the one that declares as true what is simultaneously
 true in all the stable models. Of course, the same program $\Pi(\p{P})$ can be used with different queries.

\begin{example} (example \ref{ex:trans01} continued) In
order to
 obtain \p{P1}'s PCAs to the query ${\cal Q}_1(x,y)\!:
 R^1(x,y)$,  the
 rule ~$\nit{Ans}_1(x,y) \leftarrow R^1(x,y,\tss)$ has to be added to
 $\Pi(\p{P}; D(\p{P1}) \cup D(\p{P2}))$. The PCAs are the ground
 $\nit{Ans}_1$-atoms in the intersection of all stable models of the combined program.
For the query ${\cal
 Q}_2(x)\!: \exists y R^1(x,y)$, the query rule is: \
 $\nit{Ans}_2(x) \leftarrow R^1(x,y,\tss)$. \boxtheorem
 \end{example}

In general, given a conjunctive query $\mc{Q}$ (for which we want PCAs), first its rewriting $\mc{Q}^N$ is produced (as in Definition \ref{def:pseudosat}). Next, $\mc{Q}^N$
is rewritten in its turn as a query program with annotation $\tss$. The query program is added to the peer's solution program, and the combination is run under
 the skeptical semantics.

Deciding the truth of ground atomic queries by means of the solution program for a peer under the skeptical semantics amounts to
deciding membership of the local core, which, by Corollary \ref{cor:PCAGivenCore}, is $\Pi^P_2$-complete in data. This is exactly the data complexity
of deciding skeptical entailment for atomic queries for disjunctive logic programs under the stable model semantics \cite{DEGV97}. So, the solution programs
have the right expressive power for the problem at hand.

Using solution programs, our semantics could be naively implemented as
follows. When \p{P} is posed a query,  \p{P} has to run its program,
for which it needs as  facts those in the intersections of the
solutions of its neighbors. So, \p{P} sends  to each neighbor \p{Q}
queries of the form
 ${\cal Q}\!: R(\bar{x})$, where
$R \in {\cal S}(\p{Q})$ and appears in $\Sigma(\p{P},\p{Q})$. Peer
\p{P} expects to receive from \p{Q}
the  PCAs to ${\cal Q}$, because they corresponds to the
extension of $R$ in the intersection of solutions for \p{Q}.

In order to return to \p{P} the PCAs to its queries, each of its
neighbors \p{Q} has to run its own program $\Pi(\p{Q})$.  As before, they need PCAs
from their own neighbors, etc.  This recursion  eventually
reaches peers that have no DECs, and assuming its local ICs are
satisfied, they  return query answers to its neighbors directly from their original instances. This is the start of the
backward propagation of PCAs process, which goes on  until reaching $\p{P}$. Eventually, \p{P} gets all the facts to run
its program and obtain the PCAs to the original query. If the local instance of a terminal peer is not consistent \wrt its local ICs,
the local solution program can still be used to restore consistency \wrt the local ICs, as in consistent query answering. The data
to be sent back to the neighbors  comes from the intersection of the repairs, in the form of consistent query answers \cite{ABC99}.

\begin{example} \label{ex:trans6} (example
\ref{ex:trans0} continued) Consider an instance $\mathfrak{D}=\{D(\p{P1}),D(\p{P2}),D(\p{P3})\}$ with
$D(\p{P1})=\{R^1(a,2)\!\}$, $D(\p{P2})=\{$ $R^2(c,4),R^2(d,5)\}$, and
$D(\p{P3})=\{R^3(c,4)\}$. A user poses the query ${\cal Q}_0\!:
R^1(x,y)$ to \p{P1}, expecting its PCAs.

To run its program,  \p{P1}
needs the intersection of the solutions of peer \p{P2}. So, \p{P1}
sends to \p{P2} the queries
 ${\cal Q}^1_1\!: R^2(x,y)$ and ${\cal Q}^1_2\!: S^2(x,y)$ (actually,
 \p{P1} does not need the latter, $S^2$ is not relevant to \p{P1}).

In order to peer-consistently answer these queries, \p{P2} needs
from \p{P3} the PCAs to ${\cal Q}^2_3\!: R^3(x,y)$. Since \p{P3} has
no neighbors, it returns to \p{P2} the entire extension in
its local database, i.e. $D(\p{P3}) = \{R^3(c,4)\}$. Now, $\p{P2}$
runs its  program $\Pi(\p{P2}; D(\p{P2}) \cup D(\p{3}))$. It contains, among others, the
facts $R^2(c,4), R^2(d,5), R^3(c,4)$; and (assuming $\Sigma(\p{P2},\p{P2}) = \emptyset$) the main rule:

\vspace{1mm}
\centerline{$R^2\!\_(x,y,\fa) \vee
R^3\!\_(x,y,\fa) \leftarrow R^2\!\_(x,y,\trs),
R^3\!\_(x,y,\trs), x \not= \nn, y \not =
\nn.$}

\vspace{1mm} \p{P2} has two solutions, $\{R^2(d,5)\}$ and $\{R^2(c,4),
R^2(d,5)\}$, whose intersection, $\nit{Core}(\p{P2},\mathfrak{D})
= \{R^2(d,5)\}$, is given to \p{P1}.

Finally, the program $\Pi(\p{P1};
D(\p{P1})\cup \nit{Core}(\p{P2},\mathfrak{D})$ (cf.  Example \ref{ex:trans01}) is run.
It has only
one solution, namely $\{R^1(a,2),$ $R^1(d,5)\}\}$; and the
PCAs to ${\cal Q}_0$ are $\langle a,2\rangle$ and $\langle d,5\rangle$.
\boxtheorem
\end{example}

\subsection{The import case revisited} \label{sec:uicRev}

We first consider the unrestricted case introduced in Section
\ref{sec:ui}. Here, each peer $\p{P}$ has a solution program $\Pi^{-}(\p{P})$ as in Definition
\ref{def:solProgIII}, with rules of the form:

\vspace{1mm}
2.~~~~~
$S\!\_ (\bar{y}_{j}, \mathbf{t}) \ \leftarrow \ \bigwedge_{i=1}^{n} R_{i}\_ (\bar{x}_{i},\trs),~
S\!\_(\bar{y}_{j},\fs),~\bigwedge_{x_l \in \nit{RelV}(\psi)}
x_l\neq \nn , ~\bar{\varphi}.$

\vspace{1mm}
Here, $S \in {\cal S}(\p{P})$ and the $R_i \in {\cal S}(\p{Q})$, for $\p{Q}
\in \N^\circ(\p{P})$.

\vspace{1mm} 5.(b)~~~~~
$Q\!\_(\bar{x}',\overline{\nn},\mathbf{t}) \leftarrow
R\_(\bar{x},\mathbf{t^{\star}}),  \n \
\nit{aux}_\psi(\bar{x}'),~
 \bar{x}'\neq \nn.$

\vspace{1mm}  Here, $Q \in {\cal S}(\p{P})$ and $R \in {\cal S}(\p{Q})$, for $\p{Q}
\in \N^\circ(\p{P})$.

\vspace{1mm}We also need the auxiliary rules (\ref{eq:aux1}), (\ref{eq:aux2}), and only
for predicates in ${\cal S}(\p{P})$ that appear in consequents of RDECs.

Now,  for a predicate $Q \in {\cal S}(\p{P})$,  a solution program like this never generates a tuple of the
form $Q(\bar{a},\ta)$. In consequence, the negative literals can be eliminated from the bodies of both
(\ref{eq:aux1}) and (\ref{eq:aux2}). The resulting program still contains negation, in rules 6. and 8. However, the
solution program becomes a (non-disjunctive) stratified normal program \cite{AHV1995}. In consequence, for every set of facts, it has only one stable model that
can be computed in polynomial time in data.
It corresponds to the only solution for peer \p{P} mentioned in Section  \ref{def:solProgIII}.

If the query ${\cal Q}$ (or, better, its rewriting $\mc{Q}^N$) posed to \p{P} can also be represented as a stratified normal program
$\Pi({\cal Q})$
(which is the case when ${\cal Q}$ is first-order, for example), then the combined program $\Pi(\p{P}) \cup \Pi({\cal Q})$
is also normal and stratified. In consequence, computing PCAs can also be done in polynomial time. This is under the assumption that
the peer has collected, for each of its neighbors, the intersection of its (local) solutions. We have obtained the following result.

\begin{propositionS}\label{cor:PCAimp} \red{For an unrestricted import PDES  schema $\mathfrak{P} = \langle
\mc{P}, \mf{S}, \Sigma,\trust\rangle$, an instance $\mathfrak{D}$ for $\mathfrak{P}$,  and a peer $\p{P}\in\P$, \
deciding answers to a conjunctive query posed to  \p{P} that are true in $\bigcap \nit{NS}(\p{P},\bar{D})$, where
$\bar{D} = D(\p{P}) \cup \bigcup_{\p{Q} \in \N^\circ(\p{P})}\nit{Core}(\p{Q},\mathfrak{D})$,
can be done in polynomial time in the size
of $\bar{D}$.} \boxtheorem
\end{propositionS}

\ignore{\comlb{In the previous paragraph we make a claim about polynomial time complexity, but it is assuming that the
local peer already gathered the intersection of the solutions of the neighbors. Any change if the whole
system is considered? This has to do with claim we have below in this section.
}}

In the restricted import case of Section \ref{sec:ri}, we allowed peers to have local ICs. In this case, as
illustrated in Examples \ref{ex:NoSol2} and \ref{ex:sevSols} the situation may change: Due to the local ICs, the solution program may be
disjunctive, and have none or several stable models, which is also reflected in the peers' solutions.

\ignore{
\comlb{Loreto: Puedes poner las condiciones con nulos en la reglas de abajo?}
\comlore{Done}
}

\begin{example} (example \ref{ex:NoSol2} continued) In this case we have no solution. This is captured by the
corresponding program through the program constraint \ $\leftarrow R^1\!\_(x,y,\ta), R^1\!\_(x,y,\fa)$. In each of the two models
of the program without the constraint, there will be the atoms $R^1(a,b,\ta), R^1(a,c,\ta),  R^1(a,b,\fa)$ or the atoms $R^1(a,b,\ta), R^1(a,c,\ta),  R^1(a,c,$ $\fa)$. Both models will be discarded for violating the constraint. \boxtheorem
\end{example}

\begin{example} (example \ref{ex:sevSols} continued)
\label{ex:sevSols2} The solution program for \p{P} has as main
rules, the following:
$$P\!\_(x,y,\mathbf{t}) ~\leftarrow~ Q\!\_(x,y,\trs), P\!\_(x,y,\fs),
x\neq \nn,y\neq \nn.\vspace{-3mm}$$
\begin{eqnarray*}
\!\!\!\!\!\!\!\!\!\!\!\!\!\!P\!\_(x,w_1,\fa)\!\!\!\!\!&\vee&\!\!\!\!\!P\!\_(x,w_2,\fa) \vee P\!\_(x,w_3,\fa)
 \ \leftarrow \  P\!\_(x,w_1,\trs), P\!\_(x,w_2,\trs),P\!\_(x,w_3,\trs), \nonumber\\
&&\hspace*{2cm}\bigwedge_{i\neq j} w_i \neq w_j, x \neq \nn,
w_1 \neq \nn, w_2 \neq \nn, w_3 \neq \nn.\label{eq:disy}
\end{eqnarray*}
We have a disjunctive solution program; actually one with two
stable models, as expected according to Example \ref{ex:sevSols}. \boxtheorem
\end{example}
\ignore{
\vspace{-1mm} \ni This result still holds under rather weak
conditions, e.g. if the ICs in \IC(\p{P}): (a) have a consequent that
contains at least one database predicate (not a built-in); or (b) if
only built-ins appear in the consequent, e.g. {\bf false}, there is a
predicate in the antecedent of the IC that does not appear in any of
\p{P}'s DECs.
If a PDES has no local ICs, then it is easy to see that the solution
program is head-cycle free \cite{DEGV97}, and we obtain

\begin{propositionS}  In the unrestricted import case,
the solution program for a peer with an empty set of local ICs is
equivalent to a non-disjunctive program. \boxtheorem
\end{propositionS}
}
\ni In \cite{BBB03,iidb06}, in the context of consistent query answering, several classes of ICs have been identified for which the logic
program becomes {\em head-cycle free}, in which  case, it can be replaced by and equivalent
non-disjunctive, i.e. normal, program \cite{DEGV97}. Cautious reasoning from normal logic programs is {\em
coNP}-complete \cite{DEGV97}. In consequence, for those classes of ICs,  peer consistent query answering is in {\em
coNP} in data complexity.\footnote{Again, under the assumption that
the peer has collected, for each of its neighbors, the intersection of its (local) solutions.} Among those constraints we find denial constraints, such as constraint (\ref{eq:four}) in Example
\ref{ex:trans0}. 
Even in the restricted import case with denial constraints as local constraints, peer consistent query answering is {\em coNP}-complete. (This
can be obtained from CQA under denial constraints \cite{BBB03}. Cf. \cite{bertossi11} for a survey of complexity results and references.)

\ignore{\comlb{El mismo comentario inmediatamente arriba. Aqui estamos en cierto modo
suponiendo que el peer recolecto las instancias de los otros pares, transitivamente.
Cambia algo si no?}  }

\ignore{
\comlb{OLD: Loreto: Do we still get, under the special null semantics
*and REF-acyclicity*, $\Pi^P_2$-completeness for PCQA? And in the
import case with local and RIC-acyclic ICs?}

\comlore{OLD: La demostración de $\Pi^P_2$-hard en el caso general requiere una FD
y una Inclusion Dependency que es cíclica. O sea, la demostración que tenemos
de hardness para CQA (que modifica una reducción de Chomicki \cite{CM05}) no
sirve si nos restringimos a ref-acyclic...

Sin embargo, puedo reducir el problema de de Chomicki \cite{CM05} al local
PCA (es decir el problema de la Proposición \ref{prop:PCAGivenCore}). Se
puede demostrar que sigue siendo $\Pi^P_2$-hard incluso si nos restringimos
al caso ref-acyclic. Lo agregue como una proposición en el apéndice:
Proposition \ref{prop:localAcycPCA}. No se si lo quieres poner en el cuerpo
del documento }

\comlb{Bueno, lo que tendria que ir aqui en el cuerpo no es exactamente la proposicion
\ref{prop:localAcycPCA}, sino el correspondiente al teorema \ref{theo:CheckingIfPCA} que resultaria de ella.

Lo que estariamos concluyendo entonces es que en el caso aciclico todavia el problema de PCA es $\Pi^P_3$-complete?}
}

\section{\red{Related Work}} \label{sec:related}

Our work can be placed among those on {\em semantic peer data management
systems}, a research direction that was started, to the best of our
knowledge, with \cite{HIST03} (cf. also \cite{HIM+04}). Mappings relate two
conjunctive queries that  are expressed in terms of the schemas of two
disjoint sets of peers. Already in those papers problematic cases of cyclic
dependencies, which implicitly involve a cyclic accessibility graph, were
identified. For example, we may have ${\cal P} = \{\p{P1}, \p{P2}, \p{P3}\}$,
with relations $R^1(\cdot), R^2(\cdot), R^3(\cdot)$, resp., and DECs
$\Sigma(\p{P1}) = \{\forall x(R^2(x) \rightarrow R^1(x))\}$, $\Sigma(\p{P2})
= \{\forall x(R^3(x) \rightarrow R^2(x))\}$, $\Sigma(\p{P3}) = \{\forall
x(R^1(x) \rightarrow R^3(x))\}$, each of them satisfied only by importing
data into the peer who owns the DEC, i.e. there is the implicit trust
relation $\{(\p{P1},\nit{less},\p{P2}),$ $(\p{P2},\nit{less},\p{P3}),$
$(\p{P3},\nit{less},\p{P1})\}$. This is an unrestricted and cyclic import
case. In the presence of cycles like this, peer consistent query answering
becomes undecidable.

In \cite{HIST03}, mappings are represented and given a semantics using classical first-order
logic (FOL). This choice requires consistency of peers (i.e. their
data, local ICs and local mappings) \wrt the system as a whole. More precisely,
certain answers from a peer are defined as those true in every global
instance that is consistent with the local data and metadata, which is rigid.
Some criticisms for the use of first-order logic  as a basis for the semantics
of PDESs were raised in \cite{CGL+04}, motivating a new research direction
(cf. below).

For comparison, our approach
is inconsistency tolerant, and does not provide a first-order semantics to the
global system. Instead, we appeal to a non-monotonic semantics for only
neighborhoods of peers. Our mappings are between two peers only, but in
that case, more expressive than those considered in \cite{HIST03}. \ignore{It would
not be difficult to smoothly and uniformly extend or work to include mappings
between a peer and a set of other peers.}

\red{It is worth noticing that all approaches adopting
a FOL interpretation for PDESs, e.g. \cite{HIM+04,HIST03},
do not consider peer mappings as constraints, but as logical
implications from a peer to another. This in fact coincides with having
trust relationships of the form $(\p{P}, \nit{less}, \p{Q})$ for any mapping specified
from \p{P} to \p{Q}.}

Among the approaches to  PDESs that are the closest to ours, we should mention
\cite{CGL+04,CDL+05,CDL+08,FKLZ04}. The DECs are of the ``import kind", and there are no explicit trust relationships. For example,
in \cite{CDL+08}, DECs are of the form ~$\nit{CQ}_i\rightarrow
        \nit{CQ}_{\!j}$, where $\nit{CQ}_i, \nit{CQ}_{\!j}$ are conjunctive
        queries
        over \p{Pi} and \p{Pj}'s schemas, resp. This kind of DECs keep the schemas separate, on different sides of the
        implication. The latter has to be interpreted appropriately. Actually, the
        semantics of the system, in particular of the DECs,  is given by an  {\em epistemic logic}.
 Local ICs violations are
        avoided by ignoring a peer that is inconsistent \wrt its local
                ICs.
New atoms are added into a peer by
                interaction with other peers only if this
                does not produce a local IC violation. In consequence, a peer trust other peers more than itself, as long
                as no local inconsistencies are produced. There is not consistency restoration process involved.

\begin{figure}\centering
\hspace*{1cm}\includegraphics[width=8cm]{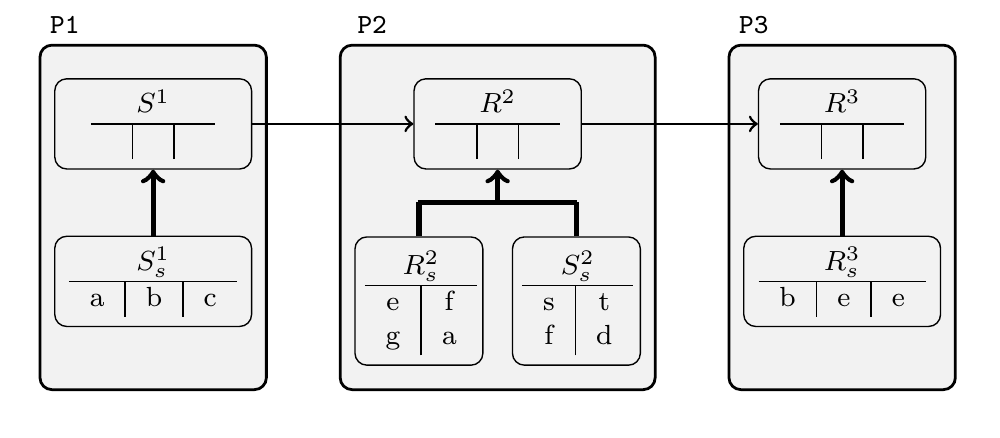}
\vspace{-3mm}\caption{Integration systems as peers}\label{fig:tanos}
\end{figure}



 As shown in Figure \ref{fig:tanos}, in \cite{CDL+08} each peer is considered as a local virtual
data integration system that follows the {\em global-as-view} approach
\cite{lenzerini02}. In this example, the three peers are locally defined by the GAV mappings
 $\forall x \forall y \forall z (S^1_s(x,y,z)
\rightarrow S^1(x,y,z))$, $\forall x \forall y \forall z (R^2_s(x,y) \wedge
S^2_s(y,z) \rightarrow R^2(x,y,z))$, and $\forall x \forall y \forall z
(R^3_s(x,y,z) \rightarrow R^3(x,y,z))$, resp. The predicates of the
form $P_s$ correspond to local sources, and those of the form
$P^i$ correspond to the mediated schema provided by peer
\p{Pi}. In addition, the DECs establish mappings between the peers. In this
case: $\forall x \forall y(R^2(x,y,z) \rightarrow \exists w
S^1(x,y,w))$, $\forall x \forall y (R^3(x,y,y) \rightarrow \exists u \exists v
R^2(u,x,v) )$. These DECs are used to pass data from the peer mentioned in the antecedent
to the one mentioned in the consequent. This activity is  based on a theory written in epistemic logic which is generated from the local and inter-peer
mappings.  More specifically,
 the following epistemic theory is used for query answering:

 \ignore{\comlb{Could you check this? Reviewer 1 says about this example: ``page 46:
There are some errors in the formula of the example ($R^1$ should be
$S^1$, and $R^3$ in the right-hand side of the second axiom should be $R^2$, maybe)}\\
\comlore{Fixed. In green. The text was using different tables than the figure!}
}

\begin{center}
\ni$\left.\begin{minipage}{8cm} $\mathbf{K_1}(\forall x \forall y \forall z
(S^1_s(x,y,z) \rightarrow S^1(x,y,z))) $

$\forall x \forall y (\mathbf{K_2}(R^2(x,y,z)) \rightarrow
\mathbf{K_1}(\exists w S^1(x,y,w)))$
\end{minipage}\right\}$ Specification of $\p{P1}$\\

\vspace{2mm} \ni$\left.\begin{minipage}{8cm}
$\mathbf{K_2}(\forall x \forall y\forall  z (R^2_s(x,y) \wedge S^2_s(y,z)
\rightarrow R^2(x,y,z)))$

$\forall x \forall y (\mathbf{K_3}(R^3(x,y,y)) \rightarrow
\mathbf{K_2}(\exists u \exists v R^2(u,x,v)))\}$

\end{minipage}\right\}$ Specification of $\p{P2}$\\

\vspace{2mm} \ni$\left.\begin{minipage}{8cm}
$\mathbf{K_3}(\forall x \forall y \forall z (R^3_s(x,y,z) \rightarrow
R^3(x,y,z)))$
    \end{minipage}\right\}$ Specification of $\p{P3}$\\
\end{center}

\ni $\mathbf{K_i}\varphi$ is interpreted as $\varphi$
        is known by peer $\p{Pi}$. The idea behind using the
        epistemic theory is that data that is known (or certain)
        is passed from local sources to mediated schemas and from peers to other peers.
A tuple $\bar{t}$ is a peer consistent answer to a
        query $\mathcal{Q}$ posed to peer $\p{Pi}$  if
        $\mathbf{K_i}\mathcal{Q}(\bar{t})$ is a logical consequence
        of
    the epistemic theory.

An advantage of this approach is that the semantics can be applied
in the presence of
      cycles. However, possibly the whole  epistemic theory has to be used by a peer \p{Pi}
to do query answering, which requires not only data, but also
the mappings and DECs; and this  not only of its
        neighbors, but of all accessible peers. (This semantics is similar in spirit to the one called
        {\em ``send all"} in Section \ref{sec:trans}.)

Our approach can be easily and uniformly adapted in order to make
each peer a local data integration system. For this, the
 answer set programs that specify the legal instances of a virtual data
integration introduced in \cite{IJCAI03,JournalVienna04,BeBra04} can be used in combination
with those presented here.

\red{The epistemic theory has also been extended in \cite{CDL+08} in order to make the PDES
``inconsistency tolerant". This is done by using additional modal operators and extending the epistemic theory, achieving that:
(a) A peer whose local data is inconsistent \wrt its ICs, is not considered for
data exchange, that is, the PDES behaves as if that peer didn't exist. In our case, however, to such a peer, whenever possible, a repair
semantics is applied. (b) Inconsistencies due to data imported from other peers, referred as P2P inconsistencies, are solved by
removing a minimal amount of data imported from other
peers in order to preserve consistency. In other words, atoms are imported into a peer by
interaction with other peers only if this does not produce a violation of an integrity constraint of a peer.
In our case, under the {\em same} or {\em less} trust conditions, the
data is accepted and the peer applies again a local repair
semantics.}


In order to answer a query \cite{CDL+05}, a peer traverses
the network eventually collecting at its site all DECs, ICs and data
of other logically related peers. With these elements, the peer
can construct its epistemic theory, that is used for
query answering.
An accessibility cycle can be detected by using request identifiers.
 The use of epistemic logic makes sure that {\em certain data}, the one
a peer really knows, is passed to another peer. In our case, a peer
collects only data from its neighbors; and certainty is achieved by
using the PCAs of a peer, or more generally, the intersection of its
solutions. A more detailed comparison can be found in \cite{Bravo07}.

\ignore{\begin{verbatim}In (Calvanese at al 2008,2005) there is consistency restoration, to
repair violations of peer constraints by data coming from
neighbors. This is obtained through a non-monotonic epistemic operator
(additional to the one mentioned at page 46).\end{verbatim}
\comlore{Done. See paragraph above in red. What the reviewer was pointing out was that
the inconsistency tolerance requires modifications of the epistemic theory  shown above. This is,
the theory shown is NOT inconsistency tolerant. The extended theory is more complex and I don't think it would add more
value to this explanation. But if you want I can write it down... }   }

The semantics in \cite{FKL+03,FKLZ04} coincides with the epistemic
semantics in \cite{CDD+04}. The former also provide a distributed algorithm,
where  peers' data is updated by instruction of a super peer. When a
query is posed to a peer, it can answer the query right away with its
data, because the PDES is already updated.

The data exchange problem among
distributed independent sources has been investigated in \cite{CarGre*06,ISMIS2008,IDEAS2011}. In \cite{CarGre*06} the authors define a declarative semantics for P2P systems that
allows one to import, into each peer, maximal subsets of atoms that do not violate the local integrity constraints.
The framework has been extended in   \cite{ISMIS2008} with a mechanism to set different degrees of reliability for neighbor peers, and in \cite{IDEAS2011}, where ``dynamic"  preferences can be used to import data in the case of conflicting sets of atoms, depending on the properties of the peers' data.
This extended  framework allows one to model preferences like ``in the case of conflicting information, it is preferable to import data from the neighbor peer that
can provide the maximum number of tuples" or ``in the case of conflicting information, it is preferable to import data from the neighbor peer such that the sum of the values of an attribute is minimum", without having to select
preferred peers a-priori. To enforce this preference mechanisms the  P2P framework  has been enriched with aggregate functions.

In \cite{Fux06}, the case of two peers that belong to a PDES is analyzed.
From a source peer, \p{P}, to a target peer, \p{Q}, there may be both source-to-target dependencies, as in data exchange \cite{kolaitis}, and also target-to-source
dependencies. The former are used to transport data from \p{P} to \p{Q}, and the second, to filter the received data in \p{Q}. In addition,
the target may have an instance and local constraints. The  existence of solutions for the target peer is investigated.
The semantics we have introduced, but using labeled nulls instead of \nn, could be adapted to a situation like this (cf. Section \ref{sec:mixed}). It is also
possible to use the solution programs in the case of  \cite{Fux06}. In this scenario, the source-to-target dependencies would
give rise to program rules that do the data exchange, and both target-to-source dependencies and target dependencies would become
hard program constraints.

More material and references on peer data exchange systems can be found in \cite{aberer}.
\red{Consistency issues that appear in our PDES scenario may also appear in the context of ontologies,
when they have to be aligned in the presence of {\em bridges} between them. When ontologies are merged, it
becomes necessary to solve inconsistencies between
them \cite{serafini}. However, in relation to mappings, we go beyond the cases considered
in that  area. The mappings, or bridges, between
ontologies are in general much simpler than the general case considered in our work.}


In this work we have concentrated on null values that are handled as in SQL relational
databases. Actually, in that case, there is a single null, namely the constant {\tt NULL}.
It can be interpreted in different ways in a database, e.g. as a representative for an existing
but unknown value, as a missing or non-existent value, as a withheld or hidden value, etc.
In our case, we do not make any ontological assumption about this constant. Our purpose
consists in logically capturing the way it is handled in data management operations by an SQL database. More
specifically, \wrt query answering, integrity checking and interaction with other data elements.

Most of the work in database theory bypasses the issue of SQL null, mainly because it is considered to be an
undesirable oddity that could be replaced by a logically cleaner solution. However, SQL {\tt NULL} exists, is used, and
will most like stay with us. It would be better to accept it, but trying to clarify its logical and operational
status. This is our approach in this work. From this point of view, we accept that  SQL {\tt NULL} appears, not only in databases,
but could also appear explicitly in data exchange constraints and integrity constraints, and in their semantics. For example, a referential constraint or data exchange
constraint could
be of the form \ $\forall x y (S(x,y) \rightarrow T(x,{\tt NULL}))$, i.e. requiring that $T$'s second attribute should be filled
with {\tt NULL}.\footnote{In our case, given our repair semantics of Section \ref{sec:nullsRep}, even a constraint of the form $\forall x y (S(x,y) \rightarrow \exists z T(x,z))$ will be enforced
by giving to $z$ the value ${\tt NULL}$ (unless some other constraint prevents this).}

A similar approach is adopted in \cite{franconi12}, where an alternative reconstruction of SQL {\tt NULL} is
attempted. They propose a corresponding relational algebra that takes care of the special status of {\tt NULL},
but in the end, after the reconstruction, this constant disappears from the data domain. We keep it, but rewrite queries
and constraints in such a way that it can be treated as any other constant.

There is a large literature on nulls in relational databases (see, e.g. \cite{LeveneBook}). However, to the best of
our knowledge those nulls are ``ideal" nulls, and not the real SQL {\tt NULL}. There are also some logical approaches,
starting with Reiter's treatment of nulls \cite{reiter}, which can be multiple, existential values not subject to the {\em unique names assumption}.
They are also different from the SQL {\tt NULL}.
More recent work extending Reiter's approach can be found in \cite{gelfond94,lifschitz12}.  See also \cite{libkin,libkin16} for a recent work on incomplete information in relational databases.

\section{Conclusions} \label{sec:concl}

We have introduced a general framework for peer data exchange with trust
relationships.  The methodology is flexible and {\em inconsistency tolerant} in that each peer solves
its data and semantic conflicts at
query time, when querying its own and other peers' data.

The general semantic framework can be specialized in several ways, and we presented some possibilities.
In particular, we developed a specific semantics based on universal and referential data exchange constraints, and
on the use on SQL null values to deal with incomplete information. In this scenario,
logic programs can be used to specify the solution instances for a peer and to
obtain peer consistent answers.

Techniques for optimizing the logic programs and their execution could be applied for query answering. Among them
we find  the partial evaluation of programs and the solution instances they specify, since we are not interested in the
solution
per se, but in the PCAs. More specifically, techniques used in CQA, such as {\em magic sets}
for stable model semantics \cite{magic}, and identification of
predicates that are relevant to queries and constraints, could also
be used in this setting. In this way, the number of rules and the
amount of data that are needed to run the program are reduced
\cite{CB10,eiter}.

The problem  of query evaluation from disjunctive programs is
$\Pi^P_2$-complete in data \cite{DEGV97}, which matches the complexity of
PCA, as we established here. In spite of this, we have also identified syntactic classes
of PDESs for which peer consistent query answering has a lower
complexity. For them, specifically tailored mechanisms to solve this
problem could be developed, as for CQA.

\ignore{Semantics for PDESs have been introduced and analyzed in
\cite{HIST03,FKL+03,KAM03,HIM+04,FKLZ04,CGL+04,CDD+04,CDL+05,Fux06},
but  without considering trust relationships. In them, if there is a
DEC from $\p{P}$ to $\p{Q}$, it is implicitly assumed that $\p{P}$
trusts itself less than $\p{Q}$. Also, all the research so far, has
concentrated on the unrestricted import case. In our setting,  a DEC
may also restrict the data that can belong to a peer.}

We should emphasize that the DECs we can handle are more general that
those found in related work on peer data exchange.

\ignore{
+++
\burg{A PDES can be seen as a set of
information  agents, each of them being the owner  of an ontology.
For a peer $\peer{P}$, its ontology consists of a relational
database instance $D(\p{P})$ and  its set
$\Sigma(\peer{P}) = \bigcup_{\peer{P'}\in {\cal P}}
\Sigma(\peer{P},\peer{P'})$ of DECs, and its trust
relationships. These ontologies may be pairwise inconsistent due
to the DECs and the database facts. \ignore{We could easily extend our
framework in order to include locally defined predicates (views) for a peer and  DECs involving those views.}
Our notion of DEC largely extends the  inclusion mappings between concepts that are commonly considered in the ontological scenario (cf. the related work section for more on this).
Actually, the DECs we can handle can be much more complex.
Actually, we are able to
show in detail how to deal with DECs that express intertwined relationships between schemas.
However, we should emphasize that our goal is not to materially change ontologies in order to make them mutually consistent,
but to solve (relevant) inconsistencies at query time and for query answering.}
+++}

Our semantics allows for inconsistent peers and inconsistencies
between peers, without unraveling logical reasoning or having
to exclude peers whose data participate in inconsistencies. Actually, our semantics for
peer data exchange systems smoothly extend the repair semantics for consistent query answering from inconsistent
databases.

Along the way, and interesting on its own merits, we developed a  semantics of conjunctive query answering and constraint satisfaction in terms of classical FO predicate
logic. It puts on a solid ground the  mechanisms implemented in SQL databases and specifications in the SQL Standard for handling null values. 

\vspace{2mm}

\vspace*{1mm} \noindent {\bf Acknowledgements:} ~ Research
supported by  NSERC DG (250279-2011) and a CITO/IBM-CAS Student Internship.   Loreto Bravo is funded
by CONICYT grant PSD-57.  We appreciate
valuable and detailed comments from  the anonymous reviewers.


\label{lastpage}

\newpage
\appendix



\vspace*{0.5cm}
\begin{center}
\large \sc \bf ELECTRONIC APPENDIX
\end{center}




\section{ \ Discussion}\label{app:disc}

\subsection{On cycles and their assumptions}\label{sec:cycles}

In this section, unless stated otherwise, we refer to the special semantics introduced in Section
\ref{sec:nulls}.

We have  assumed that $\G(\mathfrak{P})$ is acyclic. However, the
peers, not being aware of being in a cycle in $\G(\mathfrak{P})$,
could attempt to do data exchange as described above. In order to
detect an infinite loop, for a query ${\cal Q}$ posed by a peer \p{P}, a unique identifier $\nit{id}(\p{P},{\cal Q})$ can be
created and kept in all the  queries that have origin in ${\cal Q}$. If an identifier comes back to a peer, it will realize that
it is in a cycle and act accordingly.

The assumption of acyclicity of the accessibility graph is quite
cautious, in the sense that it excludes cases where a reasonable
semantics could still be given and the logic programs would work correctly. This is
because the cycles in $\G(\mathfrak{P})$ are not necessarily relevant.

\begin{exampleApp}\ignore{[{\bf cyclic accessibility graph, but reasonable semantics}]} \label{ex:irrelcycle}
Consider ${\cal S}(\p{P1}) = \{R^1(\cdot), S^1(\cdot)\},$ ${\cal
S}(\p{P2})$ $=$ $\{R^2(\cdot),$ $S^2(\cdot)\},$ $\Sigma(\p{P1},$
$\p{P2})$ $=$ $\{\forall x(R^2(x)$ $\rightarrow$ $R^1(x))\},$
$\Sigma(\p{P2},\p{P1}) = \{\forall x(S^1(x) \rightarrow S^2(x))\}$,
$\trust = \{(\p{P1},$ $\nit{less},$ $\p{P2}),$ $(\p{P2},$
$\nit{less},$ $\p{P1})\}$. If a query is posed to \p{P1}, it will
request from \p{P2} the PCAs to query $R^2(x)$, but not those to
query $S^2(x)$. Peer \p{P2} can realize it does not need data from \p{P1}
and will simply return $D(\p{P2})\rs \{R^2\}$ to \p{P1}, who will run its
solution program and answer the original query. Even though there is
a cycle in $\G(\mathfrak{P})$, there is no infinite loop in the query answering process. \boxtheorem
\end{exampleApp}

As we mentioned before,
 if there are ref-cycles in $\Sigma(\p{P})$, the stable models
 of the solution program for \p{P}
may correspond to a strict superset of the solutions. This is shown in the next example. In such a case,
post-processing that deletes models corresponding to non-minimal
``solutions" is necessary.

\begin{exampleApp}\ignore{[{\bf ref-cycles, complete but unsound program}]}
\label{ex:cyclicSame} Consider $D(\p{P1})$ $= \{R^1(a,b)\}$, $D(\p{P2})$
$=$ $\{R^2(a,c)\}$, $\Sigma(\p{P1},\p{P2}) = \{\forall x \forall z(R^1(x,z)$ $
\rightarrow$ $ \exists y$ $ R^2(x,y)),$ $ \forall x \forall z(R^2(x,z) $ $
\rightarrow$ $ \exists y$ $ R^1(x,y)\}$, which is ref-cyclic; $\Sigma(\p{P2}) = \Sigma(\p{P1},\p{P1})  =
\emptyset$; and $(\p{P1},\same,\p{P2}) \in \trust$. Here, \p{P1} has only one
solution, namely $\{R^1(a,b)\}$.

However, $\Pi(\p{P1})$
 has two models. The most relevant part of the program
 consists of the facts \
   $R^1(a,b), ~R^2(a,c)$, and the following rules:

\vspace{-3mm}\hspace{-.8cm}\begin{minipage}{10cm}\begin{eqnarray*}
R^1\!\_(x,y,\fa) \vee R^2\!\_(x,\nn,\ta)&\leftarrow& R^1\!\_(x,y,\trs), \n \ \aux_1(x), x \not= \nn\\
\aux_1(x) &\leftarrow& R^2(x,\nn), \n R^2\!\_(x,\nn,\fa)\\
\aux_1(x) &\leftarrow& R^2(x,y,\trs), \n R^2\!\_(x,y,\fa), x
\not=
\nn, y \not = \nn\\
\ni R^2\!\_(x,y,\fa) \vee R^1\!\_(x,\nn,\ta) &\leftarrow& R^2\!\_(x,y,\trs), \n \ \aux_2(x), x \not= \nn\\
\aux_2(x) &\leftarrow& R^1(x,\nn), \n R^1\!\_(x,\nn,\fa)\\
\aux_2(x) &\leftarrow& R^1(x,y,\trs), \n R^1\!\_(x,y,\fa), x
\not= \nn, y \not = \nn\\
\end{eqnarray*}\end{minipage}

\vspace{-3mm}The two models correspond to the neighborhood
``solutions" $\{R^1(a,b),R^2(a,c)\}$ and $\emptyset$, producing in their turn, the  instances
$\{R^1(a,b)\}$ and $\emptyset$, resp., for \p{P1}. Only
the former is a solution instance. The second model
 is the result of cycles through weak negation ($\nit{not}$). The cycle
creates the self justification of facts as follows: (i) If we choose
$R^2\!\_(a,c,\fa)$  to be true, then by the second and third rules above,
$\aux_1(a)$ is false. (ii) Then, the first rule can be satisfied, by
making $R^1\!\_(a,b,\fa)$ true. (iii) By the fifth and sixth rules,
$\aux_2(a)$ is false. (iv) This justifies making $R^2\!\_(a,b,\fa)$
true, thus, closing the cycle. Notice, that in the whole
justification the changes where not determined by inconsistencies.
\boxtheorem
\end{exampleApp}

\ni There are also cases with an acyclic $\G(\mathfrak{P})$, but with
ref-cycles in the DECs, where the logic programming counterpart of
the semantics is correct due to the role of the trust relationships.

\begin{exampleApp}\ignore{[{\bf ref-cycles, complete and sound program}]} \label{ex:re-cycle}  (example \ref{ex:cyclicSame} continued)
If we replace $(\p{P1},\same,\p{P2}) \in \trust$ by
$(\p{P1},\less,\p{P2}) \in \trust$, the relevant part of
$\Pi(\p{P1})$ now is:   $R^1(a,b), ~R^2(a,c)$, plus
 \begin{eqnarray*}
\ni R^1\!\_(x,y,\fa) &\leftarrow& R^1\!\_(x,y,\trs), \n \ \aux_1(x), x \not= \nn.\\
\aux_1(x) &\leftarrow& R^2(x,\nn), \n R^2\!\_(x,\nn,\fa).\\
\aux_1(x) &\leftarrow& R^2(x,y,\trs), \n R^2\!\_(x,y,\fa), x \not=
\nn, y \not = \nn.\\
\ni R^1\!\_(x,\nn,\ta) &\leftarrow& R^2\!\_(x,y,\trs), \n \ \aux_2(x), x \not= \nn.\\
\aux_2(x) &\leftarrow& R^1(x,\nn), \n R^1\!\_(x,\nn,\fa).\\
\aux_2(x) &\leftarrow& R^1(x,y,\trs), \n R^1\!\_(x,y,\fa), x \not=
\nn, y \not = \nn.
\end{eqnarray*}
Since \p{P1} trusts more peer \p{P2} than itself, it will modify only
its own data. This program computes exactly the solutions for peer
\p{P1}, i.e. $\{R^1(a,b)\}$, even though the DECs exhibit ref-cycles.
\boxtheorem
\end{exampleApp}

\ni It becomes clear that it is possible to find more relaxed conditions,
both on the accessibility graph and ref-cycles, under which a
sensible semantics for solutions and semantically corresponding logic
programs can be given. Also, for cyclic accessibility
graphs, {\em super peers} \cite{YG03} could be used, to detect cycles
and prune certain DECs, making the graph acyclic if necessary; and
then our semantics could be applied.

\subsection{Query sensitive query answering}\label{sec:queSen}

Our definition of the solution semantics and the peer consistent answers might suggest that, in order
to answer a particular query $\mathcal{Q}$, a peer \p{P} has to import the full intersection of the solutions of each of its neighbors, which in their turn have to do the same, etc. If we do this, most likely most of the data imported by \p{P} will be irrelevant for the query at hand, and is not needed.

It is possible to design query answering methodologies that are more sensitive to the query at hand, in the sense
that only the relevant data is transitively imported into \p{P} before answering $\mathcal{Q}$. A full treatment
of this subject is beyond the scope of this paper. However, we can give some indications as to how to proceed.

In \cite{CB10}, the {\em dependency graph} of database predicates \wrt a set of ICs was introduced and
used to capture the notion of possibly transitive relevance of a predicate for another, which is useful
in consistent query answering. Here we can use similar graphs for each peer in relationship with its
neighbors through a set of DECs, also taking into account the local ICs. These graphs would give us a better upper bound on what to import from other
peers (as opposed to bringing the full intersection of solutions).

More precisely, if a query $\cal{Q}$ to \p{P} contains $\cal{S}(\p{P})$-predicates $P_1, \ldots, P_n$, with relevant
$\cal{S}(\p{Q})$-predicates $Q_1^1, \ldots, Q_{m_1}^1, \ldots, Q^n_1, \ldots, Q^n_{m_n}$, resp., at a neighbor \p{Q},
then \p{P} will request from \p{Q} the PCAs to each of the constant-free, atomic queries $Q^i_j(\bar{x})$. The corresponding sets
of answers will form the (most likely smaller) instance provided by \p{Q} to \p{P}, who will prune its
repair program by keeping only the relevant rules, i.e. those that are related to $\cal{Q}$, the $P_i$ and the
 $Q^i_j$. This idea can be illustrated by means of
an example.

\begin{exampleApp}
Consider a schema $\mathfrak{P} = \langle
\mc{P}, \mf{S}, \Sigma,\trust\rangle$ with  $\Pe = \{\p{P1}, \p{P2},
\p{P3}\}$, $\mf{S} = \{{\cal S}(\peer{P1}), {\cal S}(\peer{P2}),$ $ {\cal S}(\peer{P3})\}$,
${\cal S}(\peer{P1}) = \{R^1(\cdot,\cdot),$ $S^1(\cdot,\cdot),$ $T^1(\cdot,\cdot)\}$, ${\cal S}(\peer{P2}) =
\{R^2(\cdot,\cdot),$ $S^2(\cdot,\cdot), T^2(\cdot,\cdot)\}$, ${\cal S}(\peer{P3}) =
\{R^3(\cdot,\cdot), S^3(\cdot,\cdot)\}$, ~$\trust ~= \{(\peer{P1},\less,\peer{P2}),$ $(\peer{P2},\same,\peer{P3})\}$.
Let $\mathfrak{D}$ be an arbitrary instance for the PDES.

The sets of DECs are: \
$\Sigma(\p{P1},$ $\p{P1})= \{\forall x \forall y ( R^1(x,y) \wedge S^1(x,y)
\rightarrow {\bf false})\},$ $\Sigma(\peer{P1},\peer{P2}) =$ $\{\forall x \forall y(R^2(x,y) \rightarrow
R^1(x,y)),$ $\forall x \forall y(S^2(x,y) \rightarrow
S^1(x,y))\}$, and $\Sigma(\peer{P2},$ $\peer{P3})$ $ =\{\forall x \forall y ( R^2(x,y) \wedge
R^3(x,y) \rightarrow {\bf false})\}$.

 If \p{P1} is posed the query $\mathcal{Q}_1(x): \ \exists y R^1(x,y)$, then the relevant predicates in
$\mathcal{S}(\p{P1})$ are $R^1, S^1$ (due to the DEC in $\Sigma(\p{P1})$). Then, through the DECs also, it follows that the  predicates that are relevant to  $\p{P1}$ are $R^2, S^2$
at \p{P2}. So, \p{P1} poses to \p{P2} the queries $R^2(x,y), S^2(x,y)$. The only relevant predicate at \p{P3} is
$S^3$. So, \p{P2} will pose to \p{P3} the query $S^3(x,y)$.

In this case, \p{P3} will return $D(\p{P3})\rs \{R^3\}$ to \p{P2}, which, due to the UDEC in $\Sigma(\peer{P2},\peer{P3})$, will subtract it from  $D(\p{P2})\rs \{R^2\}$, because
$(\bigcap \nit{Sol}(\p{P2},\mathfrak{D}))\rs \{R^2\} = (D(\p{P2})\rs \{R^2\} \smallsetminus D(\p{P3})\rs \{R^3\})$.
%
%
%
Peer  \p{P2}
will send this difference to \p{P1} as it is the answer to query $R^2(x,y)$. Peer  \p{P2} will also return
to \p{P1} the entire $D(\p{P2})\rs \{S^2\}$  as its answer to the query $S^2(x,y)$.

Finally, \p{P1} will answer the original query with a solution program containing as facts the tuples in
$D(\p{P1})\rs \{R^1\}, D(\p{P1})\rs \{S^1\}, D(\p{P2})\rs \{S^2\}, ((D(\p{P2})\rs \{R^2\}) \smallsetminus (D(\p{P3})\rs \{R^3\}))$. The last set, as
an extension for $R^2$ in the program. \boxtheorem
\end{exampleApp}
The methodology sketched in this example will be certainly more efficient
than computing and shipping the full intersection of a peer's solutions. It
is natural to expect that additional optimizations can be developed.

A particularly
appealing, but provably less general, approach to peer-consistent query answering is {\em first-order query rewriting}, which we illustrate by means of an example.

\begin{exampleApp} \label{ex:peer} Consider an instance $\mathfrak{D}=\{D(\p{P1}),D(\p{P2}),D(\p{P3})\}$ for the schema $\mathfrak{P} = \langle
\mc{P}, \mf{S}, \Sigma,$ $\trust\rangle$ with ${\cal P} = \{\peer{P1}, \peer{P2}, \peer{P3}\}$, $\mf{S} = \{{\cal S}(\peer{P1}), {\cal S}(\peer{P2}),$ $ {\cal S}(\peer{P3})\}$,
 ${\cal S}(\p{Pi}) = \{R^i(\cdot,\cdot)\}$, $\trust = \{ (\peer{P1},\less,$ $ \peer{P2}), (\peer{P1},\same,
\peer{P3})\}$, $D(\peer{P1}) =
\{R^1(a,b),$ $ R^1(s,$ $t)\}$, $D(\peer{P2}) = \{R^2(c,d),$ $
R^2(a,e)\}$, $D(\peer{P3}) = \{R^3(a,f),$ $ R^3(s,u)\}$ and the DECs:
\begin{eqnarray*}
\Sigma(\peer{P1},\peer{P2}) &=& \{~ \forall x \forall
y(R^2(x,y) \rightarrow R^1(x,y))~\};\\
\Sigma(\peer{P1},\peer{P3}) &=& \{~\forall x \forall y \forall z(R^1(x,y)
\wedge R^3(x,z) ~~\rightarrow~~ y = z)~\}.
\end{eqnarray*}
We are interested in \p{P1}'s
solutions.
Since \p{P2}, \p{P3} are sink peers  in the graph ${\cal G}(\mathfrak{P})$, we have the extended instance
$D = \{R^1(a,b), R^1(s,t), R^2(c,d),$
$R^2(a,e),$ $ R^3(a,f),$ $R^3(s,u)\}$, from which we have to obtain the solutions for \p{P1}.

The solutions for \peer{P1}
are obtained by first repairing $D$ \wrt $\Sigma(\p{P1},\p{P2})$,  obtaining
$D_1 = \{R^1(a,b), R^1(s,t),$ $
R^1(c,d), R^1(a,e),$ $ R^2(c,d), R^2(a,e), $ $R^3(a,f),$
$R^3(s,u)\}$. We have only one repair at this stage, which now
has to be repaired in its turn \wrt $\Sigma(\peer{P1},\peer{P3})$
(but keeping the relationship between \peer{P1} and \peer{P2}
satisfied). There are two sets of tuples violating
$\Sigma(\peer{P1},\peer{P3})$ in $D_1$: $\{R^1(s,t),R^3(s,u)\}$
and $\{R^1(a,b),R^1(a,e),$ $R^3(a,f)\}$. The first violation can be
repaired by deleting  any, but only one, of the two tuples. The
second one, by deleting tuple $R^3(a,f)$ only  (otherwise we would
violate the relationship between \peer{P1} and \peer{P2}).

As a
consequence, we obtain two neighborhood solutions: $D' = \{R^1(a,b), R^1(s,t),$ $R^1(c,d),$ $ R^1(a,e), R^2(c,d), R^2(a,e)\}$,
~and~ $D'' = \{R^1(a,b), R^1(c,d), R^1(a,e),$ $R^2(c,d),R^2(a,$ $e),$
$R^3(s,u)\}$. The solutions for \p{P1} are: $D(\p{P1})' = \{R^1(a,b), R^1(s,t),$
$R^1(c,d),$ $ R^1(a,e)\}$ and $D(\p{P1})'' = \{R^1(a,b),$ $ R^1(c,d),$ $R^1(a,e)\}$.

If the query  ${\cal Q}: R^1(x,y)$ is posed to \peer{P1}, the PCAs are
$\langle a,b\rangle, \langle c,d\rangle, \langle a,e\rangle$, because
those are $R^1$-tuples found in  in the intersection of
\peer{P1}'s solutions.

Now, let us try an alternative method for peer consistently answering the same query. We
first rewrite the query using the DEC in
$\Sigma(\peer{P1},\peer{P2})$,  obtaining~ ${\cal Q}': R^1(x,y) \vee
R^2(x,y),$ with the effect of bringing \peer{P2}'s data into
\peer{P1}. Next, considering
$\Sigma(\peer{P1},\peer{P3})$, this query
is rewritten as
\begin{equation}\label{eq:rew2} {\cal Q}'': [R^1(x,y)
\wedge \forall z_1((R^3(x,z_1) \wedge \neg \exists z_2 R^2(x,z_2))
~\rightarrow~ z_1 = y)] ~~\vee~~ R^2(x,y).
\end{equation}
 To answer this query, \peer{P1}  first issues a query to \peer{P2}
to retrieve the tuples in $R^2$, since they will be essentially in $R^1$ in all the
solutions, due to $\Sigma(\peer{P1},\peer{P2})$. Next, a query is
issued to \peer{P3} to leave aside from the answers those tuples of $R^1$ that have
the same first but not the same second argument in $R^3$. This filtering is performed  as long
as there is no  tuple in $R^2$ that ``protects" the
tuple in $R^1$. For example, the tuple $R^1(a,b)$ is protected by $R^2(a,e)$
because, as $R^1(a,e)$ belongs to all the solutions, the only way
to repair a violation \wrt $\Sigma(\peer{P1},\peer{P3})$ is
by deleting the tuple from $R^3$. In this case, the $R^1$-tuple will
be in the answer.

We can see that  answering  query (\ref{eq:rew2}) amounts to
issuing from \p{P1}  \red{queries to \p{P2}, \p{P3} about the contents of their relations, $R^2$ and $R^3$, resp.}, which
are answered by the latter by  classical query evaluation over their local
instances. After those data have been gathered by \p{P1}, it proceeds to evaluate (\ref{eq:rew2}), which contains an implicit repair process. 


Now, the answers to (\ref{eq:rew2}) are $\langle a,b\rangle,
\langle c,d\rangle, \langle a,e\rangle$, precisely the PCAs we obtained above, considering all
the explicit solutions for \p{P1}. \boxtheorem
\end{exampleApp}
The FO rewriting methodology we just illustrated is bound to have limited applicability. If this was a general mechanism,  PCAs to conjunctive queries could be obtained  in polynomial time in data,
i.e. in the size of the union of the instances of a peer and those of its neighbors (or the local intersections of their
solutions). However, Corollary \ref{cor:PCA} tells us that the complexity is  higher than this (if $P \neq
\nit{N\!P}$).

\subsection{A semantics based on arbitrary data elements}\label{sec:mixed}

The purpose of this section is twofold.  First, we will present an alternative special semantics that fits into the general semantic
framework of Section \ref{sec:semantics}. Second, we will show that this general semantics (and also the one in Section \ref{sec:nulls}) can handle data mappings
that are more complex that those usually considered
in the related  work on peer data exchange. All this will be done on the basis  of an extended example.

Consider a PDES  $\mathfrak{P} = \langle
\mc{P}, \mf{S}, \Sigma,\trust\rangle$ with $\Pe = \{\peer{P1},
\peer{P2}\}$, $\mf{S} = \{{\cal S}(\p{P1}), {\cal S}(\p{P2})\}$, ${\cal S}(\p{P1}) = \{R^1(\cdot,\cdot),$ $T^1(\cdot,\cdot)\}$,
${\cal S}(\p{P2}) = \{T^2(\cdot,\cdot),$ $S^2(\cdot,\cdot)\}$, $\trust = \{(\peer{P1},\less,\peer{P2})\}$, and
$\Sigma(\p{P1},\p{P2})$ consists of the following DEC:
\begin{equation}\label{eq:mixed} \forall x \forall y \forall z  (R^1(x,y) \wedge T^2(z,y) ~\rightarrow~
\exists w(T^1(x,w) \land S^2(z,w))).
\end{equation}
This DEC, which falls within our general syntactic class of DECs, mixes
tables of the two peers on each side of the implication. This kind of mapping is more general
than those typically considered in virtual data integration \cite{lenzerini02} or data exchange \cite{kolaitis}.\footnote{Cf. \cite{BB04a} for some connections between
PDESs and virtual data integration under the {\em local-as-view} approach. Also \cite{DeGiacomo}, for relationships between PDESs, virtual data integration, and data exchange.}

If (\ref{eq:mixed}) is not satisfied by the data in \peer{P1} and \peer{P2}, \red{which happens when the join in the antecedent is satisfied, but not
the one in the consequent,} then
solutions for \peer{P1} have to be found, keeping \peer{P2}'s data
fixed in the process, due to the trust relationship. Now, in this section we will depart
from the solution semantics introduced in Section \ref{sec:nulls}, by restoring consistency
\wrt (\ref{eq:mixed}) through the introduction of arbitrary elements of the data domain
${\cal U}$.\footnote{This $\mathcal{U}$ could be a finite superset of the union of the active domains involved or infinite. The latter case is
also covered by our semantics. The logic programming semantics is also perfectly defined in this case.}  Those elements become witnesses for the existentially quantified variables in the DECs. That is,
these values come from the data domain, and are not replaced by \nn \ or by labeled nulls as in data exchange
\cite{kolaitis}. Since we have alternative choices for them, we may obtain different solutions for a peer. However,
by definition of solution, they have to stay as close as possible to the original instance. In this case, the general comparison
relation $\preceq_D$ between neighborhood instances of Section \ref{sec:semantics} is given by: ~$D_1 \preceq^\Delta_D D_2$ iff $\Delta(D,D_1) \subseteq \Delta(D,D_2)$.

%

We will specify the solutions for this example directly in (or using) logic programs.
We will also show the most relevant part of the program $\Pi^{-}(\p{P1})$. Since we have to restore
consistency \wrt (\ref{eq:mixed}), the main rules are (\ref{eq:rep1})-(\ref{eq:disj}) below.
   \begin{eqnarray}  R^1\!\_(x,y,\fa) &\leftarrow&
R^1\!\_(x,y,\trs), T^2(z,y), \n \ \aux_1(x,z), \n \
aux_2(z).\label{eq:rep1}\\ \aux_1(x,z) &\lea& T^1\!\_(x,w,\trs), S^2(z,w). \label{eq:rep2}\\
\aux_2(z) &\lea& S^2(z,w). \label{eq:rep3}
\end{eqnarray}
That is, $R^1(x,y)$ is deleted if it participates in a violation
of (\ref{eq:mixed}) (what is captured by the first three literals
in the body of (\ref{eq:rep1}) plus rule (\ref{eq:rep2})), and
there is no way to restore consistency by inserting a tuple into
$T^1$, because there is no possible matching tuple in $S^2$ for
the possibly new tuple in $T^1$ (what is captured by the last
literal in the body of (\ref{eq:rep1}) plus rule (\ref{eq:rep3})).
In case there is such a tuple in $S^2$,  we can either delete a
tuple from $R^1$ or insert a tuple into $T^1$:
\begin{eqnarray} R^1\!\_(x,y,\fa) \vee T^1\!\_(x,w,\ta)
&\lea& R^1\!\_(x,y,\trs), \ T^2(z,y), \ \n \ \aux_1(x,z),\nonumber \\ && S^2(z,w), \
{\it choice}((x,z),w). \label{eq:disj}
\end{eqnarray}
That is, in case of a violation of (\ref{eq:mixed}), when there is
tuple of the form $S^2(a,t)$ in $S^2$ for the combination of values
$\langle d,a\rangle$, then the {\em choice operator} \cite{GPSZ91} non-deterministically
chooses a unique value for $t$, so that the
tuple $T^1(d,t)$ is inserted into $T^1$ as an alternative to deleting
$R^1(d,m)$ from $R^1$. The {\em choice} predicate can be eliminated and replaced
by another predicate that can be specified by means of extra but standard rules \cite{GPSZ91}.

If, instead, we had $\trust = \{(\p{P1},\same,\p{P2})\}$,
\peer{P2}'s relations would also be flexible when searching for solution instances.
In this case, \red{the program becomes more involved in terms of presentation (but not conceptually), in the sense that more relations can be updated, and
corresponding repair rules have to be added.}

Notice that in this example, the values that are chosen as witnesses for the existential quantifier
in the DEC are taken from the active domain of the database, namely from the set of values for
the second attribute of relation $S^2$. In other cases, for example with a DEC of the form
$\forall x \forall y (T^2(x,y) \rightarrow \exists z R^1(x,z))$, we have to consider arbitrary
values from an underlying domain $\nit{dom}$ when inserting tuples into $R^1$. In this case,
\nit{dom} requires a specification as a finite predicate in the program.

\red{Some of the ideas presented above (such as the insertion of elements from the active domain and the use of the choice operator) have been fully developed and applied by the authors
\cite{BeBra04,IJCAI03,JournalVienna04} to inconsistency management in virtual data integration systems under the {\em local-as-view approach}
\cite{lenzerini02}.}

\ignore{
\comlb{Loreto: We should say that the example above can also be formulated
in terms of null values, either \nn, as done before in this paper, or labeled nulls. How? (if yes)}
\comlore{Puede ser formulado en terminos de null ya que este tipo de
ICs esta considerado ahora como una REC}
}

\subsection{Data transport and semantics}\label{sec:trans}

  The data distributed across different peers has to
be appropriately gathered  to build solution instances for a peer,
and different semantics may emerge as candidates, depending on the
granularity of the data sent between peers. In the context of the general
semantic framework introduced in Section \ref{sec:semantics},
we developed a particular semantics in Section \ref{sec:nulls}, according
to which a peer \p{Q} passes back to
a neighbor \p{P}, who is building its solutions, this is (part of) its certain
data. This is the one that holds in all of \p{Q}'s solutions.

In \cite[chapter 7]{Bravo07} also two other alternative semantics
are fully developed and compared, in particular establishing some
conditions under which they coincide or differ. Those other semantics
assume that more detailed information, such as mappings and trust
relationships,  can be sent between peers. We briefly describe them.

\paragraph{1. Send all.} The first one assumes that data, DECs and trust
relations can be sent between peers. So, we can think that we have a
possibly large database instance that has to  be virtually
repaired in order to satisfy all the relevant DECs (obtained from
the accessibility graph) and at the same
time
accommodating the trust relationships. In this case, the DECs are
treated as traditional ICs on the integrated instance. This
semantics is similar to the repair semantics for consistent query
answering. Preferences imposed on repairs can be used
to capture the trust relationships. In this case, it is not necessary to
require the acyclicity of the accessibility graph.

\paragraph{2. Send solutions.}
The second one assumes that only solutions can be send
between neighboring peers. In this case, a peer \p{P} requests the
solutions of the neighbors in order to calculate its own solutions.
Here, the database consists of  the data at \p{P} plus
all the solutions of \p{P}'s neighbors; and the constraints are the DECs
between \p{P} and its neighbors.
As in Section \ref{sec:semantics}, this is a recursive definition, and  assumes
an acyclic accessibility graph.

We think that the semantics we developed in Section \ref{sec:semantics}, which could be called {\em ``send cores''}, is more natural (a peer passes over what it
is certain about), and
also simplifies reasoning at the local level, i.e. at each peer's site, because at most one instance peer neighbor has to be considered.

Now, under our official semantics, if we want to use local solution programs, each neighbor of a peer \p{P} has to run its program, and then send the (relevant part of the)
intersection of the stable models
to \p{P}, who runs its local solution program. This means that different programs have to be fed externally.

At least under the {\em ``send solutions"} and {\em ``send cores"}
semantics (which assumes an acyclic peer graph), we could imagine having a single program that does all this output/input concatenation,  {\em internally}.
Actually, it is possible to build a {\em single program for a peer}, say $\Pi^\nit{man}(\p{P})$, that acts as a combination of
solution programs as given in Section \ref{sec:P2Ptransit}. For each peer \p{Q} that is accessible from \p{P}, the program locally runs a solution program, produces \p{Q}'s solutions (its stable models), or the intersection thereof; and, without
leaving $\Pi^\nit{man}(\p{P})$, passes them to its preceding neighbors \p{Q'}, who uses them to locally compute its own solutions by means of its own, local solution program, etc., until reaching \p{P}. Such a program $\Pi^\nit{man}(\p{P})$ can be a {\em manifold program} \cite{faber}.

Actually, within a manifold program (MP), a program can pass certain or possible query answers as an input to another program. Conceptually, MPs offer  a nice logical
solution, by means of a single program, to this form of program combination that, otherwise, would require external intervention  (the efficient implementation of MPs is still an open problem).

\subsection{On trust} \label{sec:trust}

We introduced trust relationships in the process of peer data exchange already in \cite{BB04a}; and here we have further developed this idea, in
a general semantic framework. However, the notion of trust we have in this work is still rather simple. Actually, it can be represented
as an annotated binary relation between peers. It would be interesting to impose a more sophisticated and rich model of trust
on top of the DECs-bases network of peers. Our concern is not about computing or updating trust in a P2P overlay network \cite{xiong04,marsh06}, but about logical specifications of trust.
We envision a logic-based model of trust that can be integrated with/into the DECs. Such a model could express some higher
level properties, e.g. symmetry or transitivity of trust relationships.
A logic-based representation of trust could allow us to infer non-explicit trust relationships whenever necessary.

Trust modeling is an active and important area of research nowadays, most notably in the context of the semantic web. See  \cite{sabater05,artz07}
for surveys. The integration of trust models, including related notions, like reputation, provenance, etc., into peer data exchange
is still an open area of research. This is specially the case for logic-based models of trust \cite{herzig}. See \cite{rousset08} and references therein for
probabilistic approaches.


In Definition \ref{def:solProgIII}, the trust relationships between
peers were implicitly and rigidly captured in the specifications of
solutions by means of the disjunctive heads of the repair rules.
Although this is a simpler way of presenting things, it has some
drawbacks: (a) The approach is not modular in the sense that trust
is built-in into the rules; (b) Changes in trust relationships
requires changing heads of repair rules; and (c) In case no solution
exists due to the rigid and conflicting trust relationships, no
alternative, but possibly less desirable solution can be obtained as
a stable model of the solution program.

One way of addressing these issues is through the use of {\em
preference programs}, which are answer set programs that express
different forms of  preference, which essentially amounts to
preferring and keeping only certain stable models of the program.
For example, in a disjunctive rule of the form $A \vee B \leftarrow
\nit{Body}$, one could prefer to make $A$ true instead of $B$. If
this is possible, only that stable model would be chosen. However,
if that is not possible (due to the other rules and facts in the
program), making $B$ true is still good enough. More complex
preferences could also be captured. Preferences can be explicitly
and declaratively expressed, and the resulting programs can be
compiled into usual answer set programs with their usual stable
model semantics (cf. \cite{brewka04} and references therein).

Here we briefly outline how {\em weak program constraints}
\cite{buca00,DLV} declaratively capture the kind of preferences that
address our needs. (Cf. \cite{brewka04} for connections between
preferences in logic programs and weak constraints.).

\begin{exampleApp} Consider Example \ref{ex:trans01}, where $(\peer{P1},\less,\peer{P2}) \in \trust$ is captured
by the non-disjunctive repair rule
$$R^1\!\_(x,y,\ta) \leftarrow R^2\!\_(x,y,\trs),R^1\!\_(x,y,\fs), x \not= \nn, y \not =
\nn.$$ The same effect, and more, could be obtained by uniformly
using disjunctive rules followed by appropriate weak constraints. In
this case,
\begin{eqnarray}
R^1\!\_(x,y,\ta) \vee R^2\!\_(x,y,\fa) &\leftarrow&
R^2\!\_(x,y,\trs), R^1\!\_(x,y,\fs), x \not= \nn, y \not = \nn.
\nonumber \\ &\Leftarrow& R^2\!\_(x,y,\fa).
\label{eq:weak}
\end{eqnarray}
Here, the weak constraint (\ref{eq:weak}) expresses a preference for
the stable models of the program that minimize the number of
violations of the condition expressed its body, in this case, that,
when restoring the satisfaction of the DEC $\forall x \forall y(R^2(x,y)
\rightarrow R^1(x,y))$, the tuple  $R^2(x,y)$ is not deleted. These
weak constraints are used by a peer \p{P} to ensure that, if possible, the tuples in
the peers that it trusts more than itself are not modified. \boxtheorem
\end{exampleApp}
If the original solution program has solutions, then the new program
would have the same solutions. However, the latter could have
solutions when the former does not. This would make the semantics of
the system more flexible \wrt unsatisfiable trust requirements. It is
also clear that the weak constraints could be easily derived from
the trust relationships and the DECs. The solution program with weak
constraints can be run in the {\em DLV} system \cite{DLV} to obtain the
solutions and peer consistent answers of a peer.

Notice that the new repair programs, except for the weak program
constraints, are now of the same kind as those for specifying repairs
of single databases \wrt local ICs \cite{iidb06}. Actually, if in the
new program the weak program constraints are replaced by (hard)
program constraints, e.g. (\ref{eq:weak}) by $\leftarrow
R^2\!\_(x,y,\fa)$, the solutions coincide with those of the
programs in Definition \ref{def:solProgIII}.

\red{We should mention that in \cite{arenas03}, weak constraints were used, as a part of
a repair program, to specify the preference for {\em cardinality repairs}, i.e. repairs that
minimize the {\em number} of tuples that are inserted or deleted to restore consistency, as opposed
to minimality (with respect to subset-inclusion) of sets of inserted/deleted tuples. }


\ignore{
\subsection{Nulls}

In this paper we have concentrated on null values that are handled as in SQL relational
databases. Actually, in that case, there is a single null, namely the constant {\tt NULL}.
It can be interpreted in different ways in a database, e.g. as a representative for an existing
but unknown value, as a missing or non-existent value, as a withheld or hidden value, etc.
In our case, we do not make any ontological assumption about this constant. Our purpose
consists in logically capturing the way it is handled in data management operations by an SQL database. More
specifically, \wrt query answering, integrity checking and interaction with other data elements.

Most of the work in database theory bypasses the issue of SQL null, mainly because it is considered to be an
undesirable oddity that could be replaced by a logically cleaner solution. However, SQL {\tt NULL} exists, is used, and
will most like stay with us. It would be better to accept it, but trying to clarify its logical and operational
status. This is our approach in this work. From this point of view, we accept that  SQL {\tt NULL} appears, not only in databases,
but could also appear explicitly in data exchange constraints and integrity constraints, and in their semantics. For example, a referential constraint or data exchange
constraint could
be of the form \ $\forall x y (S(x,y) \rightarrow T(x,{\tt NULL}))$, i.e. requiring that $T$'s second attribute should be filled
with {\tt NULL}.\footnote{In our case, given our repair semantics of Section \ref{sec:nullsRep}, even a constraint of the form $\forall x y (S(x,y) \rightarrow \exists z T(x,z))$ will be enforced
by giving to $z$ the value ${\tt NULL}$ (unless some other constraint prevents this).}

A similar approach is adopted in \cite{franconi12}, where an alternative reconstruction of SQL {\tt NULL} is
attempted. They propose a corresponding relational algebra that takes care of the special status of {\tt NULL},
but in the end, after the reconstruction, this constant disappears from the data domain. We keep it, but rewrite queries
and constraints in such a way that it can be treated as any other constant.

There is a large literature on nulls in relational databases (see, e.g. \cite{LeveneBook}). However, to the best of
our knowledge those nulls are ``ideal" nulls, and not the real SQL {\tt NULL}. There are also some logical approaches,
starting with Reiter's treatment of nulls \cite{reiter}, which can be multiple, existential values not subject to the {\em unique names assumption}.
They are also different from the SQL {\tt NULL}.
More recent work extending Reiter's approach can be found in \cite{gelfond94,lifschitz12}.
}

\section{ \ Proofs of  Results}\label{app:proofs}

\defproof{Proposition \ref{prop:Exists}}{ Let $D_0$ be an empty instance
for the schema ${\cal S}({\cal N}(P))$. By being empty,  $D_0$ satisfies
$\bigcup_{\p{Q} \in {\cal N}(\p{P})} \Sigma(\peer{P},\peer{Q})$ (condition
(i) for a neighborhood solution). Also, since all the trust relationships are
of the
``same" kind, condition (ii) on neighborhood solutions is satisfied by $D_0$.
Then, either $D_0$ is a neighborhood solution, or there exists a neighborhood
solution $D''$ such that $D'' \preceq_{\bar{D}} D_0$.}

\defproof{Corollary \ref{cor:Exists}}{All we need is notice that the possibly inconsistent sink peers in the accessibility graph always have local repairs
under the kind of DECS considered (ICs in that case). A solution for a peer \p{P} can then be obtained by recursively propagating back neighborhood solutions (that always exist by Proposition
\ref{prop:Exists}) for peers along the paths that contain
\p{P}.}

\defproof{Proposition \ref{prop:SolutionGivenCore}}{Membership of $\nit{coN\!P}$  is established by
directly appealing to Definition \ref{def:localsolution} of neighborhood
solution. In fact, given neighborhood instance $J$, after having checked (in polynomial time) if $J \subseteq \rch(\bar{D},\Sigma^{-}(\p{P}))$, a non-deterministic algorithm to test that $J$ is {\em not} a
neighborhood solution for \p{P} and the neighborhood instance $\bar{D}$ \ignore{$\bigcup_{\p{Q} \in \N(\p{P})}J_{\p{Q}}$}
checks if one of the following holds:
\begin{itemize}
    \item [1.] $J  \not \models 
        \Sigma(\peer{P})$.
    \item [2.] $J\rs \{R\} \neq \bar{D} \rs \! \{R\}$, for some $\p{Q} \in
        \mathcal{N}(\p{P})$ and predicate $R \in {\cal S}(\peer{Q})$ with
        $(\p{P}, \nit{less}, \p{Q}) \in \trust$.
    \item [3.]  There is  an instance $J'$ for $\S(\N(\p{P}))$ (the non-deterministic choice) that
        satisfies conditions (i) and (ii) of Definition
        \ref{def:localsolution}, but  $J' <_{\bar{D}}^{\Sigma(\sp{P})} J$.\ignore{, \green{and $J' \neq J$}.} 
\end{itemize}
\vspace{1mm}These conditions are the basis for  a non-deterministic algorithm: First
conditions 1. and 2. can be checked deterministically in polynomial time. If
they are passed by $J$ (i.e. the answer is negative), then an instance $J'$
with size polynomially bounded  by the size of $J$ is guessed. Next,
conditions 1.-2. are checked for $J'$, and 3., for the pair $(J,J')$. The
three tests can be performed in polynomial time in $|J| + |\bar{D}|\ignore{\sum_{\p{Q} \in
   \N(\p{P})}|J_\p{Q}|}$.
\ If the answer to any of the tests is \nit{yes}, $J$ is not a neighborhood
solution.

Hardness can be proved by reduction of satisfiability of propositional formulas in CNF, which is $\nit{coN\!P}$-complete.
The reduction is a modification of that used  in Theorem 4.4 of \cite{CM05} to prove that repairs obtained through deletions are
$\nit{coN\!P}$-complete. In our case we have to deal with trust relationships and the possible insertion of tuples with \nn. Actually, in our proof
the former will
be used to exclude the latter.

We now show that the satisfiability of a propositional formula $\varphi\!: \ \varphi_1 \wedge \varphi_2 \wedge \dots \varphi_m$ in CNF
(i.e. the $\varphi_i$ are clauses)  can be reduced to checking if a particular neighborhood instance is a neighborhood solution for a given peer. 

Consider  the fixed PDES  schema (it does not depend on $\varphi$): \ $\mathfrak{P}= \langle
\mc{P}, \mf{S}, \Sigma,\trust\rangle$, with $\mc{P}=\{\p{P1},\p{P2}\}$, $\mf{S}=\{S(\p{P1}), S(\p{P2})\}$, $S(\p{P1})=\{R^1(\cdot,\cdot,\cdot,\cdot)\}$, $S(\p{P2})=\{R^2(\cdot,\cdot,\cdot,\cdot)\}$,
$\trust=\{(\p{P1},$ $\same,\p{P1}),(\p{P1},\less,\p{P2})\}$, and  $\Sigma=\{\Sigma(\p{P1},\p{P1}), \Sigma(\p{P1},\p{P2})\}$,  with:

\begin{tabbing}
$\Sigma(\p{P1},\p{P1})=\{$\=$\forall x_1 y_1 y_2 z_1 z_2 w_1 w_2 (R^1(x_1,y_1,z_1,w_1) \wedge R^1(x_1,y_2,z_2,w_2) \rightarrow y_1=y_2),$\\
\>$\forall x_1 y_1 z_1 w_1 (R^1(x_1,y_1,z_1,w_1) \rightarrow  \exists x_2 y_2 z_2 R^1(x_2,y_2,z_2,z_1) )  \}$.\\
$\Sigma(\p{P1},\p{P2})=\{\forall x y z w (R^1(x,y,z,w) \rightarrow  R^2(x,y,z,w) )\}$.
\end{tabbing}

Now, consider a propositional formula $\varphi$ as above, on which the instances for the peer system will depend.
Let $\mathfrak{D}_\varphi=\{D(\p{P1}), D(\p{P2})\}$ be the instance  for $\mathfrak{P}$, with:
\begin{eqnarray*}
D(\p{P1}) &=& \{R^1(p_j, 0, \varphi_i, \varphi_{i+1}) \ |\  p_j \mbox{ occurs negatively in }
\varphi_i\} \cup\\ && \{R^1(p_j, 1, \varphi_i, \varphi_{i+1}) \ | \ p_j \mbox{ occurs positively in } \varphi_i\}.\\
D(\p{P2}) &=& \{R^2(a, b, c, d) \ |\ R^1(a,b,c,d) \in D(\p{P1})\}.
\end{eqnarray*}
(The addition $i + 1$ is meant to be modulo the number $m$ of clauses in $\phi$.) Notice that tables $R^1$ and $R^2$ for peers \p{P1} and \p{P2}, respectively,
have the same rows. Intuitively, the UDEC  in $\Sigma(\p{P1},\p{P1})$ ensures
that, for every proposition in the first attribute of $R^1$, the truth assignment, if any,
is unique; whereas the RDEC in it, ensures that there are
assignments that make all clauses in the formula true.

We now show
that the neighborhood instance $\bar{D}$, with the empty relation for $R^1$ plus the original contents of $D(\p{P2})$, is the neighborhood solution for \p{P1} with initial neighborhood
instance $D(\p{P1}) \cup D(\p{P2})$  if and only if $\varphi$ is {\em not}
satisfiable. In this case, $\bar{D}$ would be obtained through the deletion of all tuples from $R^1$ in $D(\p{P1})$. Notice that due to  the
trust relations and the DEC in $\Sigma(\p{P1},\p{P2})$, only tuple deletions from peer $\p{P1}$'s instance are admissible updates.

To prove that $\bar{D}$ being a neighborhood solution for \p{P1} implies that $\varphi$ is not
satisfiable, assume by contradiction that $\varphi$ is satisfiable. Then, there is an assignment $\sigma$ that makes $\varphi$ true.

The instance $\bar{D}' := \{R^1(p, 0, \varphi_i, \varphi_{i+1}) \in D(\p{P1}) \ | \ \sigma(p)=0\}$ $\cup$ $\{R^1(p, 1, \varphi_i, \varphi_{i+1}) \in D(\p{P1}) \ | \ \sigma(p)=1\}$   $ \cup$ $D(\p{P2})$
\ is a subinstance of $D(\p{P1}) \cup D(\p{P2})$, $\bar{D}'\rs \mc{S}(\p{P1}) \neq \emptyset$, satisfies the DECs, and does not modify the more trusted relations, i.e. $\bar{D}'\rs \mc{S}(\p{P2}) = D(\p{P2})$.
 Thus,  $\bar{D}$ cannot be a neighborhood solution since  $\bar{D}' <_{D(\p{P1}) \cup D(\p{P2})}^{\Sigma(\p{P1})} \bar{D}$.

Now we show that if $\varphi$ is not satisfiable, then $\bar{D}$ is a neighborhood solution for \p{P1} when starting with neighborhood instance $D(\p{P1}) \cup D(\p{P2})$.
Assume by contradiction that $\bar{D}$ is not a neighborhood solution. Since $\bar{D}$ satisfies all the DECs and respects the trust relationships, $\bar{D}$ cannot be a neighborhood solution only if there is  a
 neighborhood instance $\bar{D}'$, such that:  $\bar{D}' \models \Sigma(\peer{P1})$;  $\bar{D}'\rs S(\p{P2}) = D(\p{P2})$,  and  $\bar{D}' <_{D(\p{P1}) \cup D(\p{P2})}^{\Sigma(\p{P1})} \bar{D}$. Since $\bar{D}'$ can be obtained only through tuple deletions, it holds:  $\bar{D} \subsetneqq D'\subseteq D(\p{P1}) \cup D(\p{P2})$. Thus, there is at least one tuple $R^1(\bar{t}) \in (\bar{D}' \cap D(\p{P1}))$. Due to the UDEC in $\Sigma(\p{P1},\p{P2})$, we conclude that, for every $i \in [1,m]$, there exists a $p$ and $v$ with $R^1(p,v,\varphi_i,\varphi_{i+1}) \in (D'\cap D(\p{P1}))$. Using these tuples we can define the following assignment $\sigma'$:

\vspace{1mm}
\hspace*{3.5cm}$\sigma'(p)= \left\{ \begin{minipage}{7cm}
1 {\hfill if $R^1(p,1,\varphi_i,\varphi_{i+1})\in D'$ with $i \in [1,m]$}\\
0 {\hfill if $R^1(p,0,\varphi_i,\varphi_{i+1})\in D'$ with $i \in [1,m]$}
\end{minipage}\right.$

\vspace{1mm}
The assignment is well defined, because the functional dependency in $\Sigma(\peer{P1},\peer{P1})$ ensures that only one value exists for each proposition. By construction,
$\sigma'$ is an assignment that satisfies $\varphi$. We have reached a contradiction, which completes the proof.}

\defproof{Proposition \ref{lemma:PCAGivenCore}}{First we prove membership of $\Pi^P_2$. An
atom $R(\bar{t})\in\mathit{localCore}(\p{P},\bar{D})$ if for every $D' \in
\NS(\p{P},\bar{D})$, $D' \models R(\bar{t})$. Thus, a non-deterministic algorithm
that
checks if $R(\bar{t})\not\in\mathit{localCore}(\p{P},\bar{D})$ guesses
an instance $J$ of $\S(\N(\p{P}))$, next checks if it is a neighborhood
solution for \p{P} and $\bar{D}$, and finally, if $R(\bar{t}) \not \in J$. By
Proposition
\ref{prop:SolutionGivenCore}, the first
of these two tests is in $\nit{coN\!P}$;  and the second one is in
polynomial time. Thus,
the problem is in $\Pi^P_2$.

Hardness holds by a reduction  from
satisfiability
of a quantified propositional formulas (QBF) $\beta$ of the form \ $\forall
p_1\cdots
\forall p_k \exists q_1\cdots \exists q_l \psi$, where $\psi$ is a
quantifier-free propositional formula in CNF, i.e. of the form \
$\psi_1 \wedge \dots
\wedge \psi_m$, where the $\psi_i$ are clauses.  This problem is
$\Pi^P_2$-complete \cite{comp,papa}. (The reduction is adapted from that for Theorem 4.7 in \cite{CM05}.)

We
 construct a PDES  schema (independent from $\beta$) $\mathfrak{P}_0 = \langle
\mc{P}_0, \mf{S}_0, \Sigma_0,\trust_0\rangle$, with $\mc{P}_0 = \{\p{P1}, \p{P2}\}$, $\mf{S}_0=\{ \S(\p{P1}), \S(\p{P1})\}$, $\S(\p{P1})=\{R(\cdot,\cdot,\cdot),T(\cdot)\}$,
$\S(\p{P2})=\{\nit{Clause}(\cdot),\nit{Var}(\cdot)\}$,
 $\trust_0$ $=\{(\p{P1},\nit{less},\p{P2})\}$, and the DECs
$\Sigma_0=\{\Sigma(\p{P1},\p{P2}),$ $\Sigma(\p{P1},\p{P1})\}$ with:

\vspace{-2mm}\hspace*{-1.2cm} \begin{minipage}{7cm}
\begin{eqnarray*}
\Sigma(\p{P1},\p{P2})&=&\{\forall x(\nit{Clause}(x) \rightarrow \exists y z
R(y,z,x)), \\&&~~~\forall x(\nit{Var}(x) \rightarrow R(x,1,a) \vee
R(x,0,a))\},\\
\Sigma(\p{P1},\p{P1})&=&\{\forall x y_1 y_2 z_1 z_2(R(x,y_1,z_1) \wedge
R(x,y_2,z_2)) \rightarrow y_1=y_2,\\&&\hspace*{-1.1cm} \forall xyzw (R(x,y,z) \wedge
T(w) \rightarrow w \neq \mathbf{sat} \vee \mathit{IsNotNull}(x) \vee
\mathit{IsNotNull}(y))\}.\\
\end{eqnarray*}\end{minipage}

Now, given a QBF $\beta$, we construct
an instance
$\bar{D}_\beta$ for the neighborhood schema $\S(\N(\p{P1}))$ around \p{P1}, such that:  $T(\mathbf{sat}) \in
\mathit{localCore}(\p{P1},\bar{D}_\beta)$ \ iff \
 $\beta$ is true.

Now, for $\beta = \forall
p_1\cdots
\forall p_k \exists q_1\cdots \exists q_l(\psi_1 \wedge \dots
\wedge \psi_m)$,
$\bar{D}_\beta := D_{\!\beta}(\p{P1}) \cup D_{\!\beta}(\p{P2})$, with:
\begin{eqnarray*}
D_{\!\beta}(\p{P1})&:=&\{R(\nit{var}_{\!j},1,\psi_i) \ | \
 \nit{var}_{\!j} \mbox{ occurs positively in } \psi_i \}  \cup\\
&&\{R(\nit{var}_{\!j},0,\psi_i) \ | \ \nit{var}_{\!j} \mbox{ occurs negatively in }
\psi_i \}  \cup\\ && \{\nit{Var}(p_i) \ | \ p_i \mbox{ is
universally
quantified in } \psi\} \cup \{T(\mathbf{sat})\}.\\
 D_{\!\beta}(\p{P2})&:=&\{\nit{Clause}(\psi_1),\ldots,
\nit{Clause}(\psi_m)\}.
\end{eqnarray*}

Intuitively, relation $R(x,y,z)$ is used to provide a truth value $y$ to
variable $x$ in conjunct $z$. This truth value will be unique across $\psi$
due to the integrity constraints on $R$ contained in
$\Sigma(\p{P1},\p{P1})$.
 The first DEC in $\Sigma(\p{P1},\p{P2})$
ensures that, for every clause $\psi_i$, there is, if possible, a literal
which is true in it. If it is not possible (the formula is not true), it
inserts
a tuple of the form $R(\nn,\nn,\psi_i)$. The second DEC ensures that
all possible assignments for the universally quantifies variables are tested
in different solutions. It uses a constant, $a$, which is different from all
$\psi_i$. The first IC in $\Sigma(\p{P1},\p{P1})$  enforces that, in each solution, each propositional
variable takes a unique value. Finally, the second IC in $\Sigma(\p{P1},\p{P1})$ ensures that if
$R(\nn,\nn,\psi_i)$ is true, then predicate $T(\mathbf{sat})$ should not be
part of the neighborhood solution.
In this way, formula $\beta$ is true if and only of $T(\mathbf{sat})
\in \mathit{localCore}(\p{P1},\bar{D}_\beta)$. To conclude the proof, we illustrate the reduction
with the following example.

\begin{figure}
\hspace*{0.1cm}\includegraphics[width=8cm]{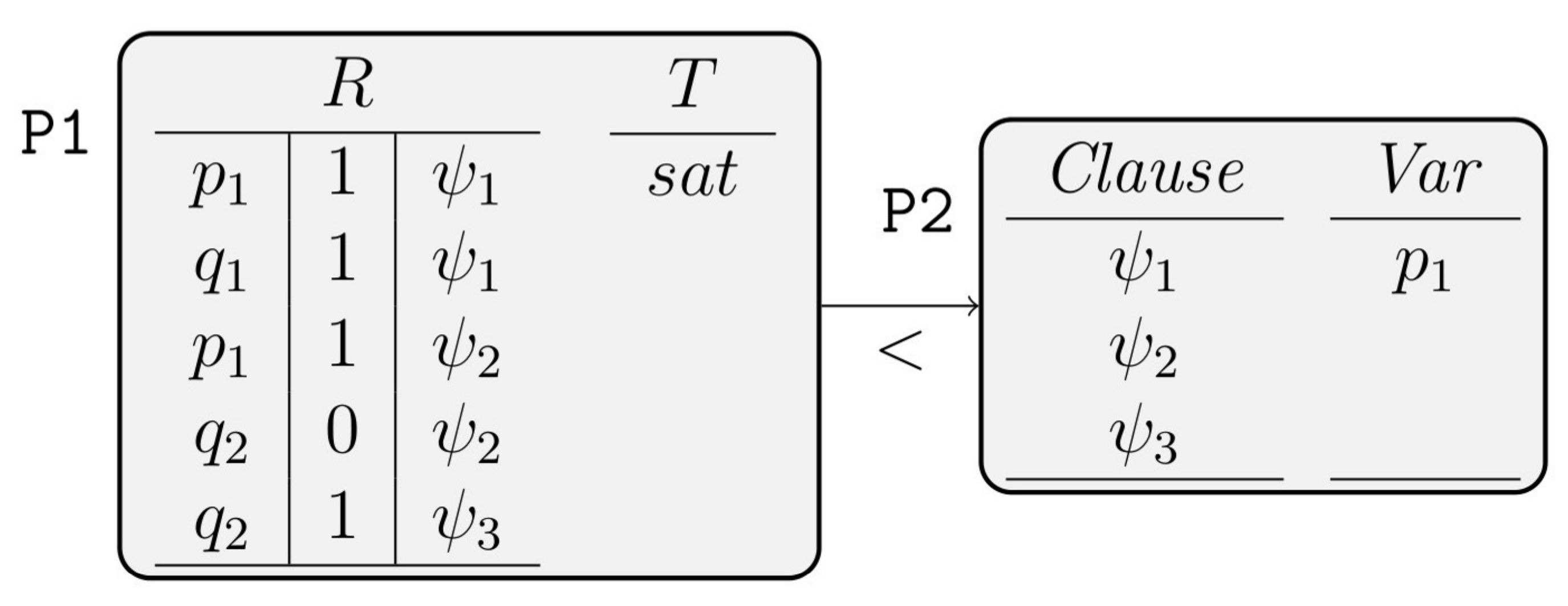}\vspace{-3mm}
  \caption{Instances for a peer system}\label{fig:instances}
\end{figure}

\begin{exampleApp} \label{ex:reduc}
Consider the QBF $\forall p_1 \exists q_1 \exists
q_2(\psi_1 \wedge \psi_2 \wedge \psi_3)$, with $\psi_1\!: \ (p_1 \vee q_1)$,
$\psi_2\!: \ (p_1 \vee \neg q_2)$, and $\psi_3\!: \ q_2$.
\ Instance
 $\bar{D}_{\!\beta}$ is the union of the instances in  Figure
\ref{fig:instances}.

On this basis,
the neighborhood solutions for $\p{P1}$ and $\bar{D}_{\!\beta}$ are:

\ni $\ignore{\nit{N\!S}_1\!\!} D_1 =\!
\{R(p_1,1,a),R(p_1,1,\psi_1),R(q_1,1,\psi_1),R(p_1,1,\psi_2),R(q_2,1,\psi_3),T(\mathbf{sat})
\} \cup D(\p{P2})$,

\ni $\ignore{\mathit{N\!S}_2\!\!} D_2 =\!
\{R(p_1,0,a),R(q_1,1,\psi_1),R(q_2,1,\psi_3),R(\nn,\nn,\psi_2)\}\cup
D(\p{P2}),$ and

\ni $\ignore{\mathit{N\!S}_3\!\!} D_3 =\!
\{R(p_1,0,a),R(q_1,1,\psi_1),R(q_2,0,\psi_2),R(\nn,\nn,\psi_3)\}\cup
D(\p{P2})$.

Since $T(\mathbf{sat})  \not \in \mathit{localCore}(\p{P1},\bar{D}_\beta)
:= (D_1 \cap D_2 \cap D_3)\rs \mc{S}(\p{P1})$, the QBF formula is
false. 
\end{exampleApp}}

\defproof{Corollary \ref{cor:PCA}}{Membership is established with a test
similar to that in the proof
of Proposition \ref{lemma:PCAGivenCore}. Hardness follows from  Proposition \ref{lemma:PCAGivenCore}, because it is about a
 particular kind of conjunctive queries, namely atomic of the form $\mc{Q}(\bar{x})\!: R(\bar{x})$, where $R$ is a predicate for a peer \p{P}. The peer consistent
answers to this query are exactly the $\bar{c}$s, such that $R(\bar{c})$ is in the local core of \p{P}.}

\begin{algorithm}[ht!]
\TitleOfAlgo{{\ImportSolution}}
     \KwIn{An instance $\mathfrak{D}$ for a PDES  schema $\mathfrak{P} = \langle
\mc{P}, \mf{S}, \Sigma,\trust\rangle$ and a peer $\p{P}\in \mc{P}$}
    \KwOut{The unique solution of \p{P}}
     \lIf{\p{P} has no outgoing edges}{\Return $D(\p{P})$}
     \Else{
        \ForEach{$\p{Q} \in \N^\circ(\p{P})$}
            {
            $\nit{Sol}(\p{Q},\mathfrak{D}) \leftarrow \ImportSolution(\P,\p{Q})$\;
            }
            $D' \leftarrow D(\p{P}) \cup \bigcup_{\p{Q} \in \N^\circ(\p{P})}\nit{Sol}(\p{Q},\mathfrak{D})$\;
            $\nit{N\!S} \leftarrow \red{\mbox{\nit{minimal model of Datalog import program} }\mathfrak{I}(\p{P},D')}$ \;
            \Return $\nit{N\!S}\rs S(\p{P})$\;
        }
    \caption{Computing the solution for a peer in the import case} \label{alg:SolImport}
\end{algorithm}

\defproof{Proposition \ref{prop:Import}}{The existence and uniqueness is straightforward
since there are no local restrictions (existence), and there is no
non-determinism involved (uniqueness). The unique solution for a peer \p{P}
can be computed by means
 of Algorithm \ImportSolution shown in Figure \ref{alg:SolImport}. It recursively computes the solutions for
all the peers that are accessible from \p{P}. The base case occurs  when a peer
\p{Q} has no DECs (line 4). In that case, its unique solution is its own
database $D(\p{Q})$. \red{ Otherwise (lines 5-10), the algorithm  recursively requests the solutions of its neighbors (lines 6-7) and uses them to
construct instance $D'$ (line 8).} Then, the unique neighborhood solution for the peer consists
of the minimal model of $\mathfrak{I}(\p{P},D')$  (line 9).
\ignore{
whose extensional database is $D'$ and whose
rules are as follows:
\begin{itemize}
  \item For every IUDEC in $\Sigma(\p{P})$ of the form
      (\ref{eq:importUDEC}), the rule: $$Q(\bar{y}) \leftarrow
      R_1(\bar{x}_1), \dots R_n(\bar{x}_n), \neg \varphi.$$
  \item For every import RDEC in $\Sigma(\p{P})$ of the form
      (\ref{eq:importRDEC}), the rule: $$Q(\bar{y},\overline{null})
      \leftarrow R_1(\bar{x}_1), \dots R_n(\bar{x}_n).$$
\end{itemize}
We obtain a Datalog program whose only minimal model corresponds to \p{P}'s unique neighborhood solution.}
By restricting \p{P}'s neighborhood
solution to \p{P}'s schema we get \p{P}'s solution (line 10).}

\defproof{Proposition \ref{theo:corresp}}{
We will prove this result for the case where the central  peer trusts its neighbors as much as itself, which is more general in some sense than
that where it trusts neighbors' data more, because more alternatives for repairs come up, using the full power of disjunctive
programs. Below $D$ is \p{P}'s neighborhood instance for which neighborhood solutions are specified by means of the  program in Definition \ref{def:solProgIII}.
$\mc{C}$ is the set of constraints, i.e. UDECs and RDECs, for the neighborhood.

Given the trust assumptions, the program takes a special form, as follows. For
exchange constraints in $\mc{C}$ of the forms:
\begin{itemize}
\item[(a)] \emph{Universal constraint} \ (UDEC):
\begin{equation}\label{eq:format}
 {\forall}\bar{x}(\bigwedge_{i = 1}^{m} P_i(\bar{x}_i) ~\longrightarrow~
\bigvee_{j=1}^n Q_{j}(\bar{y}_j) \vee
    \varphi).
\end{equation}
\item[(b)]
\emph{Referential
constraint}: \ (RDEC) \ 
\begin{equation}\label{eq:formatRIC}
\forall \bar{x}~(P(\bar{x}) \longrightarrow \exists
\bar{y}~Q(\bar{x}',\bar{y})).\footnote{To simplify  the presentation, we are assuming that
the existentially quantified variables appear in the last
attributes of $Q$.}
  \end{equation}
  \end{itemize}
the neighborhood solution program $\Pi(\p{P},D)$ becomes:
\begin{enumerate}[leftmargin=2em] \item
\label{it:dom3}$\nit{dom}(c)$ for every $c \in  \nit{Adom}(D) \smallsetminus \{\nn\}$.

\item \label{it:fact3} The fact $P(\bar{a})$ for every atom $P(\bar{a}) \in \DB$.

  \item \label{it:uic3}For every UDEC $\psi$ of the form
(\ref{eq:format}), the rule:

$\bigvee_{i=1}^{n} P_{i}\_(\bar{x}_{i},\fa) \vee
\bigvee_{j=1}^{m} Q_{j}\_(\bar{y}_{j}, \ta)
~\longleftarrow~ \bigwedge_{i=1}^{n}
P_{i}\_(\bar{x}_{i},\mathbf{t}^{\star}), \bigwedge_{j=1}^{m}
Q_{j}\_(\bar{y}_{j},\mathbf{f}^{\star}),$\\ \hspace*{7.2cm}$
\bigwedge_{x_l \in \nit{RelV}(\psi)} \nit{dom}(x_l), \bar{\varphi}.$

\noindent  where $\nit{RelV}(\psi)$ is the set of relevant attributes for $\psi$,
$\bar{x}=\bigcup_{i=1}^n x_i$, and $\bar{\varphi}$ is a
conjunction of built-ins that is equivalent to the negation of
$\varphi$.

\item \label{it:ric3} For every RDEC $\psi$ of the form
(\ref{eq:formatRIC}), the rules:\footnote{Literal
$\nit{dom}(\bar{x})$ denotes the
conjunction of the atoms $\nit{dom}(x_j)$ for $x_j \in \bar{x}$.}\\
 $P\!\_(\bar{x},\fa) \vee
Q\!\_(\bar{x}',\overline{\nn},\ta) \leftarrow
P\!\_(\bar{x},\mathbf{t^{\star}}),  \n \
\nit{aux}_\psi(\bar{x}'),$ $
 \nit{dom}(\bar{x}').$\\ And for every $y_i \in \bar{y}$:\\
 $\nit{aux}_\psi(\bar{x}') \leftarrow Q\!\_(\bar{x}',\bar{y},\trs),$ $\n \
Q\!\_(\bar{x}',\bar{y},\fa),$ $\nit{dom}(\bar{x}'),$ $\nit{dom}(y_i)$.\\
 $\nit{aux}_\psi(\bar{x}') \leftarrow Q(\bar{x}',\overline{\nn}),$ $\n \
Q\!\_(\bar{x}',\overline{\nn},\fa),$ $\nit{dom}(\bar{x}')$.

\item \label{it:ts3} For every predicate $P \in {\cal S}(\mc{N}(\p{P}))$, the rules:

$P\!\_(\bar{x},\mathbf{t^{\star}}) \leftarrow
P(\bar{x}).
~~~~~~~~P\!\_(\bar{x},\mathbf{t^{\star}}) \leftarrow
P\!\_(\bar{x},\ta).$

$P\!\_(\bar{x},\mathbf{f^{\star}})
 \leftarrow P(\bar{x},\fa).~~~~~\,~~~P\!\_(\bar{x},\mathbf{f^{\star}})
 \leftarrow  \nit{dom}(\bar{x}), \n \ P(\bar{x}).$

\item \label{it:tss3} For every predicate $P \in {\cal S}(\mc{N}(\p{P}))$, the {\em interpretation rules}:


$P\!\_(\bar{x},\mathbf{t^{\star\star}})~\leftarrow~P\!\_(\bar{x},\ta). \ \ \
~~~~~P\!\_(\bar{x},\mathbf{t^{\star\star}})~\leftarrow~P(\bar{x}),\n \
P\!\_(\bar{x},\fa).$

\item \label{it:denial3} For every predicate $P \in {\cal S}(\mc{N}(\p{P}))$, the {\em coherence constraints}:

$\leftarrow \ \ P\!\_(\bar{x},\ta), \
P\!\_(\bar{x},\fa).$
\end{enumerate}
The claim is: \  If $\mm$ is a stable model
of $\Pi(\p{P},D)$, then $D_M$ is a neighborhood solution repair of $\DB$. Furthermore,  the neighborhood
solutions obtained in this
way are all the neighborhood
solutions  of $\DB$. \ We recall that for a stable model of $\Pi(\p{P},D)$,
\begin{equation}
D_M =\{ P(\bar{a}) \ \ | \ \ P \in \mc{S}(\mc{N}(\p{P})) \mbox{ and } P\_(\bar{a},\tss) \in M \}. \label{eq:ins}
\end{equation}
The proof follows directly
from Propositions \ref{theo:rep-to-mod-st} and
\ref{theo:mod-to-rep-st} below, which require in their turn  some
lemmas and intermediate definitions.}

In the following, for a disjunctive program $\Pi$ and a set of ground atoms $M$, $\Pi^M$ is the positive ground
program obtained by the Gelfond-Lifschitz transformation \cite{GL91}:
$$\Pi^M = \{H \leftarrow B  \ \ | \ \  H \leftarrow B, \n A_1, \ldots, \n A_m \in \nit{ground}(\Pi), \mbox{ and } A_1, \ldots A_m \notin M\}.$$

\begin{lemma}  \label{lemma:casesst} Given an instance $D$ and a
RDEC-acyclic set of UDECs and RDECs, if $M$ is a stable model of
$\Pi(\dbic)$,
then exactly one of the following cases holds:
\begin{enumerate}[leftmargin=2em]

\item $P(\bar{a})$, $P\_\oatstar$ and $P\_\oatstarr$ belong to $M$, and no other
$P\!\!\_(\bar{a},v)$, for $v$ an annotation, belongs to
$M$. \item $P(\bar{a})$, $P\_\oatstar$ and $P\_(\bar{a},\fa)$ belong to
$M$, and no other $P\!\!\_(\bar{a},v)$, for $v$ an annotation, belongs to $M$.
\item $P(\bar{a})\not \in M $, and
$P\_(\bar{a},\ta)$,  $P\_\oatstar$, $P\_\oatstarr$ belong to $M$, and no
other $P\!\!\_(\bar{a},v)$, for $v$ an annotation,
belongs to $M$. \item $P(\bar{a})\not \in M $, and no
$P\!\!\_(\bar{a},v)$, for $v$ an annotation, belongs to
$M$.
\end{enumerate}
\end{lemma}

\dproof{For an atom $P\oa$,  there are two
possibilities:
\begin{itemize}
\item [(a)]$P(\bar{a}) \in M$. Then, from rule \ref{it:ts3}., $P\_\oatstar \in M$. Two
cases are possible now: $P\_(\bar{a},\fa) \not \in M$ or $P\_(\bar{a},\fa) \in
M$.  In the first case, since $M$ is minimal, $P\_(\bar{a},\ta) \not
\in M$ and $P\_\oatstarr \in M$. In the second case,
due to rule \ref{it:denial3}., $P\_(\bar{a},\ta) \not \in M$. These
cases cover the first two  in the statement of the lemma.

\item[(b)] $P(\bar{a}) \not \in M$. Two cases are possible now: $P\_(\bar{a},\ta) \in M$ or
$P\_(\bar{a},\ta) \not \in M$. In the first one, it also holds
$P\_\oatstarr$, $P\_\oatstar \in M$, by  rules
\ref{it:ts3}. and \ref{it:tss3}.; and $P\_(\bar{a},\fa) \not \in M$
by rule \ref{it:denial3}. In the second case,
$P\_\oatstar \not \in M$ (because $M$ is minimal), $P\_(\bar{a},\fa) \not
\in M$ (because $P\_\oatstar \not \in M$, and $M$ is
minimal). These cases cover the last two in the statement of the lemma.
\end{itemize}
}

\noindent From two database instances we can define a
structure.

\begin{definitionApp}  \label{def:dbtomm} For two database instances $D_{1}$ and $D_{2}$
over the same schema and domain and a set of constraints $\ICc$,
$M^{\star}_{\ICc}(\D_{1},\D_{2})$ is the Herbrand
structure $\langle \mc{U}, I_{\cal P},$ $I_{\cal B}\rangle$, where
$\cal U$ is the underlying domain,\footnote{In this case it can be restricted to the active domain of the neighborhood instance $D$ (or the union of the active domains of $D_1, D_2$) plus the constant \nn.} and $I_{P}$, $I_{B}$ are the interpretations for the
database predicates (extended with annotation arguments), and
the built-ins, respectively. $I_{P}$ is inductively defined as
follows:

\begin{enumerate}[leftmargin=2em]
    \item If $P\oa \in D_{1}$ and $P\oa \in D_{2}$, then $P(\bar{a})$,
    $P\_\oatstar$ and $P\_\oatstarr \in I_{P}$.
    \item If $P\oa \in D_{1}$ and $P\oa \not \in D_{2}$, then $P(\bar{a})$,
    $P\_\oatstar$ and $P\_(\bar{a},\fa)  \in I_{P}$.
    \item If $P\oa \not \in D_{1}$ and $P\oa \not \in D_{2}$, then $P\_(\bar{a},v) \not \in
    I_{\cal P}$ for every annotation $v$.
    \item If $P\oa \not \in D_{1}$ and $P\oa \in D_{2}$,
    then $P\_(\bar{a},\ta)$, $P\_\oatstar$ and $P\_\oatstarr \in I_{\cal P}$.
    \item For every RDEC $\psi \in \ICc$ of the form  $\forall \bar{x} (P(\bar{x}) \rightarrow
    \exists \bar{y} Q(\bar{x}',\bar{y}))$: If $P\_\oatstarr \in I_{\cal P}$ and $Q\oaobtstarr
        \in I_{\cal P}$, with $\bar{a} \neq
        \nn$ and at least one $b
        \in \bar{b}, b \not = \nn$, then $\nit{aux}_\psi(\bar{a}') \in I_{\cal
        P}$.

\end{enumerate}
The interpretation $I_{\cal B}$ is defined as expected: if $Q$
is a built-in, then $Q\oa \in I_{\cal B}$ iff $Q(\bar{a})$ is
true in classical logic, and $Q\oa \not \in I_{\cal B}$ iff
$Q(\bar{a})$ is false. \boxtheorem
\end{definitionApp}

Notice that the database instance associated to
$M^{\star}_\ICc(D_{1},D_{2})$ through (\ref{eq:ins}) corresponds exactly to
$D_2$, i.e. ~$D_{M^{\star}_\ICc(D_{1},D_{2})}=D_2$.

\begin{lemma} \label{lemma:rep-to-mod-st} Given an instance $D$ and a
set $\mc{C}$ of UDECs and RDECs, if $D' \schn \ICc$, then there is a model
$M$ of the program $\Pi(\dbic)^{M}$ with
$D_{M}=D'$. Actually,
$M^{\star}_\ICc(D,D')$ is such a model.
\end{lemma}

\dproof{Since $D_{\Mdd}=D'$, we only need to show that
$\Mdd$ satisfies all the rules of $\Pi(\dbic)^{\Mdd}$.
First, by construction, it is
clear that rules \ref{it:fact3}., \ref{it:ts3}.
and \ref{it:tss3}. are satisfied by $\Mdd$.

For every UDEC in
$\ICc$, the program has the rule in \ref{it:uic3}. If its
body is satisfied, then the atoms
$P_i\!\_(\bar{a}_i,\trs) \in \Mdd$ and $Q_i\!\_(\bar{b}_i,\fa)
\in \Mdd$ or $Q_i(\bar{b}_i) \not \in \Mdd$. Also, since the
constraint is satisfied, at least one of the $P_i(\bar{a}_i)$
is not in $D'$ or
 one of the $Q_i(\bar{b}_i)$ is in $D'$. By construction of $\Mdd$,
  at least one of $P_i\!\_(\bar{a}_i,\fa)$ or
 $Q_i\!\_(\bar{b}_i,\ta)$ is in $\Mdd$. Therefore, the head of
the rule is also satisfied.

For
every RDEC in $\ICc$, there are the rules
\ref{it:ric3}. By construction  of $\Mdd$, for every  $\psi \in \ICc$, those that
define $\nit{aux}_\psi(\bar{x})$  are
satisfied.

If the body of the first rule in \ref{it:ric3} is
true in $\Mdd$, it means that the constraint is not
satisfied in the initial instance or at some point along the repair
process. Since the constraint is satisfied by $D'$, the
satisfaction is restored by adding $Q\_(\bar{x},
\overline{\nn})$ or by deleting $P(\bar{x})$. This implies that
$Q\!\_(\bar{x}, \overline{\nn},\ta) \in \Mdd$ or
$P\!\!\_(\bar{x},\fa) \in \Mdd$. As a consequence, the first
(or second) rule is satisfied. \ignore{OJO:NNC For every NNC in
$\ICc$ there is a rule of the form \ref{it:nnc}. If the body of
the rule is true, e.g $P\!\!\_(\bar{a},\nn,\trs) \in \Mdd$,
the constraint is not satisfied at some point in the repair
process. Since it is satisfied in $D'$,  $P\_(\bar{a},\nn) \not
\in \D'$.} Then, by construction of $\Mdd$,
$P\!\!\_(\bar{a},\fa) \in \Mdd$, and the head of the rule
is  satisfied.}

\vspace{3mm}\noindent The next lemma shows that if $M$ is a
minimal model of the program $\Pi(\dbic)^{M}$, then $D_M$
satisfies the constraints.

\begin{lemma} \label{lemma:mod-to-rep-st}  Given a database $D$ and a
set of constraints $\mc{C}$, if $M$ is a stable model of  the
program $\Pi(\dbic)$, then $D_{M} \schn \ICc$.
\end{lemma}

\dproof{We want to show that $D_{M} \schn \psi$, for
every constraint  $\psi \in \ICc$. There are two cases to
consider:
\begin{itemize}
\item[A.]IC $\psi$ is a UDEC. Since $M$ is a model of $\Pi(\dbic)^{M}$,  $M$ satisfies rules \ref{it:uic3}. of $\Pi(\dbic)$.
Then, at least one of the following cases holds:

\begin{itemize}
\item[(a)] $P_i\_(\bar{a},\fa) \in M$. Then, $P_{i}\oatstarr \notin M$ and
$P\oa \not \in D_{M}$ (by Lemma~\ref{lemma:casesst}). Hence,
$P_{i}\oa \notin D_{M}$. Since the analysis was done for
an arbitrary value $\bar{a}$, $D_{M} \schn
\bigwedge_{i=1}^{n} P_{i}\oxi \rightarrow \bigvee_{j=1}^{m}
Q_{j}\oyj \vee \varphi$ holds.

\item[(b)] $Q_j\_(\bar{a},\ta) \in M$. It is symmetric to the previous one.

\item[(c)] It is not true that $M \schn \bar{\varphi}$. Then $M \schn
\varphi$. Hence, $\varphi$ is true, and $D_{M} \schn
\bigwedge_{i=1}^{n} P_{i}\oxi \rightarrow \bigvee_{j=1}^{m}
Q_{j}\oyj \vee \varphi$ holds.

\item[(d)] $P_{i}\_\oatstar \notin M$. Given that $M$ is minimal, just the
last item in Lemma \ref{lemma:casesst} holds. This means $P_{i}\oatstarr \notin M$, $P_{i}\oa \not \in D_{M}$ and
$P_{i}\oa \notin D_{M}$. Since the analysis was done for
an arbitrary value $\bar{a}$, $D_{M} \schn
\bigwedge_{i=1}^{n} P_{i}\oxi$ $ \rightarrow$ $
\bigvee_{j=1}^{m} Q_{j}\oyj \vee \varphi$ holds.

\item[(e)] $Q_j\_(\bar{a},\fa) \notin M$ or $Q_{j}(\bar{a}) \in M$. Given that $M$ is minimal,
just the first item in Lemma \ref{lemma:casesst} holds. Then,
$Q_{j}\_\oatstarr \in M$, $Q_{j}\oa \in D_{M}$ and $D_{M}
\schn Q_{j}\oa$. Since the analysis was done for an arbitrary
value $\bar{a}$, $D_{M} \schn \bigwedge_{i=1}^{n} $ $
P_{i}\oxi \rightarrow \bigvee_{j=1}^{m} Q_{j}\oyj \vee \varphi$
holds.
\end{itemize}

\item[B.] Formula $\psi$ is a RDEC. Since $M$ is a model of $\Pi(\dbic)^{M}$,
$M$ satisfies rules \ref{it:ric3}. of $\Pi(\dbic)$. Then, at
least one of the following cases holds:
\begin{itemize}
\item[(a)] $P\_(\bar{a},\fa) \in M$. Then, $P_{i}\_\oatstarr \notin M$ and $P\oa
\not \in D_{M}$ (by Lemma~\ref{lemma:casesst}). Hence,
$P_{i}\_\oa \notin D_{M}$. Since the analysis was done for
an arbitrary value $\bar{a}$, $D_{M} \schn (P\ox \rightarrow
Q\oxpy)$ holds.

\item[(b)] $Q\_(\bar{a}',\overline{\nn},\ta) \in M$. It is symmetric to the previous one.

\item[(c)] $P\_\oatstar \notin M$. Given that $M$ is minimal, just the last
item in Lemma \ref{lemma:casesst} holds. This means $P\_\oatstarr \notin M$, $P\oa \not \in D_{M}$ and $P\oa \notin D_{M}$. Since the analysis was done for an arbitrary value
$\bar{a}$, $D_{M} \schn (P\ox$ $\rightarrow$ $ Q\oxpy)$
holds.

\item[(d)] $\nit{aux}_\psi\opa \in M$. This means that  $P\_\oatstar \in M$ and
there exists $\bar{b} \not = \overline{\nn}$ with
$Q\oaobtstar \in M$, $Q\_(\bar{a},\fa) \not \in M$, and then,
that $P\oa \in D_M$ and $Q\oaob \in D_M$. Then, the
constraint is satisfied. 
\end{itemize}
\end{itemize} }

\begin{lemma} \label{lemma:satisfaction}  Let
$D$ and $D'$ be instances over the same schema and domain. If $M$ is a
minimal model of $\Pi(\p{P},D)^{M^{\star}_{\mc{C}}(D,D')}$ with
$M \subsetneqq M^{\star}_{\mc{C}}(D,D')$, then there exists $M'$
that is  a minimal model of $\Pi(\p{P},D)^{M'}$ with
$D_{M'} <_\D D'$. \ignore{ $\Delta(D,D_{M'}) \subsetneqq
\Delta(D,D')$.}
\end{lemma}

\dproof{Since $M$ is a minimal model of
$\Pi(\p{P},D)^{\Mdd}$, $P(\bar{a}) \in M$ iff $P\oa \in D$. By
definition of $\Mdd$ and  $M \subsetneqq M^{\star}_{\mc{C}}(D,D')$,
the only two ways that both models can differ is that, for some
$P\oa \in D$, $P\_(\bar{a},\fa) \in M^{\star}_{\mc{C}}(D,D')$ and neither
$P\oa$ nor $P\_(\bar{a},\fa)$ belong to $M$, or for some $P\oa \not \in
D$, $\{P\_(\bar{a},\ta),$ $ P\oatstar,$ $P\_\oatstarr\} \subseteq \Mdd$
and none of $P\oa$, $P\_(\bar{a},\ta),$ $ P\oatstar$, $P\_\oatstarr$  belong to
$M$. \ignore{Now, some of the atoms in $M$ may have not
received an interpretation in terms of $\mathbf{t^{\star
\star}}$ and $\mathbf{f^{\star \star}}$, {\em i.e.} $M$ is
not a minimal model of $(\Pi(D,\ICc))^{M}$. Anyway,} Now,  we can
use the interpretation rules over $M$ to construct
$M'$ that  is a minimal model of $\Pi(\p{P},D)^{M'}$, as  follows:
\begin{enumerate}
\item \noindent If $P(\bar{a}) \in M$ and $P\_(\bar{a},\fa) \not \in M$, then $P(\bar{a}),
P\_\oatstar$ and $P\_\oatstarr \in M'$. \item \noindent If
$P(\bar{a}) \in M$ and $P\_(\bar{a},\fa) \in M$, then $P(\bar{a}), P\_\oatstar$
and $P\_(\bar{a},\fa)  \in M'$. \item \noindent If $P(\bar{a}) \not \in
M$ and $P\_(\bar{a},\ta) \not \in M$, then nothing is added to
$M'$. \item \noindent If $P(\bar{a}) \not \in M$ and $P\_(\bar{a},\ta)
\in M$, then $P\_(\bar{a},\ta)$, $P\_\oatstar$ and $P\_\oatstarr  $ $\in
M'$.
\end{enumerate}

It is clear that $M'$ satisfies the coherence constraints, and is a minimal model of
$\Pi(\p{P},D)^{M'}$.

It just rests to prove that $D_{M'}
<_\D D'$. First, we prove that $D_{M'} \leq_\D D'$. Let us
suppose $P\oa \in \Delta(D,D_{M'})$. Then, either $P\oa \in
D$ and \enlargethispage{\baselineskip} $P\oa \not \in D_{M'}$
or $P\oa \not \in D$ and $P\oa \in D_{M'}$. In the first
case, $P(\bar{a})$, $P\_\oatstar$ and $P\_(\bar{a},\fa)$ belong to $M'$. These
atoms are also in $M$ and, given the only two ways in which
$M$ and $\Mdd$ can differ, they are also in $\Mdd$. Hence,
$P\oa \in \Delta(D,D')$. In the second case, $P\_(\bar{a},\ta)$ and
$P\_\oatstar$ belong to $M'$. These atoms are also in $M$ and,
given the only two ways in which $M$ and $\Mdd$ can differ,
they are also in $\Mdd$. Hence, $P\oa \in \Delta(D,D')$.

We  now prove that $D_{M'} <_D D'$. We know that, for some
fact $P\oa$,  $P\!\_(\bar{a},\ta) \in \Mdd$ and
$P\!\_(\bar{a},\ta) \not \in M$,  or $P\!\_(\bar{a},\fa) \in
\Mdd$ and $P\!\_(\bar{a},\fa) \not \in M$. If $P\_(\bar{a},\fa)$ is in
$\Mdd$ and not in $M$, then, $P\oa \in \Delta(D,D')$, but
$P\oa \not \in \Delta(D,D_{M'})$. Alternatively, if $P\_(\bar{a},\ta)$
and $P\_\oatstar$ belong to $\Mdd$ but not to $M$, then $P\oa
\in \Delta(D,D')$, but $P\oa \not \in \Delta(D,D_{M'})$.
Therefore, $D_{M'} <_D D'$.}

\begin{proposition}  \label{theo:rep-to-mod-st} Given a neighborhood instance $D$ and a
set $\ICc$ of UDECs and RDECs, if $D'$ is a neighborhood solution for $D$ with
respect to $\ICc$, then there is a stable model $M$ of the
program $\Pi(\dbic)^{M}$ with $D_{M}=D'$.
Furthermore, the model $M$ corresponds to $\Mdd$.
\end{proposition}

\dproof{By Lemma \ref{lemma:rep-to-mod-st},
 $\Mdd$ is a  model of
  $\Pi(\dbic)^{\Mdd}$. We now show that
it is minimal. Assume, by contradiction, that there
exists a model $M$ of $\Pi(\dbic)^{\Mdd}$ with  $M
\subsetneqq \Mdd$. We can
assume, without loss of generality, that $M$ is a minimal model. Since $M \subsetneqq \Mdd$, the model
$M$ contains the atom $P(\bar{a})$ iff $P\oa \in D$.

By Lemma \ref{lemma:satisfaction}, there exists model $M'$
such that $\D_{M'} <_\D \D'$ and $M'$ is a minimal model of
$\Pi(\p{P},D)^{M'}$. By Lemma \ref{lemma:mod-to-rep-st},
$D_{M'} \schn \ICc$. This contradicts that $D'$ is a neighborhood solution.}

\begin{proposition} \label{theo:mod-to-rep-st} If $M$ is a  stable model of
$\Pi(\dbic)$, then $D_{M}$ is a neighborhood solution for $D$ with respect to
$\ICc$.
\end{proposition}

\dproof{From Lemma
\ref{lemma:mod-to-rep-st}, it holds $D_{M} \schn \ICc$. We have
 to prove that it is $\leq_{D}$-{\it minimal}. Let us
suppose there is a neighborhood solution $D'$ (that satisfies $\ICc$) with $D' <_{D}D_{M}$.
From
Proposition \ref{theo:rep-to-mod-st}, $\Mdd$ is a stable model
of $\Pi(\dbic)$ and $D_{\Mdd}=D'$ \ (we denote it simple with $M^\star$ in the rest of the proof).

If $D' <_{D} D_{M}$, there is an atom
$P\oa \in \Delta(D,D_{M})$, with  $P\oa \notin \Delta(D,D')$, or there is
an atom $P\oaob \in \Delta(D,D_{M})$, with $\bar{a}',
\bar{b} \neq \nn$,  and an atom
$P(\bar{a},\overline{\nn}) \in \Delta(D,D')$. We analyze both cases:
\begin{enumerate}
    \item $P\oa \in \Delta(D,D_{M})$ and $P\oa \not \in
    \Delta(D,D')$:

     Since $P\oa \in \Delta(D,D_{M})$,  $P\_(\bar{a},\ta)$ or $P\_(\bar{a},\fa)$ belong to $M$. By
    Lemma \ref{lemma:casesst}, there are two cases:
        \begin{itemize}
    \item[(a)] \label{it:prop4item1} $P(\bar{a})$, $P\_\oatstar$ and $P\_(\bar{a},\fa)$ belong to
    $M$, and no other $P\_(\bar{a},v)$, for $v$
    an annotation, belongs to $M$. $P(\bar{a})$, $P\_\oatstar$ and
    $P\_\oatstarr$ belong to $M^{\star}$, and, for any other annotation $v$,
    $P\_(\bar{a},v) \notin M^{\star}$.
    \item[(b)] $P\_(\bar{a},\ta)$, $P\_\oatstar$ and $P\_\oatstarr$
    belong to $M$, and no other $P\_(\bar{a},v)$, for $v$ an annotation, belongs to $M$. No  $P\_(\bar{a},v)$, for $v$ an annotation, belongs to $M^{\star}$.
        \end{itemize}
     If an atom belongs
    to a model $M_1$, e.g. $P\_(\bar{a},\fa)$, and there is another model $M_2$ in
    which it is not present, then there must be in $M_2$
    an atom annotated with $\ta$ or $\fa$ in order to
    satisfy the rule that was satisfied in $M_1$ by
    $P\_(\bar{a},\fa)$.     This implies that $M^{\star}$ has an atom annotated with
    $\ta$ or $\fa$ that does not belong to $M$. This implies
    that there is an atom that belongs to $\Delta(D,D')$ and
    that does not belong to $\Delta(D,D_M)$. We have reached a
    contradiction,
    because $\Delta(D,D')$ is a proper subset of $\Delta(D,D_M)$.

    \item $P\oaob \in \Delta(D,D_{M})$ and $P(\bar{a},\overline{\nn}) \in \Delta(D,D')$:

If $P\oaob \notin M $, then $P\_(\bar{a},\bar{b},\ta) \in M$,
    $P(\bar{a},\overline{\nn}) \notin M$, $P\_(\bar{a},\overline{\nn},\ta) \notin M$.

    Since $P(\bar{a},\overline{\nn}) \in \Delta(D,D')$ and $P(\bar{a},\overline{\nn}) \notin M$,
     $P\_(\bar{a},\overline{\nn},\ta) \in M^\star$.

     Since $P\_(\bar{a},\overline{\nn},\ta) \in M^\star$,
      there must be a rule representing a RDEC
    in $\Pi(\D,\ICc)$ such that $P\_(\bar{a},\overline{\nn},\ta)$ is the only true atom in the head.
    For that rule to be also satisfied by $M$, there must be another
    atom in the head of that rule that is true in $M$ but not in
    $M^\star$. This means that there is a $P(\bar{b}) \in \Delta(D,D_{M})$ and
    $P(\bar{b}) \not \in \Delta(D,D')$, which brings us back to
    case 1. above. Again we obtain a contradiction.
\end{enumerate}
Therefore, it is not possible to have $D' <_\D D_{M}$; and
$D_{M}$ is a neighborhood solution for $D$.}


\end{document}